\documentclass[prd,preprintnumbers,floatfix,
nofootinbib,superscriptaddress]{revtex4}

\usepackage{float}
\usepackage{nicefrac}
\usepackage{mathtools}
\usepackage{amsfonts} % AMS
\usepackage{amssymb} % AMS
\usepackage{amsmath} % AMS
\usepackage{bbold}
\usepackage{graphicx} % Include figure files
\usepackage{subfigure} % Include figure files
\usepackage{array} % array
\usepackage{dcolumn} % Align table columns on decimal point
\usepackage{bm} % bold math
\usepackage{latexsym} % latex symbols
\usepackage{longtable} % long tables
\usepackage{hyperref} % hypertext links 
\usepackage{verbatim}
\usepackage{epsfig}
\usepackage{slashed}
\usepackage{color}

\usepackage{glossaries-extra}
\setabbreviationstyle[acronym]{long-short}
\glssetcategoryattribute{acronym}{nohyperfirst}{true}
\renewcommand{\vec}[1]{\mathbf{#1}}
\newcommand\bra[1]{\langle #1 \vert}
\newcommand\ket[1]{\vert #1 \rangle}

\newcommand{\cB}[0]{\mathcal B}

\newcommand{\cD}[0]{\mathcal D}
\newcommand{\cI}[0]{\mathcal I}
\newcommand{\cK}[0]{\mathcal K}

\newcommand{\cM}[0]{\mathcal M}
\newcommand{\cO}[0]{\mathcal O}

\newcommand{\cS}[0]{\mathcal S}
\newcommand{\cT}[0]{\mathcal T}
\newcommand{\cY}[0]{\mathcal Y}

\newcommand{\Eqref}[1]{(\ref{#1})}

\newcommand{\wh}[0]{\widehat}
\newcommand{\wt}[0]{\widetilde}

\newcommand{\df}[0]{\mathrm{df}}

\newcommand{\uu}[0]{{(u,u)}}

\newcommand{\PV}[0]{{\mathrm{PV}}}

\newcommand{\BHSQC}[0]{Briceno:2017tce}
\newcommand{\BHSnum}[0]{Briceno:2018mlh}
\newcommand{\BHSK}[0]{Briceno:2018aml}
\newcommand{\HSQCa}[0]{Hansen:2014eka}
\newcommand{\HSQCb}[0]{Hansen:2015zga}
\newcommand{\dwave}[0]{Blanton:2019igq}
\newcommand{\largera}[0]{Romero-Lopez:2019qrt}
\newcommand{\HSrev}[0]{Hansen:2019nir}

\newcommand{\HHanal}[0]{Blanton:2019vdk}
\newcommand{\isospin}[0]{Hansen:2020zhy}
\newcommand{\BSQC}[0]{Blanton:2020jnm}
\newcommand{\BSequiv}[0]{Blanton:2020gha}

\newcommand{\Akakia}[0]{Hammer:2017uqm}
\newcommand{\Akakib}[0]{Hammer:2017kms}

\newcommand{\MDpi}[0]{Mai:2018djl}
\newcommand{\HH}[0]{Horz:2019rrn}

\newcommand{\MD}[0]{Mai:2017bge}

\newcommand{\Akakinum}[0]{Doring:2018xxx}
\newcommand{\Maiisobar}[0]{Mai:2017vot}
\newcommand{\isobar}[0]{Jackura:2018xnx}
\newacronym{CMF}{CMF}{center-of-momentum frame}

\begin{document}

%\preprint{\vbox{\hbox{JLAB-THY-19-2945} }}

\title{Relativistic three-particle quantization condition for nondegenerate scalars}

%%%%%%%%%%
\author{Tyler D. Blanton}
\email[e-mail: ]{blanton1@uw.edu}
\affiliation{Physics Department, University of Washington, Seattle, WA 98195-1560, USA}

%%%%%%%%%%
\author{Stephen R. Sharpe}
\email[e-mail: ]{srsharpe@uw.edu}
\affiliation{Physics Department, University of Washington, Seattle, WA 98195-1560, USA}
%%%%%%%%%%

%%%%%%%%%%

%%%%%%%%%%
\date{\today}
%%%%%%%%%%

%%%%%%%%%%
\begin{abstract}
The formalism relating the relativistic three-particle infinite-volume scattering amplitude 
to the finite-volume spectrum has been developed thus far only for identical
or degenerate particles.
We provide the generalization to the case of three nondegenerate scalar particles
with arbitrary masses.
A key quantity in this formalism is the quantization condition, 
which relates the spectrum to an intermediate K matrix.
We derive three versions of this quantization condition,
each a natural generalization of the corresponding results for identical particles.
In each case we also determine the integral equations relating the intermediate K matrix
to the three-particle scattering amplitude, $\cM_3$.
The version that is likely to be most practical involves a single 
Lorentz-invariant intermediate K matrix, $\wt \cK_{\df,3}$.
The other versions involve a matrix of K matrices,
with elements distinguished by the choice of which
initial and final particles are the spectators.
Our approach
should allow a straightforward generalization of the relativistic approach to all other
three-particle systems of interest.
\end{abstract}
%%%%%%%%%%

%\keywords{weak decays, lattice QCD}

\nopagebreak

\maketitle

%%%%%%%%%%%%%%%%%%%%%%%%%%%%%%%%%%%%%%%
%%%%%%%%%                          PAPER                             %%%%%%%%%
%%%%%%%%%%%%%%%%%%%%%%%%%%%%%%%%%%%%%%%

\section{Introduction\label{sec:intro}}

The theoretical formalism needed to study three-particle interactions using lattice QCD (LQCD)
has advanced considerably in recent years~\cite{Polejaeva:2012ut,\HSQCa,\HSQCb,\BHSQC,\Akakia,\Akakib,\MD,\MDpi,\BHSK,Pang:2019dfe,\largera,\isospin,\BSQC,\BSequiv}.
In addition, the formalism has been shown to be a practical tool in simple systems~\cite{\MDpi,\Akakinum,\BHSnum,\dwave},
and applied to LQCD results for the $3\pi^+$~\cite{\HH,Mai:2019fba,Culver:2019vvu,Fischer:2020jzp,Hansen:2020otl}
and $3K^-$ systems \cite{Alexandru:2020xqf},
as well as to the $\phi^4$ theory~\cite{Romero-Lopez:2018rcb,Romero-Lopez:2020rdq}.
For recent reviews, see Refs.~\cite{\HSrev,Rusetsky:2019gyk}.\footnote{%
For alternative approaches, see Refs.~\cite{Briceno:2012rv,Guo:2017ism,Klos:2018sen,Guo:2018ibd,Guo:2020kph,Guo:2020spn}.
}

The relativistic formalism has so far only been developed for degenerate scalars.\footnote{%
The only nondegenerate three-particle formalism
of which we are aware is the very recent Ref.~\cite{Pang:2020pkl},
in which the $DDK$ system is studied using the nonrelativistic effective field theory approach
of Refs.~\cite{\Akakia,\Akakib}, and assuming only $s$-wave two-particle interactions.}
Within the generic relativistic effective field theory (RFT) approach, which we adopt here,
the initial development was for identical scalars with a G-parity-like $\mathbb Z_2$ 
symmetry~\cite{\HSQCa,\HSQCb}, 
with the extension to theories without the $\mathbb Z_2$ symmetry presented in Ref.~\cite{\BHSQC}, 
and that to nonidentical but degenerate scalars (e.g.~three pions with all allowed total isospins) 
given in Ref.~\cite{\isospin}.
An additional generalization to allow the inclusion of poles in the two-particle K matrices
was given in Refs.~\cite{\BHSK,\largera}.
Alternative approaches have been developed using either nonrelativistic effective field 
theory (NREFT)~\cite{\Akakia,\Akakib},
or the application to finite volume of a unitary representation of the three-particle scattering
amplitude~\cite{\MD,\MDpi}.
Both approaches have so far only considered identical scalars, and also only s-wave two-particle interactions.
Recently,  the RFT and finite-volume unitarity approaches have been shown to be equivalent~\cite{\BSequiv}.

In this work we generalize the RFT approach to nondegenerate scalar particles.
We derive three forms of the three-particle quantization condition,
Eqs.~(\ref{eq:QC1}), (\ref{eq:QC2}), and (\ref{eq:QC3}),
each with associated integral equations relating the intermediate K matrices
to the three-particle scattering amplitude, $\cM_3$.

The first form is derived
using the simplified method,
based on time-ordered perturbation theory (TOPT),
that we introduced recently in Ref.~\cite{\BSQC}, 
a reference henceforth referred to as BS1.
The quantization condition involves the nondegenerate
generalization of the asymmetric three-particle K matrix used in BS1,
and for this reason we refer to it as ``asymmetric.''
This generalization is a three-dimensional flavor matrix of K matrices, denoted $\wh \cK_{\df,3}$.
A significant disadvantage of this approach is that the K matrices are not Lorentz invariant
(although the formalism is valid for relativistic kinematics).

The second form of the quantization condition resolves this shortcoming,
as it involves a flavor matrix of Lorentz-invariant K matrices.
Its derivation follows the original RFT works~\cite{\HSQCa,\HSQCb} in using Feynman diagrams,
but, compared to those works, rearranges the order in which the diagrams are analyzed,
and the manner in which finite- and infinite-volume quantities are related.
The guiding principle is to mirror, at every step, the form of the TOPT analysis, so that the algebraic
simplifications in the latter approach carry over.
For this reason the resulting quantization condition is also asymmetric.
This leads to the major disadvantage of the resulting formalism 
(shared with the TOPT form),
namely that it depends on {\em nine} intermediate three-particle K matrices, 
collected in the matrix denoted $\wh \cK'_{\df,3}$,
which are distinguished by the choice of spectator flavors for incoming and outgoing
particles.

This disadvantage is resolved by the final form of the quantization condition.
By algebraic manipulations that generalize the (anti)symmetrization procedure introduced in BS1,
we are able to take the second form of the quantization condition and reexpress it
in terms of a {\em single}, symmetrized Lorentz-invariant three-particle K matrix, $\wt \cK_{\df,3}$.
The resulting ``symmetric'' quantization condition,
given in Eq.~(\ref{eq:QC3}), 
provides the natural generalization of that derived in Ref.~\cite{\HSQCa}
for identical particles, and indeed the two have very similar forms.
We expect that this final form will be the most useful in practice.
Given a technical assumption, it is also possible to obtain this final form by 
applying the same symmetrization procedure to the TOPT form 
of the quantization condition.
In this way one can avoid the intermediate step involving Feynman diagrams.

We stress that no truncation of the two-particle angular momenta is needed
in any of the derivations.
The only approximation made is to drop terms that are exponentially suppressed in the box size $L$.

Our main focus in this work is the presentation of
a theoretical framework that will be straightforward to generalize
to all three-particle systems of interest, e.g.~those involving multiple three-particle channels,
and ``$2+1$'' systems involving two identical particles plus a third that is different (e.g.~$N\pi\pi$).
The applications of the specific results we present here to QCD are relatively limited.
They require each particle to carry a different
combination of flavors in such a way that there is  only one allowed three-particle state,
e.g.~the $D_s^+ D^0 \pi^-$ and $D_s^+ D^0 D^+$ systems.
We do not discuss here the practical implementation of the new formalism,
which we expect to involve a straightforward  generalization of previous implementations of the 
RFT approach~\cite{\BHSnum,\dwave,\largera,\HH,Hansen:2020otl}.

The derivation we present here is lengthy, and the logic and necessity of the
various steps may be difficult to follow. Thus we provide, in a brief first section, a road map to
the derivation, which also serves to present the organization of the paper.
We only note here that conclusions and directions for future work are presented
in Sec.~\ref{sec:conc}.

\section{Summary of the steps of the derivation}
\label{sec:recipe}
Here we provide a summary of the approach that we follow in this work, which also serves to
 provide a ``recipe'' for future generalizations.
\begin{enumerate}
\item 
Choose the desired three-particle state of interest,
e.g.~$3\pi^+$, $D_s^+D^0\pi^-$, $K^+\pi^+\pi^0$, \dots.
Consider a finite-volume correlator $C_{3,L}(E,\vec P)$ with operators 
coupling to this state, restricting the overall 4-momentum $P^\mu=(E,\vec P)$ so that only said state is kinematically allowed to go on shell.
This is discussed in Sec.~\ref{sec:setup}.
\item 
Working in a generic relativistic effective field theory describing the interactions of the particles under
consideration, express $C_{3,L}$ as an infinite sum of diagrams in TOPT.
Organize them by number of ``relevant" cuts---sections consisting of the three-particle state of interest---while taking the infinite-volume limit of all sections involving ``irrelevant" cuts. The result is a simple geometric series for the correlator involving off-shell generalized infinite-volume kernels.
Project the kernels on shell to rewrite $C_{3,L}$ in terms of on-shell  two and
three-particle K-matrices, here $\cK_2$ and $ \wh{\cK}_{\df,3}$, 
and known finite-volume kinematic quantities associated with the relevant cuts.

These steps are presented here in Sec.~\ref{sec:derivation}, with some details
relegated to Appendix~\ref{app:not},
and are mostly a straightforward
generalization of the analysis of BS1 for identical particles. 
%This work culminates in the quantization condition given in Eq.~(\ref{eq:QC1}).

\item 
At this point one obtains a quantization condition relating the finite-volume energy spectrum to the K matrices, here given in  Eq.~(\ref{eq:QC1}). It will, however, involve a K matrix that is
asymmetric under particle interchange and is not Lorentz invariant.

\item 
A shortcut is now available, based on an assumption explained in Sec.~\ref{sec:symTOPT}.
This uses symmetrization identities, introduced in Sec.~\ref{sec:symmetric}, 
to rewrite the quantization condition in terms of a symmetric, three-particle K matrix 
(denoted here $\wt \cK_{\df,3}$). 
Here this form of the quantization condition is given in Eq.~(\ref{eq:QC3}).
The K matrix obtained in this way is also Lorentz invariant.

Two appendices fill in steps in the derivation of Sec.~\ref{sec:symmetric}.
Appendix~\ref{app:symid} derives the symmetrization identities,
while Appendix~\ref{app:alg}
gives further details of the steps needed to derive the quantization condition.

\begin{comment}
\item 
Although the assumption of Sec.~\ref{sec:symTOPT} is plausible, it has not been demonstrated.
An alternative approach to obtain the final quantization condition without assumptions
replaces TOPT with a method using Feynman diagrams.
While more complicated than the TOPT approach, it is simpler and more explicit than
the original method of Ref.~\cite{\HSQCa}.
This new method is presented in Sec.~\ref{sec:derivation2}, with details  of the
derivation given in the associated Appendix~\ref{app:Feynman}.
It involves an asymmetric but Lorentz-invariant K matrix, here $\wh \cK'_{\df,3}$,
and leads to an intermediate, asymmetric form of the quantization condition,
here Eq.~(\ref{eq:QC2}).
This is then converted into the symmetric form using the same symmetrization identities
as for the TOPT result, as explained in Sec.~\ref{sec:symmetric}.
\end{comment}

\item 
The K matrices can be related
to the physical scattering amplitudes $\cM_2$ and $\cM_3$
via nested integral equations. How this is done is explained in detail
for the asymmetric, non-Lorentz-invariant K matrix $\wh{\cK}_{\df,3}$ in Sec.~\ref{sec:KtoM}.
For the final form of the quantization condition the integral equations are
sketched briefly in Sec.~\ref{sec:KtoMsymm}.
\end{enumerate}

Although the assumption
mentioned in step 4
is plausible, it has not been demonstrated.
An alternative approach to obtain the final quantization condition without assumptions
replaces TOPT with a method using Feynman diagrams.
While more complicated than the TOPT approach, it is simpler and more explicit than
the original method of Ref.~\cite{\HSQCa}.
This new method is presented in Sec.~\ref{sec:derivation2}
and mirrors steps 2-5 of the TOPT recipe,
with details of the
derivation given in the associated Appendix~\ref{app:Feynman}.
It involves an asymmetric but Lorentz-invariant K matrix, here $\wh \cK'_{\df,3}$,
and leads to an intermediate, asymmetric form of the quantization condition,
here Eq.~(\ref{eq:QC2}).
This is then converted into the symmetric form using the same symmetrization identities
as for the TOPT result, as explained in Sec.~\ref{sec:symmetric}.
Both $\wh \cK'_{\df,3}$ and its symmetrized version can be related to the
physical amplitude $\cM_3$ via integral equations, as shown in Secs.~\ref{sec:KtoMFeynman} and \ref{sec:KtoMsymm}, respectively.

\section{Setup and overview}
\label{sec:setup}
For the derivations presented below we work in the following theoretical setup.
Our theory has three real scalar fields, $\phi_i$, $i=1,2,3$,
with the Lagrangian having 
the most general Lorentz-invariant form that is symmetric under $\phi_i\to-\phi_i$ for each field separately.
We describe the fields as having different
``flavors,'' although the associated symmetry is $\mathbb Z_2$
rather than the usual $U(1)$.
We label the physical masses of the particles $m_i$, and assume, without loss of generality, the
ordering $m_1 \le m_2 \le m_3$.
We determine the quantization condition from the poles in the correlator
\begin{align}
	C_{3,L}(E,\vec{P}) \equiv \int_L d^4x \; e^{i(Ex^0 - \vec{P}\cdot\vec{x})} 
	\langle0|\text{T}\sigma_{123}(x)\sigma_{123}^\dagger(0)|0\rangle \,,
\end{align}
where $\sigma_{123}^\dagger\sim \phi_1\phi_2 \phi_3$ 
is an operator that creates states having an odd number of each of the flavors.
We do not need to specify the spatial form of the operator, except to note that we allow the three fields
to be spatially separated in such a way that no rotational quantum numbers are excluded.
The theory lives in a cubic spatial box of length $L$, as indicated by the subscript on the integral sign,
and periodic boundary conditions are assumed.
The total four-momentum flowing through the correlator is $(E,\vec P)$, with 
$\vec P$ lying in the finite-volume set, $\vec P\in (2\pi/L) \mathbb Z^3$.
In the overall center of mass frame (CMF), the energy is $E^* =\sqrt{E^2-\vec P^2}$.

To ensure that the only intermediate states that can go completely on shell are those consisting of exactly
three particles (one of each flavor), we impose the following restriction:
\begin{equation}
0 < E^* < 3 m_1 + m_2 + m_3\,.
\label{eq:Estrange}
\end{equation}
Here we are using the mass ordering assumed above, such that the lowest-energy 
on-shell five-particle state involves the addition of two particles of flavor 1. 
Because of the $\mathbb Z_2^3$ symmetry, there are no possible single-particle
intermediate states, so the lower bound on $E^*$ is lower than in the identical-particle case
(where $m < E^* < 5 m$).

In the first derivation below, we follow the same strategy as in BS1.
We begin by considering all Feynman diagrams contributing to
$C_{3,L}$ and then, after an initial analysis dealing with self-energy diagrams (discussed in Appendix~A
of BS1 and carrying over essentially unchanged to the present analysis), convert to time-ordered diagrams. 
The rules for such diagrams are summarized in Sec.~IIA of BS1.
We only note here that the factor associated with a propagator of flavor $i$ and momentum $\vec p_i$ is
$1/(2\omega_{p_i})$, where $\omega_{p_i}=\sqrt{\vec p_i^2+ m_i^2}$, with $m_i$ the physical
(not bare) mass.

Given the kinematic restriction \Eqref{eq:Estrange}, 
the only singular behavior in TOPT diagrams arises from energy denominators
for intermediate states  (``cuts'') containing three particles, and thus having the 
form $1/(E-\omega_{p_1}-\omega_{p_2}-\omega_{p_3})$.
We refer to these intermediate states as ``relevant cuts.''
All other (``irrelevant") cuts lead to sums over internal momenta (which lie in the finite-volume set)
with nonsingular summands, which can therefore be replaced by integrals, 
up to exponentially suppressed corrections, 
which are typically of the form $\sim \exp(-m_i L)$.
We assume throughout this work that such corrections can be neglected.

In the second derivation below, we work entirely with Feynman diagrams.
As in Ref.~\cite{\HSQCa}, we do begin by using TOPT to justify that power-law volume effects
come only from three-particle intermediate states, but the actual analysis does not use TOPT.

We close this overview by discussing the generality of the results that we derive.
The consideration of a theory with a $\mathbb Z_2^3$ symmetry is convenient---reducing the number of
diagrams that contribute to the skeleton expansion of the correlator $C_{3,L}$---but not necessary.
Once the dust of the derivations has settled, it becomes clear
that the only necessary criterion is for there to be
a range of $E^*$ in which the only allowed on-shell intermediate state consists of one of each flavor
of particle. Thus, for example, the derivation applies also to the $D^s D^0 \pi^-$ system,
since the quark compositions, $(c \bar s) (c \bar u) (d \bar u)$, constrain the flavor such that
there are no other intermediate states until one reaches the $D^0 D^0 K^0$ threshold.
Our formalism is valid in this case for  $0<E^* <  2 M(D_0)+M(K_0) =4224\, {\rm MeV}$.
In practice, of course, aside from the possibility of bound states, the lower limit
of interest is $E^* \approx M(D^s)+M(D^0)+M(\pi^-) = 3973\, {\rm MeV} $, so
there is a small kinematic range of applicability.
Other similar examples are discussed in the conclusions.

\section{Derivation of quantization condition using TOPT}
\label{sec:derivation}

\begin{figure}[tb]
\begin{center}
\vspace{-10pt}
\includegraphics[width=\textwidth]{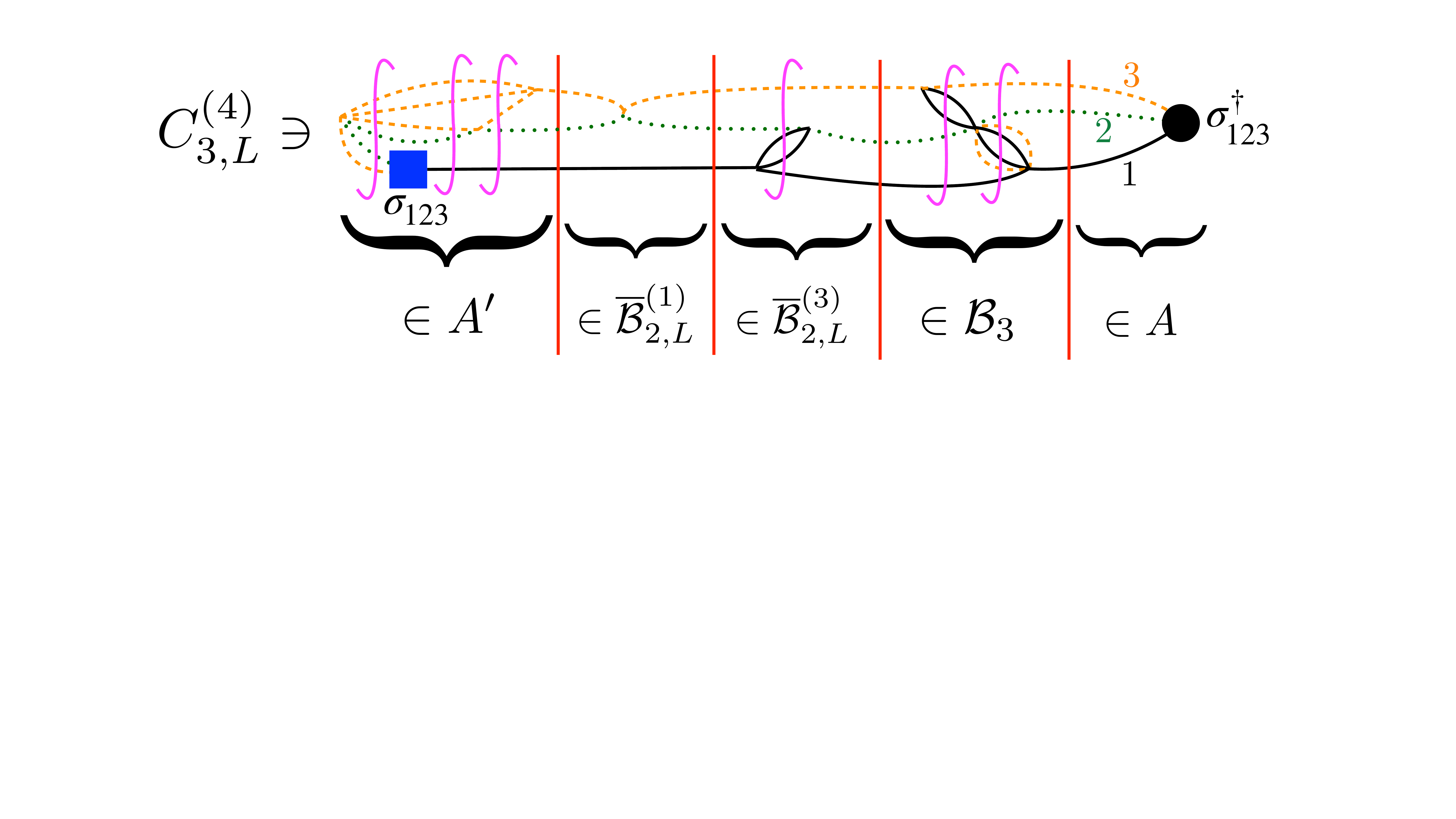}
\vspace{-2.3truein}
\caption{
Example of a contribution to the correlator $C_{3,L}$ in TOPT, illustrating the 
segments that appear. The solid (red) vertical lines indicate relevant (three-particle) cuts,
which are associated with factors of $D$. Time runs from right to left. Cuts that cannot go on shell
are indicated by the (magenta) integral signs. The three flavors are denoted, respectively, by
(black) solid lines, (green) dotted lines, and (orange) dashed lines.
We note that all vertices, which arise from the generic EFT, contain an even number of each
flavor of particle. We also stress that vertices are allowed to lie outside of the time
interval bracketed by $\sigma_{123}(t)$ and $\sigma_{123}^\dagger(0)$, an example being
the six-point vertex at the left of the diagram.
\label{fig:CL}
}
\end{center}
\end{figure}

In this section we derive the three-particle quantization condition for nondegenerate scalars
using the TOPT-based approach of BS1.

An example of the TOPT diagrams contributing to $C_{3,L}$ is shown in  Fig.~\ref{fig:CL}.
As in BS1, we can divide the diagrams into segments separated by relevant cuts.
A simplification compared to BS1 is that we do not need to keep track of symmetry or relabeling factors. 
The four types of segment that appear (all illustrated in the figure) are:\footnote{%
We make one notational change compared to BS1, namely removing the hats on $A'$ and $A$.
This allows us to reintroduce hats below as a notation for the matrix forms of the various kernels.}
\begin{align}
\begin{split}
\text{left endcap:}&\ \
 A'(\{\vec p\})\,,
\\
\text{right endcap:}&\ \
 A(\{\vec k \})\,,
\\
\text{3PIs kernel:}&\ \
i\cB_3(\{\vec p \}; \{\vec k\})\,,
\\
\text{2PIs kernels:}&\ \
i\overline\cB_{2,L}^{(i)}(\{\vec p \}; \{\vec k\})
 = 2\omega_{p_i} L^3 \delta_{\vec p_i \vec k_i} 
i\cB_2^{(i)}(\vec p_j,\vec p_k;\vec k_j,\vec k_k)\,.
\end{split}
\label{eq:TOPTkernels}
\end{align}
Here we are using the notation 
\begin{equation}
\{\vec p\} \equiv \{\vec p_1,\vec p_2, \vec p_3\} \,,\ \ 
\{\vec k\} \equiv \{\vec k_1,\vec k_2, \vec k_3\} \,, \ \ {\rm etc.}
\label{eq:momlist}
\end{equation}
for sets of three spatial momenta.
The subscript on each momentum indicates the flavor of the particle, as do the superscripts on 
$\overline \cB_{2,L}^{(i)}$ and $\cB_2^{(i)}$.
Where we use the triplet of flavor labels, $i$, $j$, and $k$, they are assumed to be in cyclic order.
The quantities $A'$, $A$, $\cB_3$, and $\cB_2$ can all be evaluated in infinite volume,
i.e.~with momentum sums replaced by integrals.
Since, by construction, $\cB_3$ has no three-particle cuts, it is three-particle irreducible in the
s channel (3PIs). Similarly, the two-to-two kernels $\cB_2^{(i)}$ are 2PIs.
Finally, we note that all kernels depend implicitly on the input energy $E$.

There are several changes in these kernels compared to those needed for identical particles in BS1.
First, there are three types of two-particle kernels, $\overline\cB_{2,L}^{(i)}$,
corresponding to the three choices of flavor $i$ of the spectator particle.
Second, the kernel $\cB_3$ is defined here without any overall factor, whereas the quantity of the
same name in BS1 (but which applies to identical particles) is defined to include a factor of $1/9$.
Third, the endcaps $A'$ and $A$ here have no overall factor, while in BS1 the corresponding quantities
(denoted $\wh A'$ and $\wh A$)  include a factor of $1/3$.
Finally, we use a redundant labeling for the allowed discrete choices of momenta,
listing all three, as in Eq.~(\ref{eq:momlist}),
although they are constrained to sum to $\vec P$.
This is a notational convenience, and avoids picking out arbitrarily one of the momenta.
Although the kernels are infinite-volume quantities, defined for all external momenta,
within the correlator the momenta are in the finite-volume set.
Thus we treat the sets $\{\vec p\}$, $\{\vec k\}$, etc.~as matrix indices,
with $ A'$ a row vector, $\overline \cB_{2,L}^{(i)}$ and $\cB_3$ matrices,
and $ A$ a column vector. 
In the following, these indices are implicitly summed, subject to the above-mentioned constraint.

The correlator can be written as a sum over terms containing different numbers of
relevant cuts. Between each cut one can have either the kernel $\cB_3$
or one of the $\overline{\cB}_{2,L}^{(i)}$.
This leads immediately to the geometric series
\begin{align}
C_{3,L} &= C_{3,L}^{(0)} +
A' iD \sum_{n=0}^\infty 
\left[(i \overline\cB_{2,L}^{(1)} + i \overline\cB_{2,L}^{(2)} + i \overline\cB_{2,L}^{(3)} + i \cB_3 ) iD \right]^n A 
\label{eq:C3Lres1}
\\
&= C_{3,\infty}^{(0)} +
A' i D\frac1{1 -
(i \overline\cB_{2,L}^{(1)} + i \overline\cB_{2,L}^{(2)} + i \overline\cB_{2,L}^{(3)} + i \cB_3 ) i D} A\,.
\label{eq:C3Lres2}
\end{align}
Here $C_{3,L}^{(0)}$ is the contribution with no relevant cuts, which can be converted into an infinite-volume
quantity up to exponentially-suppressed corrections.
The matrix $D$ is associated with relevant cuts and is given by
\begin{equation}
D(\{\vec p\};\{\vec k\}) 
= 
\mathbb 1_{\vec p,\vec k}
\frac1{L^6} \frac1{8 \omega_{p_1} \omega_{p_2}\omega_{p_3}} \frac1{E-\omega_{p_1}-\omega_{p_2}-\omega_{p_3}}
\,,
\label{eq:D}
\end{equation}
where 
\begin{equation}
\mathbb 1_{\vec p,\vec k} \equiv \delta_{\vec p_1 \vec k_1} \delta_{\vec p_2 \vec k_2} \delta_{\vec p_3 \vec k_3}\,.
\label{eq:onep}
\end{equation}
The third Kronecker delta is redundant and could be dropped---we include it to emphasize
the symmetry of the expression.
The simplicity of the result in Eq.~(\ref{eq:C3Lres1}) shows the advantage of the TOPT approach.

At this stage, the TOPT kernels are, in general, off shell, i.e.~$E\ne \omega_{p_1}+\omega_{p_2}+\omega_{p_3}$
for each intermediate cut. The next step is to include an on-shell projection.
The procedure for doing so in the degenerate case is described in detail in BS1, following the
analysis of Ref.~\cite{\HSQCa}. The only changes needed here are kinematical, and are explained below.
The nature of the on-shell projection depends on the form of the kernels adjacent to factors of $D$.
In particular, between two-particle kernels with the same spectator flavor,
i.e.~$\overline \cB_{2,L}^{(i)} D \overline \cB_{2,L}^{(i)}$, 
are relevant cuts with a common spectator particle\footnote{%
Note that there are no diagrams in $\overline \cB_{2,L}^{(i)} D \overline \cB_{2,L}^{(i)}$ where the spectator particle switches between the kernels, as this would require two of the three particles in the relevant cut to have the same flavor $i$.}
(``F cuts"),
while if the flavors differ, as in
$\overline \cB_{2,L}^{(1)} D \overline \cB_{2,L}^{(2)}$,
the relevant cuts all have the spectator particle switch between the kernels (``G cuts").
Between two- and three-particle kernels, or between a pair of three-particle kernels, one can use either type of cut,
or a linear combination thereof, and we use this freedom to give a compact expression for subsequent results.
In particular, we find it convenient to add an additional layer of matrix indices, corresponding
to the flavor of the spectator particle. 
This allows us to rewrite Eq.~\eqref{eq:C3Lres2} as
\begin{equation}
C_{3,L} - C_{3,\infty}^{(0)} =
\wh A' i \wh D \frac1{1- i \wh \cB i \wh D} \wh A \,,
\label{eq:C3Lshort}
\end{equation}
where
\begin{align}
\wh A' &\equiv \frac{A'}{3}  \bra1 \,, 
\label{eq:Aphat}
\\
\wh A &\equiv \frac{A}{3} \ket{1}\,,
\label{eq:Ahat}
\\
\wh \cB &\equiv \wh{\overline{\cB}}_{2,L} + \ket1 \frac{\cB_3}9  \bra1\,,
\label{eq:Bhat}
\\
\wh D &\equiv \ket1  D \bra1  \,,
\label{eq:Dhat}
\end{align}
with
\begin{equation}
\wh{\overline{\cB}}_{2,L} \equiv {\rm diag}(
\overline\cB_{2,L}^{(1)}, \, \overline\cB_{2,L}^{(2)}, \, \overline\cB_{2,L}^{(3)}) \,,
\label{eq:B2Lhat}
\end{equation}
and  the (unnormalized) vector $\bra1$ given by
\begin{equation}
\bra1\equiv 
\begin{pmatrix} 1, & 1, &1 \end{pmatrix} \,,
\qquad \ket1\equiv \bra{1}^T \,.
\end{equation}
We stress that all quantities still have implicit momentum indices as well as the explicit flavor indices.

The extra flavor matrix structure allows us to implement different cuts depending on the nature
of the adjacent kernels.
Specifically, on-shell projection is effected by rewriting $\wh D$ as
\begin{align}
\wh D &= \wh F_G + \wh{\delta F}_G\,,
\label{eq:Ddecomp}
\\
\wh F_G &=
\begin{pmatrix}
\wt F^{(1)} & \wt G^{(12)} P_L  & P_L\wt G^{(13)} \\
P_L \wt G^{(21)} & \wt F^{(2)} & \wt G^{(23)} P_L \\
\wt G^{(31)} P_L & P_L \wt G^{(32)} & \wt F^{(3)} 
\end{pmatrix}\,,
\label{eq:FGhat}
\\
\wh{\delta F}_G &=
\begin{pmatrix}
\wt I_F^{(1)} & \delta \wt G^{(12)} &\delta \wt G^{(13)} \\
\delta \wt G^{(21)} & \wt I_F^{(2)} & \delta \wt G^{(23)} \\
\delta \wt G^{(31)} & \delta \wt G^{(32)} & \wt I_F^{(3)} 
\end{pmatrix}\,,
\label{eq:deltaFGhat}
\end{align}
where the objects in $\wh F_G$ project adjacent kernels on shell, while those in $\wh{\delta F}_G$ are integral
operators that sew together adjacent kernels into new infinite-volume quantities.
The notation for these objects is the same as in BS1, except that here there are superscripts
indicating the flavors of the spectator particles.
In particular,
\begin{equation}
\left[\wt F^{(i)}\right]_{p_i \ell' m';k_i \ell m} =
\delta_{\vec p_i \vec k_i} \frac{H^{(i)}(\vec p_i)}{2\omega_{p_i} L^3}
\left[ \frac1{L^3} \sum_{\vec p_j}^{\rm UV} - \PV \int^{\rm UV} \frac{d^3 p_j}{(2\pi)^3} \right]
\frac{\cY_{\ell' m'}(\vec p_j^{*(p_i)})}{q_{2,p_i}^{*\ell'}}
\frac1{4\omega_{p_j} \omega_{p_k}(E-\omega_{p_1}-\omega_{p_2}-\omega_{p_3})}
\frac{\cY_{\ell m}(\vec p_j^{*(p_i)})}{q_{2,p_i}^{*\ell}}
\label{eq:Ft}
\end{equation}
is a generalized L\"uscher zeta function, and
\begin{equation}
\left[\wt G^{(ij)}\right]_{p_i \ell' m';k_j \ell m} = 
\frac1{2\omega_{p_i} L^3}
\frac{\cY_{\ell' m'}(\vec k_j^{*(p_i)})}{q_{2,p_i}^{*\ell'}}
\frac{H^{(i)}(\vec p_i) H^{(j)}(\vec k_j)}{b_{ij}^2-m_k^2}
\frac{\cY_{\ell m}(\vec p_i^{*(k_j)})}{q_{2,k_j}^{*\ell}}
\frac1{2\omega_{k_j} L^3}
\label{eq:Gt}
\end{equation}
is a generalized switch factor,\footnote{%
Here we are using the relativistic form of the energy denominator, which is an allowed choice,
as explained in BS1, and is needed when we construct the fully Lorentz-invariant form of 
three-particle K matrix below. For $\wt F^{(i)}$ we keep the nonrelativistic form of the denominator
for notational brevity; the change to the relativistic form only changes $\wt F^{(i)}$ by
exponentially suppressed contributions.}
with the four-vector $b_{ij}$ given by
\begin{equation}
b_{ij} = (E-\omega_{p_i}-\omega_{k_j},\vec P-\vec p_i-\vec k_j)\,.
\end{equation}
The parity operators are a new feature here and are given by
\begin{equation}
\left[P_L\right]_{p_i\ell'm';k_i\ell m} = \delta_{p_i\ell'm';k_i\ell m} (-1)^\ell \,.
\label{eq:PLdef}
\end{equation}
The integral operators $\wt I_F^{(i)}$ and $\delta\wt G^{(ij)}$ are then defined by the difference 
$\wh D-\wh F_G$, and explicit forms are not needed. 
The discussion of their general properties in BS1
remains valid here, and we do not repeat it.

We now explain the notation in Eqs.~(\ref{eq:Ft}) and \Eqref{eq:Gt}.
We begin with the matrix indices on both $\wt F^{(i)}$ and $\wt G^{(ij)}$,
which are of similar form to those used in all previous RFT quantization conditions. 
A key property of $\wt F^{(i)}$ and $\wt G^{(ij)}$ is that they project adjacent kernels---here 
the elements of $\wh B$, $\wh A'$, and $\wh A$---on shell.
This projection changes the matrix indices from $\{\vec k\}$ to
$\{k_i \ell m\}$, with the latter denoting an on-shell, three-particle state.
The new feature for nondegenerate particles
is that there are three choices of indices, labeled by $i=1-3$. 
Here $i$ is the flavor of the particle chosen as the ``spectator,''
and $k_i$ is shorthand for its momentum, $\vec k_i$, 
which is drawn from the finite-volume set.
The remaining two (``nonspectator'') particles,
whose flavors are denoted $j$ and $k$,
are then boosted to their center-of-mass frame (CMF), 
in which the kernel is decomposed into spherical harmonics.
Denoting a generic on-shell kernel by $X(\{\vec k\})$---this could, for example, be 
$B_3(\{\vec p\};\{\vec k\})$, with the $\{\vec p\}$ index left implicit,
and with $\{\vec k\}$ restricted so that the three particles are on shell
for the given $E$ and $\vec P$---this decomposition is
\begin{equation}
X(\{\vec k\}) = \sum_{\ell m} 
X(\{\vec k\})_{k_i \ell m} \sqrt{4\pi} Y_{\ell m}(\wh{\vec k}_j^{*(k_i)})\,.
\label{eq:basischange}
\end{equation}
Here $\wh{\vec k}_j^{*(k_i)}$ is the unit vector in the direction of $\vec k_j^{*(k_i)}$,
which itself is the spatial part of the four-vector 
obtained by boosting $(\omega_{k_j},\vec k_j)$ into the CMF
of the nonspectator pair.
Details of the boost are discussed in Appendix~\ref{app:not}; we stress that,
for an on-shell quantity, the two boosts discussed there are equivalent.
In Eq.~(\ref{eq:basischange}),
we must specify which flavor from the pair is used to define the harmonic decomposition.
This is because $\vec k_j^{*(k_i)} = - \vec k_k^{*(k_i)}$, 
implying that $Y_{\ell m}(\wh{\vec k}_j^{*(k_i)}) = (-1)^\ell Y_{\ell m}(\wh{\vec k}_k^{*(k_i)})$.
Since both even and odd waves are present for nondegenerate particles,
the decompositions with respect to flavors $j$ and $k$ differ.
Our convention for the decomposition of kernels is that it is done relative
to the direction of the particle whose flavor follows cyclically after that of the spectator.

Returning to the definition of $\wt F^{(i)}$, Eq.~(\ref{eq:Ft}), we note that
the spectator flavor is chosen to be $i$ for both incoming (right-hand) and
outgoing (left-hand) indices. We follow the convention just described in choosing the flavor
used for spherical harmonic decompositions, namely that $j$ follows cyclically after $i$.
Note that, even though $\vec p_j$ is a dummy variable, this choice has content
because it specifies which mass to use when calculating $\omega_{p_j}$.
The third momentum, needed to determine $\omega_{p_k}$,
is then given by $\vec p_k= \vec P-\vec p_i - \vec p_j$.
The boosted momentum $\vec p_j^{*(p_i)}$ is defined in the same way
as $\vec k_j^{*(k_i)}$.
Here the three particles are in general off shell, so that
the two boosts discussed in Appendix~\ref{app:not} differ.
However, as discussed in BS1, the difference leads only to exponentially suppressed
shifts in $\wt F^{(i)}$.
The harmonic polynomials are defined by
\begin{equation}
\cY_{\ell m}(\vec{a})
= \sqrt{4\pi} Y_{\ell m}(\widehat{\vec a}) |\vec a|^\ell\,,
\label{eq:harmonicpoly}
\end{equation}
with the spherical harmonics chosen to be in the real basis.
The quantity $q_{2,p_i}^*$ is given by 
\begin{align}
4 q_{2,p_i}^{*2}  &= \frac{\lambda(\sigma_i,m_j^2, m_k^2)}{\sigma_i}\,,
\\
\sigma_i &= (E-\omega_{p_i})^2 - (\vec P-\vec p_i)^2\,,
\end{align}
where $\lambda(a,b,c) = a^2+b^2+c^2 -2 ab -2 ac - 2 bc$ is the standard triangle function.
$q_{2,p_i}^*$ is the magnitude of the spatial momenta of each of the interacting pair
for fully on-shell kinematics, i.e.~if $E=\sum_i \omega_{p_i}$. This requires that the momenta
of the pair not lie in the finite-volume set. 
The superscript UV on the sum and integral in $\wt F^{(i)}$ indicate an ultraviolet regularization,
 the nature of which affects $\wt F^{(i)}$ only at the level of exponentially suppressed terms.
 The sum over $\vec p_j$  runs over the finite-volume set,
 while the integral is defined by the generalized principal-value (PV) pole prescription
introduced in Ref.~\cite{\largera}.

Turning now to $\wt G^{(ij)}$, Eq.~(\ref{eq:Gt}), here the
incoming and outgoing spectator indices differ, the former being $j$ and the latter $i$.
In this case the harmonic decompositions are done using a different convention from
that used above,
so as to conform, in the degenerate limit, with the definitions used in previous RFT works.
On the outgoing side, with spectator flavor $i$, the
decomposition is done relative to the direction of the particle of flavor $j$ in the pair CMF,
as indicated by the argument of $\cY_{\ell' m'}$ in Eq.~(\ref{eq:Gt}) being $\vec k_j^{*(p_i)}$.
Similarly, with the incoming flavor $j$, 
the decomposition is done relative to the direction of the particle of flavor $i$. 
In one of these two cases, the ordering is not cyclic, and thus there is a mismatch between 
the convention used for $\wt G^{(ij)}$ and the adjacent kernels.
The factors of $P_L$ correct this mismatch, as described in more detail shortly.
To define the arguments of the harmonic polynomials in $\wt G^{(ij)}$,
we must, at this stage, use the Wu boost, which is 
one of the two boosts discussed in Appendix~\ref{app:not}.

Both $\wt F^{(i)}$ and $\wt G^{(ij)}$ contain the cutoff functions $H^{(i)}(\vec p_i)$.
These are smooth functions whose role is to cut off the sums over spectator momenta
in the region where the three particles lie far below threshold. 
They are generalizations to nondegenerate kinematics
of the cutoff function introduced for identical particles in Ref.~\cite{\HSQCa}, 
and their technical properties are discussed in Appendix~\ref{app:not}.
We stress two features of these functions. First, their introduction (discussed in detail
in BS1) is an intrinsic part of the derivation, and not an ad hoc feature.
Second, they introduce a scheme dependence into intermediate quantities that appear in
the quantization condition that is derived below, specifically into $\cK_2$ and 
$\wh{\cK}_{\df,3}$.
This scheme dependence cancels, however, in the spectrum that is predicted by
the quantization condition, and in the relation of these intermediate quantities
to the three-particle scattering amplitude, $\cM_3$.

There are two particularly notable features of these definitions in which they differ
from the forms used in BS1 and previous RFT works.
The first is that $\wt F^{(i)}$ does not contain
the symmetry factor $1/2!$ that is present in the quantity $\wt F$ defined in BS1.
This change arises simply because we are considering nonidentical particles.

The second feature is the presence of the parity matrices $P_L$.
As indicated
by Eq.~\Eqref{eq:PLdef}, these matrices are diagonal in the $\{k_i \ell m\}$ indices---with the flavor $i$
determined by the position in the flavor matrix---and simply give a minus sign for odd values of angular momentum.
As announced above, the factors $P_L$ 
are needed to account for mismatches in the momenta used to decompose into
spherical harmonics as part of the on-shell projection.
We show how this works by considering two examples.
First, consider the following subsequence contained in $C_{3,L}$,
\begin{equation}
\overline\cB_{2,L}^{(3)} D \overline\cB_{2,L}^{(1)} D\overline\cB_{2,L}^{(1)} D\overline\cB_{2,L}^{(3)} 
\,.
\end{equation}
Among the terms that arise after inserting the decomposition of Eq.~\Eqref{eq:Ddecomp} is 
\begin{equation}
\overline\cB_{2,L}^{(3)} \wt G^{(31)} P_L \overline\cB_{2,L}^{(1)} 
\wt F^{(1)} \overline\cB_{2,L}^{(1)}P_L \wt G^{(13)} \overline\cB_{2,L}^{(3)} 
\,.
\end{equation}
The leftmost $P_L$ is needed because the leftmost $\overline\cB_{2,L}^{(1)}$ is projected on shell
on its left-hand side using $\vec p_3$ to determine the harmonic decomposition (since this decomposition
arises from $\wt G^{(31)}$), while the right-hand projection is done using $\vec p_2$ for the harmonic
decomposition (since flavor 2 follows cyclically after flavor 1).
The $P_L$ converts the left-hand decomposition into that for flavor 2, so that both decompositions match.
A mirrored explanation holds for the right-hand factor of $P_L$.

For the second example we begin with the subsequence
\begin{equation}
\left[\wh \cB \wh D \wh \cB = \wh \cB (\wh F_G + \wh{\delta F}_G) \wh \cB\right]_{ij}\,,
\end{equation}
and pick out the $B_3$ parts of both $\wh \cB$s and the $\wh F_G$ part of $\wh D$.
Since $B_3$ is distributed equally among all elements of $\wh \cB$, there are,
for a given choice $i$ and $j$, nine contributions to this quantity,
one from each of the elements of $\wh F_G$.
We focus on the contribution containing $\wt G^{(12)}$.
This is given by 
\begin{equation}
\tfrac19 \left[B_3\right]_{;p_1 \ell' m'} \wt G^{(12)}_{p_1\ell' m'; k_2 \ell m} (-1)^\ell \tfrac19 \left[B_3\right]_{k_2 \ell m;}\,,
\label{eq:PLex2}
\end{equation}
where for the sake of clarity, we have shown only the ``internal" indices of the $B_3$s.
What we need to explain is why the factor of $(-1)^\ell$, 
which arises from the $P_L$ on the right of $\wt G^{(12)}$ in $\wh D$,
is needed.
To see this we note that, according to the definition of $\wt G^{(12)}$ given above, the decomposition
into spherical harmonics with indices $\ell m$ is done relative to the momentum of the flavor 1 particle
(with the flavor 2 being the spectator).
By contrast, the harmonic decomposition of the right-hand $B_3$, which also has spectator flavor 2,
is done relative to the momentum of the particle of flavor 3 (using the cyclic convention).
This mismatch is corrected by the $(-1)^\ell$.
There is no such mismatch for the left-hand $B_3$, since with a flavor 1 spectator, 
the harmonics $\ell' m'$
are defined relative to the flavor 2 momentum both in the $B_3$ and in $\wt G^{(12)}$.
All other factors of $P_L$ in $\wh F_G$ can be understood in a similar fashion.
This example also affords an example of the difference between the TOPT approach of BS1 and
the original derivation of Ref.~\cite{\HSQCa}.
In the latter, the three-particle cut between two $B_3$ kernels is expressed entirely in terms of $\wt F$,
whereas in the TOPT approach there are both $\wt F$ and $\wt G$ contributions. 
Both are legitimate expressions, but only the latter leads to simple all-orders 
expressions for $C_{3,L}$.

As a side note, we observe that the matrix $\wh F_G$ is symmetric under the interchange
of all indices (i.e.~flavor and $\{k \ell m\}$).
This is because $[\wt G^{(ij)}]^{\rm Tr} = [\wt G^{(ji)}]$, i.e.
\begin{equation}
\left[\wt G^{(ij)}\right]_{p_i \ell' m'; k_j \ell m}
=
\left[\wt G^{(ji)}\right]_{k_j \ell m; p_i \ell' m'}\,,
\end{equation}
and because $\wt F^{(i)}$ is symmetric in its indices.
Since $\wh D$ is manifestly symmetric, this implies that $\wh{\delta F}_G$ is symmetric.
By construction, $\wh B$ is also symmetric.

We now insert the decomposition of Eq.~\Eqref{eq:Ddecomp}
into our result for the correlator, Eq.~\Eqref{eq:C3Lshort}.
After some rearrangement, this leads to
\begin{equation}
C_{3,L} - C_{3,\infty}^{(0)} =
\wh A'_F i \wh F_G \frac1{1- i \wh \cK_{\df,23,L}  i \wh F_G} \wh A_F \,,
\label{eq:C3Lf}
\end{equation}
where the new endcaps are
\begin{align}
\wh A_F = \frac1{1 - i \wh \cB\, i\wh{\delta F}_G} \wh A\,,
\label{eq:AFhat}
\\
\wh A'_F = \wh A' \frac1{1 - i\wh{\delta F}_G\, i \wh \cB}\,,
\label{eq:AFphat}
\end{align}
while\footnote{%
When evaluating the terms in the geometric series in Eq.~\Eqref{eq:AFphat} one must
do the integrals associated with the integral operators contained in $\wh{\delta F}_G$.
Those in $\wt I_F^{(i)}$ involve a PV prescription, and must be done first, leaving the
integrals implicit in the $\delta \wt G^{(ij)}$ for second. The latter do not require a pole prescription.
}
\begin{equation}
i \wh \cK_{\df,23,L}
=
i \wh \cB \frac1{1 - i\wh{\delta F}_G\, i \wh \cB}\,.
\label{eq:Kdf23L}
\end{equation}
Although these three definitions work both off and on shell,
in Eq.~\Eqref{eq:C3Lf} these quantities always appear
adjacent to factors of $\wh F_G$, 
and are thus always projected on shell. In the following we consider only the on-shell forms.

From the result for the correlator, Eq.~(\ref{eq:C3Lf}), we can read off the quantization condition
\begin{equation}
\det\left[1+ \wh \cK_{\df,23,L}   \wh F_G\right] = 0\,.
\label{eq:QC1}
\end{equation}
This has a similar form to that for identical particles obtained in BS1,
a similarity that is made clearer if one replaces $\wh F_G$ with the equivalent
$\wh F+\wh G$.
Here, however, 
the determinant runs over both the on-shell indices and the additional 3 flavor dimensions.
It is important to keep in mind that, in this expression, different decompositions of
the on-shell momenta are being used for the different indices of the matrices
$\wh\cK_{\df,23,L}$ and $\wh F_G$. If the index is $i$, then the momentum of
the corresponding flavor is the spectator.\footnote{%
This implies that, in general, the number of
values of $\vec k_i$ that lie below the cutoff depends on $i$.
}
We also note that, at this stage, we can change the boost used in defining $\wt G^{(ij)}$
to that used in Ref.~\cite{\HSQCa} (referred to below as the HS boost), 
since this change can be absorbed by a shift in
the integral operators $\delta \wt G^{(ij)}$.

In order to make the content of the quantization condition clearer, it is useful to unpack $\wh \cK_{\df,23,L}$. 
As the name suggests,\footnote{%
In BS1, we added tildes to quantities that were composed of TOPT (rather than Feynman-diagram-based) kernels,
but here we drop the tildes to lighten the notation (given the presence of hats).
Kernels defined in terms of Feynman diagrams, to be discussed below, are denoted by primes.}
this contains both two- and three-particle K matrices---real
functions of the kinematic variables that are devoid of unitary cuts.
The subscript ``df" stands for ``divergence-free,'' and indicates the absence of singularities related to exchanging a particle between two pairs.
This name is inherited from the Feynman-diagram approach
of Refs.~\cite{\HSQCa,\HSQCb}; within the TOPT framework the absence of such divergences follows
from their factorization into $\wh F_G$.
To pull out the part containing only two-particle K matrices, 
we set $\cB_3$ and the $\delta G^{(ij)}$ to zero,
leaving
\begin{align}
\wh \cK_{\df,23,L} \bigg|_{\cB_3=\delta G^{(ij)}=0}
&= 
{\rm diag}\left(
\overline \cK_{2,L}^{(1)}, \,
\overline \cK_{2,L}^{(2)}, \,
\overline \cK_{2,L}^{(3)}\right)
\equiv \wh{\overline{\cK}}_{2,L}\,,
\label{eq:K2part}
\end{align}
where
\begin{equation}
i\overline \cK_{2,L}^{(i)} = i\overline \cB_{2,L}^{(i)} \frac1{1 - i\wt I_F^{(i)} i\overline \cB_{2,L}^{(i)}}\,.
\end{equation}
As shown in Appendix B of BS1, $\overline\cK_{2,L}^{(i)}$ can be written
\begin{equation}
\left[\overline\cK_{2,L}^{(i)}\right]_{p_i \ell' m'; k_i \ell m} 
= 
2\omega_{k_i} L^3  \delta_{\vec p_i \vec k_i} \delta_{\ell' \ell} \delta_{m' m} 
\cK^{(i)}_{2,\ell}(q_{2,k_i}^*)\,,
\label{eq:K2Li}
\end{equation}
where $\cK^{(i)}_{2,\ell}$ is the $\ell$th partial wave of the infinite-volume two-particle K matrix involving
scattering of flavors $j$ and $k$.
These K matrices depend on the details of the PV pole prescription, as described explicitly in BS1.
In general, all partial waves are nonzero, unlike for identical particles,
where only even waves are present.

The remainder of $\wh \cK_{\df,23,L}$ involves all three particles, either through alternating factors
of $\overline \cK_{2,L}^{(i)}$ and $\delta G^{(ij)}$, or through factors of $\cB_3$. 
We call it $\wh \cK_{\df,3}$, and stress that all entries of this flavor matrix are nonzero.
An explicit expression can be given, but is not illuminating. 
Thus we simply define it by
\begin{equation}
\wh \cK_{\df,23,L} = \wh{ \overline{\cK}}_{2,L} + \wh \cK_{\df,3}\,.
\end{equation}
An important property of $\wh \cK_{\df,3}$ is that, for all of its elements, one can take the
$L\to\infty$ limit, up to exponentially suppressed corrections. As explained in BS1, this holds only if
one chooses the PV scheme such that there are no poles in any of the $\cK_{2,\ell}^{(i)}$ in the kinematic
region of interest. This is possible with the generalized PV prescription of Ref.~\cite{\largera}.\footnote{%
We note that the freedom to treat each value of $\ell$ differently in this prescription extends also to
allowing different prescriptions for each spectator flavor.}
We also note that $\wh \cK_{\df,3}$ is a symmetric matrix.

The elements of the matrix $\wh \cK_{\df,3}$ have a similar status to the asymmetric quantity
$\wt \cK_{\df,3}^\uu$ entering the alternate form of the RFT quantization condition derived in BS1.
It is for this reason that we refer to $\wh \cK_{\df,3}$ as an asymmetric (or, perhaps better,
``unsymmetrized'') K matrix, and the quantization condition itself as asymmetric.
The nine flavor elements of $\wh \cK_{\df,3}$ are distinguished by the nature of the external
two-particle interaction for both initial and final states---the $ij$'th element has incoming spectator flavor
$j$ and outgoing spectator flavor $i$.\footnote{%
 Note that, because of the factor of $\cB_3/9$ in the matrix $\wh\cB$
[see Eq.~\Eqref{eq:Bhat}], the external TOPT kernel can also be $\cB_3$, which is symmetric.}
By contrast, the full three-particle scattering amplitude $\cM_3$, discussed in the next section,
is obtained by summing over all choices of initial and final spectators.
This raises the question of whether the quantization condition can be written in terms of a similarly summed
K matrix. This would be the analog of rewriting the alternate form of the quantization condition in terms of a symmetrized $\wt\cK_{\df,3}$, which is achieved in BS1, 
and leads to the original form obtained in Ref.~\cite{\HSQCa}.
This is achieved below in Sec.~\ref{sec:symmetric}.

\section{Relation of $\wh \cK_{\df,3}$ to $\cM_3$}
\label{sec:KtoM}

All three-particle formalisms require a second step, in which the three-particle K matrix
that enters the quantization condition is related by integral equations to the physical
three-particle scattering amplitude $\cM_3$. 
This is necessary because the intermediate K matrix,
despite being an infinite-volume quantity, is not physical, since it depends on the cutoff function
and PV prescription.
In this section we derive the form of the integral equations,
using the method of Ref.~\cite{\HSQCb}.
This begins by considering the finite-volume scattering amplitude, $\cM_{3,L}$, expressing it in terms of
$\wh \cK_{\df,3}$, and taking an appropriate $L\to\infty$ limit in order to obtain an expression for $\cM_3$.

As for $\wh \cK_{\df,23,L}$,
 simpler expressions are obtained by using a combination of two- and three-particle amplitudes,
 specifically
 \begin{equation}
 \cM_{23,L} = \overline\cM_{2,L}^{(1)} + \overline\cM_{2,L}^{(2)} + \overline\cM_{2,L}^{(3)} + \cM_{3,L}\,,
 \end{equation}
 where 
$\overline \cM_{2,L}^{(i)}$ corresponds to the scattering of flavors $j$ and $k$ with $i$ spectating.
It is given by
\begin{equation}
i\overline\cM_{2,L}^{(i)} = i \overline\cK_{2,L}^{(i)}
\frac1{1 - i \wt F^{(i)} i  \overline\cK_{2,L}^{(i)}}\,,
\label{eq:M2LfromK2L}
\end{equation}
which is the nondegenerate generalization of Eq.~(B3) from BS1.

 \begin{figure}[tb]
\begin{center}
\vspace{-10pt}
\includegraphics[width=\textwidth]{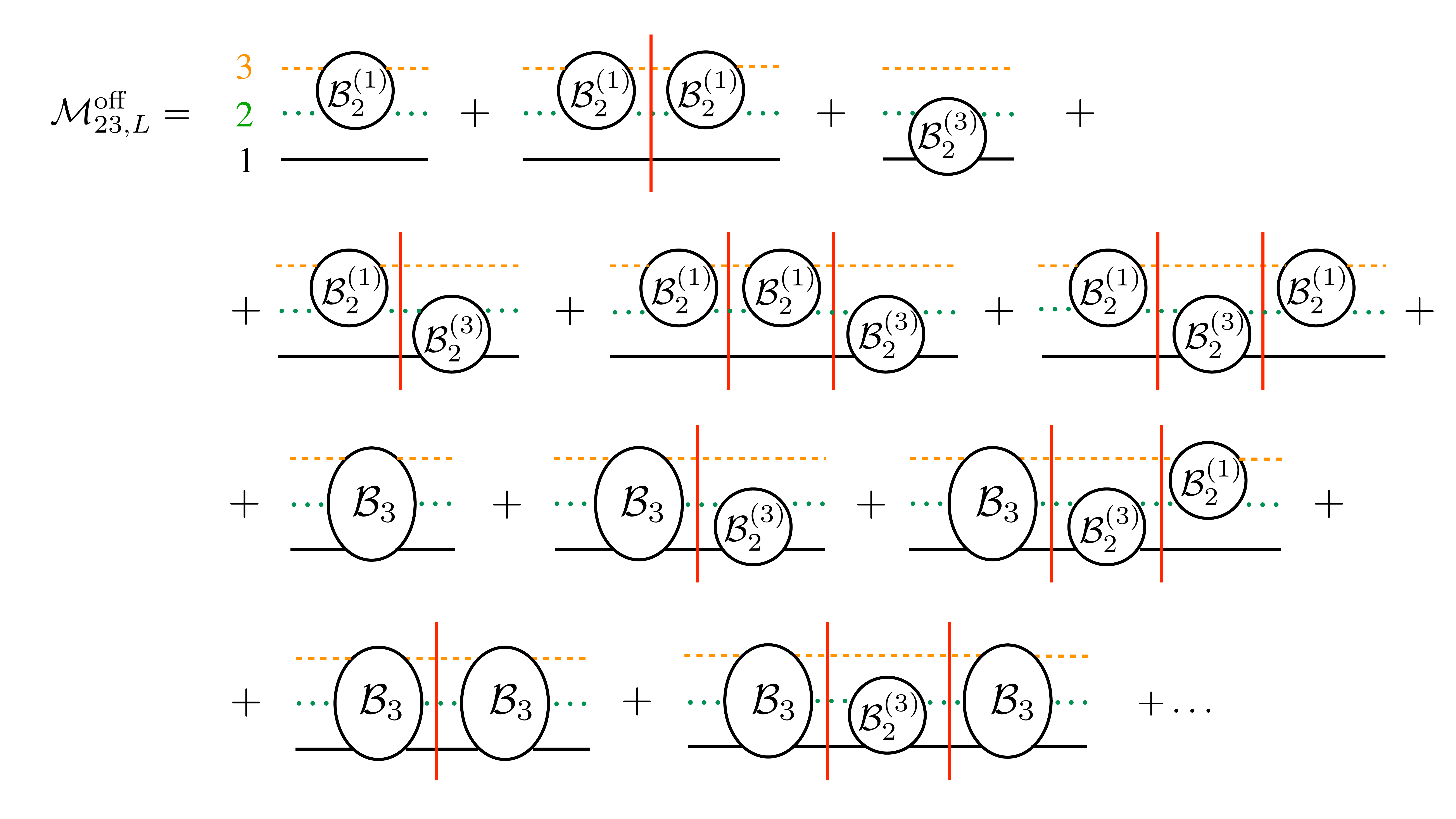}
\vspace{-0.2truein}
\caption{
Contributions to the combined two- and three-particle finite-volume scattering amplitude $\cM_{23,L}$
in TOPT.
Notation as in Fig.~\ref{fig:CL}. The absence of diagrams involving the two-particle kernel
$\cB_2^{(2)}$ is for representational simplicity---such diagrams are contained in the ellipsis.
\label{fig:M23L}
}
\end{center}
\end{figure}

The diagrams contributing to the off-shell $\cM_{23,L}$ in TOPT are shown in Fig.~\ref{fig:M23L}
and lead to
\begin{align}
i\cM_{23,L}^{\rm off} &=
i(\overline\cB_{2,L}^{(1)} + \overline\cB_{2,L}^{(2)} + \overline\cB_{2,L}^{(3)} + \cB_3 )
\sum_{n=0}^\infty
\left[i D i(\overline\cB_{2,L}^{(1)} + \overline\cB_{2,L}^{(2)} + \overline\cB_{2,L}^{(3)} + \cB_3 ) \right]^n\,.
\end{align}
This can be written compactly in matrix notation
\begin{align}
\cM_{23,L}^{\rm off} &= \bra1 \wh \cM_{23,L}^{\rm off} \ket 1 \,,
\label{eq:M23Lres2}
\end{align}
with
\begin{equation}
i \wh \cM_{23,L}^{\rm off} =
i\wh \cB \frac1{1 - i \wh D i\wh \cB}\,.
\label{eq:M23Lres3}
\end{equation}
The nine flavor elements of $\wh \cM_{23,L}^{\rm off}$ are the analogs of
the asymmetric amplitude $\overline \cM_{2,L} +\wt \cM_{3,L}^\uu$ appearing in the degenerate case
analyzed in BS1.
As for $\wh \cK_{\df,3}$, they correspond to the different choices of the initial and final spectator flavors.
Summing over the different choices, as in Eq.~\Eqref{eq:M23Lres2}, 
gives $\cM_{23,L}^{\rm off}$.

The next step is to consider the on-shell amplitude, and insert the decomposition
Eq.~\Eqref{eq:Ddecomp}. After some rearrangement this leads to the simple result
(using the notation that the amplitude is on shell unless there is an explicit superscript ``off")
\begin{equation}
i \wh \cM_{23,L} = i\wh \cK_{\df,23,L} \frac1{1 - i \wh F_G i \wh \cK_{\df,23,L}}\,,
\label{eq:M23Lres}
\end{equation}
with $\wh \cK_{\df,23,L}$ given in Eq.~(\ref{eq:Kdf23L}).
The key point here is that the same two- and three-particle K matrices enter as in the quantization condition.
We stress again that different on-shell projections are used for different flavor indices.
This means that the elements of the matrix cannot be combined as in Eq.~\Eqref{eq:M23Lres2},
to give an on-shell $\cM_{23,L}$. Indeed, even if we multiply by spherical harmonics to convert the
$\ell m$ indices back into momenta, the elements of $\wh \cM_{23,L}$ cannot be combined for finite $L$,
since the on-shell projection moves some momenta out of the finite-volume set.
Such a combination is possible only in the infinite-volume limit, which, however, is all that we require below.

We next unpack the result \Eqref{eq:M23Lres} in order to extract a result for the three-particle amplitude itself.
First, we package the finite-volume two-particle amplitudes into matrix form
\begin{align}
\wh {\overline{\cM}}_{2,L}
&\equiv
{\rm diag}(\overline \cM_{2,L}^{(1)}, \, \overline \cM_{2,L}^{(2)}, \, \overline \cM_{2,L}^{(3)}) \,.
\label{eq:M2Lhat}
\end{align}
Next, we separate $\wh F_G$, given in Eq.~(\ref{eq:FGhat}),
into its $\wt F$ and $\wt G$ parts:
\begin{align}
\wh F_G &= \wh F + \wh G\,,
\end{align}
with the diagonal terms contained in
\begin{align}
\wh F &= {\rm diag}(\wt F^{(1)}, \wt F^{(2)}, \wt F^{(3)})\,,
\label{eq:Fhat}
\end{align}
and the off-diagonal terms contained in $\wh G$.
Then Eq.~\Eqref{eq:M2LfromK2L} becomes
\begin{equation}
i \wh{\overline{\cM}}_{2,L} = i \wh{\overline{\cK}}_{2,L} 
\frac1{1 - i \wh F i \wh{\overline{\cK}}_{2,L} }\,.
\label{eq:Mhat2L}
\end{equation}
The matrix version of $\cM_{3,L}$ is given by
\begin{equation}
\wh \cM_{3,L} \equiv \wh \cM_{23,L} - \wh{\overline{\cM}}_{2,L}\,,
\end{equation}
and is related schematically to the full scattering amplitude by
$\cM_{3,L} = \bra1 \wh \cM_{3,L} \ket1$.
As noted above, this equation only makes sense in the infinite-volume limit, after multiplying by
appropriate spherical harmonics and summing over angular-momentum indices.

With this setup, the algebraic steps needed to obtain an expression for $\wh \cM_{3,L}$ are identical to
those in BS1 (and given explicitly in Appendix C of that work).
We find
\begin{align}
i\wh \cM_{3,L} &= i \wh \cD_L + i \wh \cM_{\df,3,L}\,,
\label{eq:M3Ldecomp2}
\\
i\wh \cD_L &= i \wh{\overline{\cM}}_{2,L} i\wh Gi \wh{\overline{\cM}}_{2,L}
\frac1{1- i\wh G i \wh{\overline{\cM}}_{2,L} }\,,
\label{eq:DLdef}
\\
i \wh \cM_{\df,3,L} &=
\left[1+ i \wh \cD_{23,L} i \wh F_G\right]
i \wh \cK_{\df,3}
\frac1{ 1 - [1 + i \wh F_{G} i \wh \cD_{23,L}] i \wh F_G i \wh \cK_{\df,3}}
\left[1+ i \wh F_G i \wh \cD_{23,L}\right]\,,
\label{eq:Mdf3Lhat}
\\
i\wh \cD_{23,L} &=  i\wh{\overline{\cM}}_{2,L} + i\wh \cD_L
= i \wh{\overline{\cM}}_{2,L}
\frac1{1- i\wh G i \wh{\overline{\cM}}_{2,L} }\,.
\label{eq:D23L}
\end{align}
In the appropriate $L\to\infty$ limit~\cite{\HSQCb} these results become integral equations
for $\wh \cM_3$. 
We do not give these explicitly, since their form is almost identical
to those arising in the Feynman diagram derivation, and we present
the latter in full detail in Sec.~\ref{sec:KtoMFeynman} below.

It is worth understanding the source of the various terms contributing to $\wh \cM_{3,L}$
in Eqs.~(\ref{eq:M3Ldecomp2})-(\ref{eq:D23L}).
$\wh D_L$ is the contribution to three-particle scattering arising from repeated two-particle
interactions, connected by the switch factors in $\wh G$, arising from the diagrams on the second
line of Fig.~\ref{fig:M23L}.
The off-diagonal nature of $\wh G$ enforces the switching of spectators, and the matrix structure ensures that
all possible switches occur. 
Up to kinematical factors, $\overline\cM_{2,L}^{(i)}$ goes over in the infinite-volume limit
to the Lorentz-invariant two-particle scattering amplitude involving flavors $j$ and $k$, $\cM_2^{(i)}$
(see Appendix E of BS1).
It follows that, if the relativistic form of $\wt G$ is used, the elements of $\wt \cD_L^\uu$ are
Lorentz invariant.\footnote{%
Strictly speaking, since all quantities in the quantization conditions carry indices $\{\vec k,\ell, m\}$, one
must first multiply by the appropriate spherical harmonics in order to obtain a quantity whose Lorentz
transformation properties can be studied. See Refs.~\cite{\HSQCa} and BS1 for more details.}

The remaining part of $\wh \cM_{3,L}$ is denoted $\wh \cM_{\df,3,L}$, where
the subscript $\df$ indicates the ``divergence-free'' nature of this object, since the poles corresponding to
on-shell one-particle exchange are contained in $\wh \cD^\uu$.
$\wh \cM_{\df,3,L}$ contains the contributions to three-particle scattering that involve the three-particle
K matrix, $\wh \cK_{\df,3}$. In words, the external factors in square braces correspond to repeated
two-particle interactions with switches, prior to a genuine quasilocal three-particle interaction
due to an element of $\wh\cK_{\df,3}$, after which the middle section of Eq.~(\ref{eq:Mdf3Lhat})
corresponds to repeated two-particle interactions prior to another three-particle interaction, etc.
This is all a natural and simple generalization of the interpretation of the corresponding expression
for identical particles.

We see from the result (\ref{eq:Mdf3Lhat}) that the elements of $\wh\cK_{\df,3}$ are not Lorentz invariant.
This is because, when $L\to\infty$, the set of integral equations that this matrix equation goes over to connects
it to $\wh \cM_{\df,3}$, whose elements are not Lorentz invariant because they are defined in TOPT.
As noted in the introduction, the lack of Lorentz invariance of $\wh\cK_{\df,3}$ is expected 
in the TOPT approach.
This leads to complications when implementing the formalism in practice, 
and in the next section we explain how this problem can be resolved.

We close this section by emphasizing that
we can use the expression (\ref{eq:M23Lres}) for $\wh \cM_{23,L}$ as an alternative vehicle for deriving
the quantization condition.
This possibility was first noted in Ref.~\cite{\HSQCb} in the context of identical particles.
The point is that $\cM_{23,L}$ is a type of finite-volume correlator, so its poles determine the spectrum.
Indeed, from the form of the denominator in Eq.~(\ref{eq:M23Lres}) we immediately obtain the
quantization condition obtained in the previous section, Eq.~(\ref{eq:QC1}).
One might be concerned that, since $\cM_{23,L}$ contains $\wh{\overline{\cM}}_{2,L}$, 
there will also be poles at the positions where the latter quantity diverges.
This occurs at energies of a free spectator combined with a two-particle finite-volume state, and these energies
are not in the three-particle spectrum.
It turns out, however, that these spurious poles cancel in $\cM_{23,L}$, as can be seen by
writing it as
\begin{equation}
\wh \cM_{23,L} = \wh \cD_{23,L} + \wh \cM_{\df, 3, L}\,,
%\label{eq:M23Lres2}
\end{equation}
and noting, from Eq.~(\ref{eq:D23L}), that $\wh \cD_{23,L}$ remains finite when
$\wh{\overline{\cM}}_{2,L}$ diverges.
We stress that the quantization condition arising from the poles
in $\wh \cM_{\df,3,L}$ is indeed Eq.~(\ref{eq:QC1}).
This can be most easily seen by rewriting Eq.~(\ref{eq:Mdf3Lhat}) using
\begin{equation}
1 + i \wh F_G i \wh \cD_{23,L} = \big(1 + \wh F_G \wh{\overline{\cK}}_{2,L}\big)^{-1}\,,
\end{equation}
from which it follows that 
\begin{equation}
i \wh \cM_{\df,3,L} = \left[1 + i \wh \cD_{23,L} i \wh F_G\right] i \wh \cK_{\df,3} 
\frac1{1 + \wh F_G \wh \cK_{\df,23,L}}\,.
\end{equation}

\section{Quantization condition with Lorentz-invariant $\cK_{\df,3}$}
\label{sec:derivation2}

In this section we derive the following alternate form for the quantization condition for nondegenerate scalars,
\begin{equation}
\det\left[1+ \wh \cK'_{\df,23,L}   \wh F_G\right] = 0\,,
\label{eq:QC2}
\end{equation}
where 
\begin{equation}
\wh\cK'_{\df,23,L}= \wh{\overline{\cK}}_{2,L} + \wh \cK'_{\df,3}\,.
\label{eq:Kpdf3}
\end{equation}
Here $\wh{\overline{\cK}}_{2,L}$ the same as above [see Eq.~\Eqref{eq:K2part}], 
but now
$\wh\cK'_{\df,3}$ is a (matrix of) Lorentz-invariant three-particle K matrices that differs from $\wh \cK_{\df,3}$.
In this way, we obtain a fully Lorentz-invariant formalism: one that not only is valid for
relativistic kinematics, but in which the elements of $\wh\cK'_{\df,3}$ are Lorentz scalars.
This is important for practical implementations, which typically use multiple values of $\vec P$,
and thus require the relationship between $\wh\cK'_{\df,3}$ in different Lorentz frames.

A striking feature of this result is that the quantization condition \Eqref{eq:QC2}
has exactly the same form as that derived above using TOPT, Eq.~\Eqref{eq:QC1},
differing only in the K matrix that enters.
This redundancy is of the same nature as that found in the identical-particle case in BS1, where
two identical forms of the quantization condition were established, both involving asymmetric K matrices,
one of which is Lorentz invariant while the other is not.
This was understood as being due to the intrinsic ambiguity in the definition of an
asymmetric object, since the only constraint is that by combining terms one ends up with the correct
symmetrized quantity.
An analogous understanding applies here: 
it is only by summing over the different choices of flavors of the external spectators for, say,
the elements of $\wh\cM_{3,L}$ that one obtains the physical amplitude, and thus there is some freedom
in the definition of the individual elements. The same holds for the K matrices.
Examples of this ambiguity will be seen in the subsequent discussion.

In BS1, we obtained the form of the quantization condition containing the Lorentz-invariant asymmetric K matrix
by starting from the result derived using Feynman diagrams in Ref.~\cite{\HSQCa}. 
Here there is no such result, so we must begin {\em de novo}.
Our strategy is to reorganize the original Feynman-diagram-based approach of Ref.~\cite{\HSQCa}
into a form that mirrors the TOPT result at every step, so that, after setting up the calculation, we can simply
carry over the algebra of the TOPT approach described above.
In addition, we derive the quantization condition using the Feynman diagram version of
the finite-volume amplitude $\wh \cM_{23,L}$, rather than the correlator $C_{3,L}$.
As shown in the previous section, this leads to the same quantization condition, but avoids the need to
deal with endcaps.

The starting point is the finite-volume three-particle amplitude with external spectators having flavors $i$ and $j$.
We refer to such amplitudes as ``asymmetric," as it is only after summing the nine
combinations of $\{i,j\}$ that we obtain the full amplitude.
In the previous section we considered the asymmetric amplitude $[\wh \cM_{3,L}]_{ij}$, 
with its asymmetry defined using TOPT diagrams.
Here we define the asymmetry using Feynman diagrams,  leading to a different asymmetric amplitude
$[\wh \cM^{\prime\,\rm off}_{3,L}]_{ij}(\{p\};\{k\})$,
where $\{p\}\equiv \{p_1,p_2,p_3\}$, etc.~are sets of three four-momenta,
and we are using the notation that a prime denotes quantities defined using Feynman diagrams.
The external three-momenta are drawn from the finite-volume set, but at this stage the external energies
are arbitrary, so that the four-momenta are in general off shell.
Momentum conservation implies that the four-momenta satisfy $\sum_i p_i = P = \sum_j k_j$.

 \begin{figure}[tb]
\begin{center}
\vspace{-10pt}
\includegraphics[width=\textwidth]{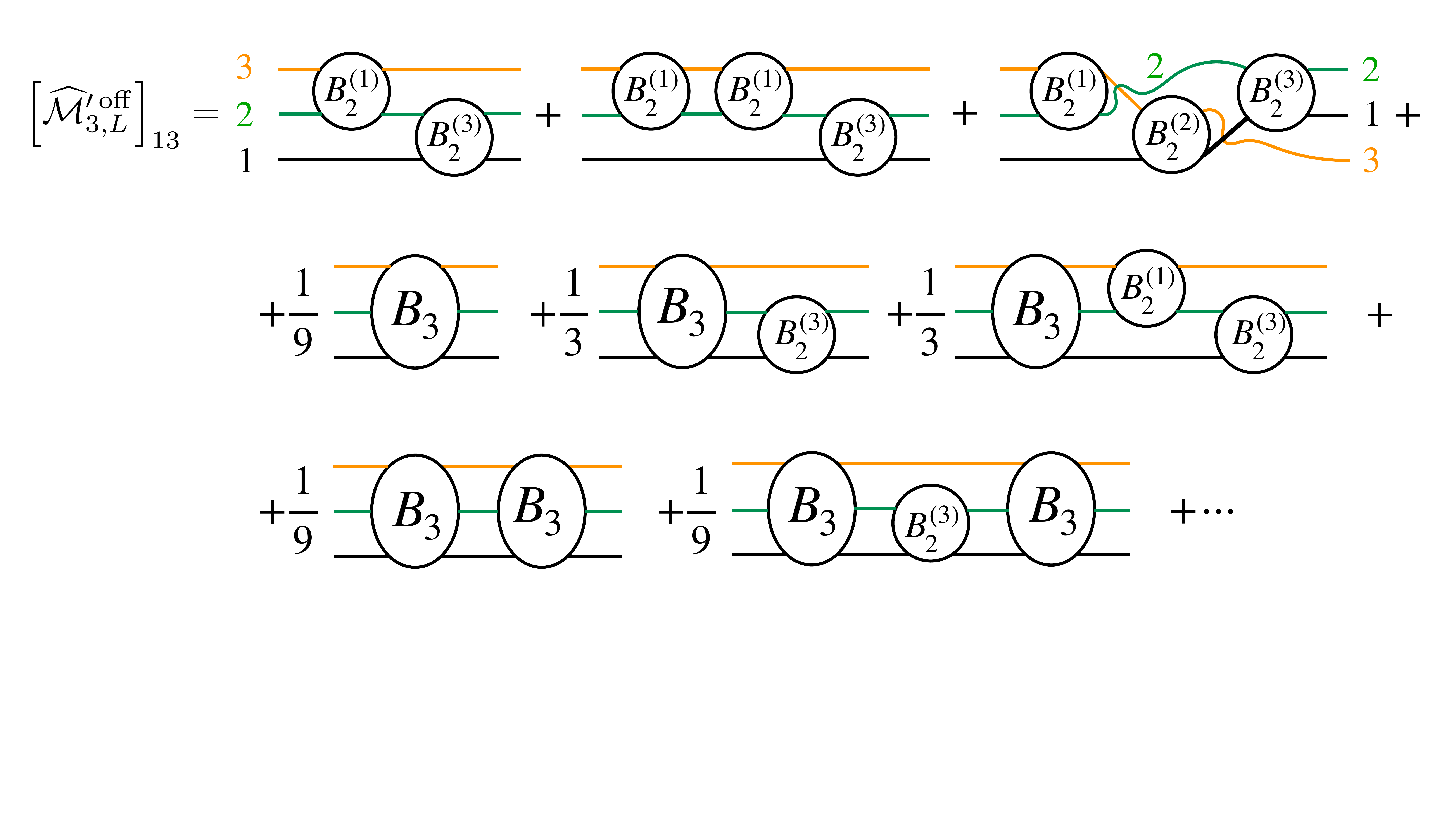}
\vspace{-1.2truein}
\caption{
Contributions to $[\wh{ \mathcal M}^{\prime\,\rm off}_{3,L}]_{13}$ in the Feynman-diagram skeleton expansion.
$B_2^{(i)}$ and $B_3$ are Bethe-Salpeter kernels, and solid lines represent fully dressed propagators. 
Unlike in earlier figures, all propagators are shown by solid lines, with the
flavors now distinguished only by colors and explicit labels. External propagators are amputated.
\label{fig:M'13L}
}
\end{center}
\end{figure}

As explained in Ref.~\cite{\HSQCa}, when using Feynman diagrams, the amplitudes are given by
a skeleton expansion in terms of the Bethe-Salpeter kernels\footnote{%
At the risk of confusion, we use the same letter for these kernels as for the corresponding TOPT objects,
but without the calligraphic font.
}
\begin{equation}
B_2^{(i)}(p_j,p_k;k_j,k_k) \ \ {\rm and}\ \
B_3(\{p\};\{k\})\,.
\end{equation}
These are, respectively, the 2PIs and 3PIs two- and three-particle kernels,
with the former having flavor $i$ as the spectator, and with flavor labels $\{i,j,k\}$ ordered cyclically.
In the skeleton expansion the kernels can be evaluated in infinite volume. 
In contrast to the TOPT kernels given in Eq.~(\ref{eq:TOPTkernels}),
$B_2^{(i)}$ and $B_3$ depend on four-momenta that are, in general, off shell.
They are connected by fully dressed, relativistic propagators, normalized to unity at the single-particle pole,
whose spatial momenta must be summed over the finite-volume set,
while the energy is integrated as usual.
External propagators are amputated.
For a given quantity, 
the set of skeleton diagrams that contributes is exactly the same as in the expansion in TOPT kernels
(see, e.g.,~Fig.~\ref{fig:M23L}), except that there is no time ordering.
As a concrete example, we show diagrams that contribute to
$[\wh \cM^{\prime\,\rm off}_{3,L}]_{13}$ in Fig.~\ref{fig:M'13L}. 
This amplitude is defined so that, if there is a two-particle Bethe-Salpeter
kernel on the left (right) end, it must be a $B_2^{(1)}$ ($B_2^{(3)}$). 
In addition, for each end with a $B_3$ kernel the contribution is multiplied by $1/3$.
The latter factors ensure that, when the flavor indices are summed, 
the contribution to the total amplitude has the correct weight.

 \begin{figure}[tb]
\begin{center}
\vspace{-10pt}
\includegraphics[width=\textwidth]{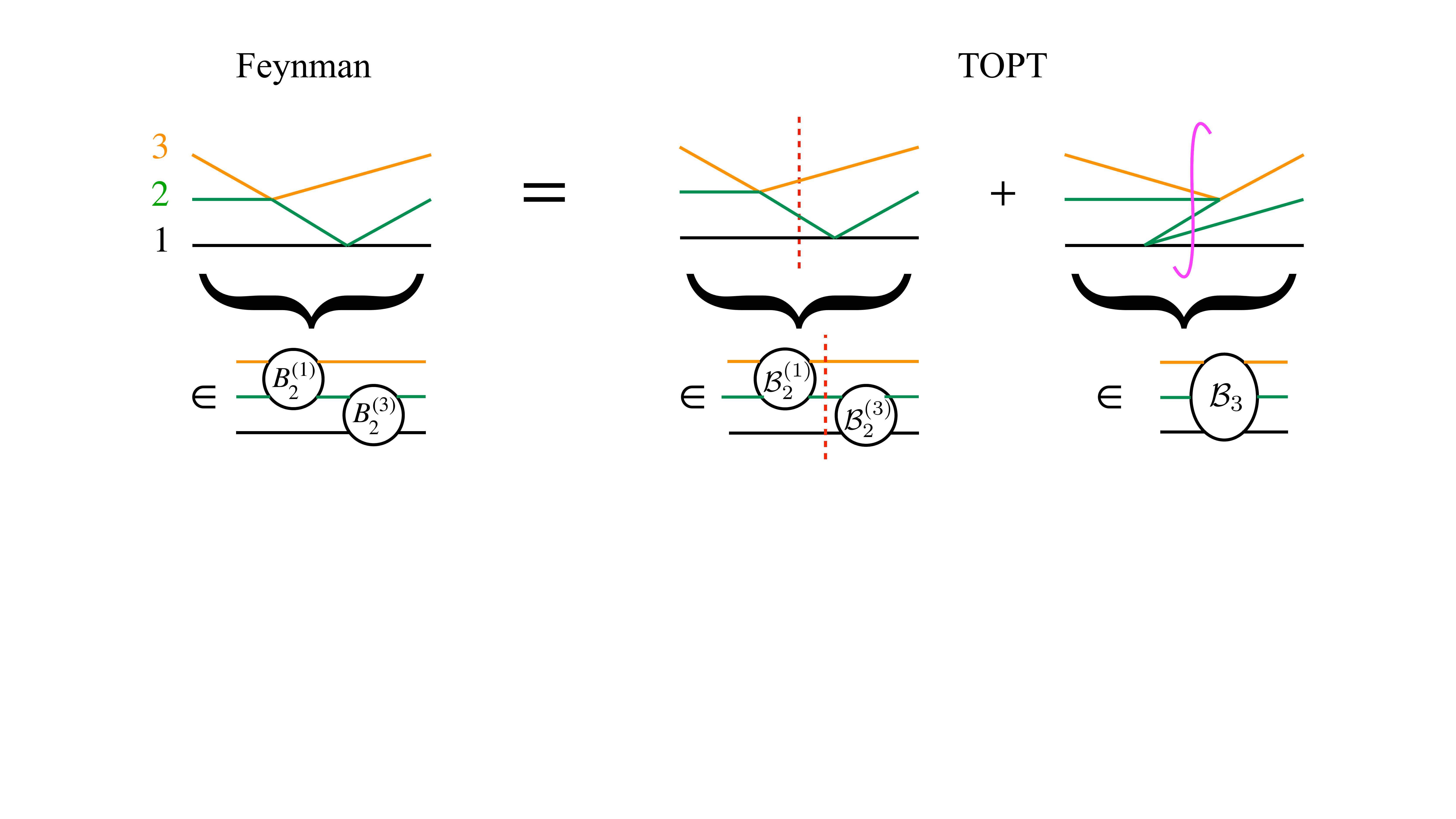}
\vspace{-1.8truein}
\caption{
Example of how a Feynman diagram, which is assigned to one term in the corresponding skeleton
expansion, breaks up into several (in this case two) TOPT diagrams, which are in turn assigned to several
(here two) terms in the TOPT skeleton expansion. As discussed in the text, this implies that the
asymmetric TOPT amplitudes $[\wh\cM_{3,L}]_{ij}$ are not Lorentz invariant.
Notation is as in Figs.~\ref{fig:CL}, \ref{fig:M23L} and \ref{fig:M'13L}, except that all propagators
are shown by solid lines.
\label{fig:FvsTOPT}
}
\end{center}
\end{figure}

To make clear that the elements of $\wh\cM'_{3,L}$ differ from those of the TOPT version, $\wh\cM_{3,L}$,
we consider in Fig.~\ref{fig:FvsTOPT} the simplest contribution to the first diagram in Fig.~\ref{fig:M'13L}.
In TOPT, it breaks into two diagrams, one of which contributes to $[\wh \cM_{3,L}]_{13}$, and the other
of which is split equally between all elements of $\wh \cM_{3,L}$ (since it contributes to $\cB_3$).
Thus only $1/9$th of the diagram is included in $[\wh \cM_{3,L}]_{13}$.
This also shows that the latter quantity is not Lorentz invariant, since it is only by adding the two TOPT
diagrams with equal weight that one regains an invariant quantity.
On the other hand, each of the elements of $\wh \cM^{\prime\,\rm off}_{3,L}$ is Lorentz invariant,
simply because it is composed of Feynman diagrams.

We now begin the analysis of the elements of $\wh\cM^{\prime\,\rm off}_{3,L}$. 
Our approach quickly diverges from that in  Refs.~\cite{\HSQCa,\HSQCb}, 
so that we cannot make a step-by-step comparison, but will rather emphasize 
global similarities and differences. 
We present an overview of the derivation in the main text,
and describe the details in Appendix~\ref{app:Feynman}.

In both approaches, the first step when analyzing a given diagram
is to do the energy integrals for all independent momenta, 
i.e.~those not constrained by four-momentum conservation.
The difference from Refs.~\cite{\HSQCa,\HSQCb} is that here we do such integrals for all diagrams before
proceeding to the second step, rather than analyzing subsets of diagrams completely and then combining.
As explained in Appendix~\ref{app:Feynman}, 
the results of the energy integrals are diagrams in
which two of the three particles in all cuts are on shell, 
i.e.~with momenta $p_i^{\rm on}=(\omega_{p_i},\vec p_i)$,
while the momentum of the third particle remains, in general, off shell.
The momentum configuration is then specified in the same way as in the TOPT analysis,
namely with the (redundant) set of three finite-volume momenta 
$\{\vec k\}$.
In order to present the result in a compact form,
we need to introduce operators that specify which pair of momenta in the kernels are placed on shell.
We call these $\overleftarrow O^{(i)}$ and $\overrightarrow O^{(i)}$,
 where the flavor label $i$ indicates that the particles of the other two flavors are set on shell,
 and the arrow indicates whether the operator acts on the kernels immediately
 to the left or right, respectively.
With this notation, we find
\begin{align}
\wh\cM^{\prime\,\rm off}_{3,L} &=\wh \cM^{\prime\,\rm off}_{23,L} 
- \wh{\overline{\cM}}{}^{\prime\,\rm off}_{2,L} \,,
\\
i \wh\cM^{\prime\, \rm off}_{23,L} &= \wh{\overrightarrow O} i \wh B \frac1{1 - i \wh D' i \wh B}
\wh{\overleftarrow O}\,,
\label{eq:Mp23Lres_off}
\\
\wh{\overrightarrow O} &= 
{\rm diag}\left(\overrightarrow O^{(3)}, \overrightarrow O^{(1)}, \overrightarrow O^{(2)} \right)\,,
\\
\wh{\overleftarrow O} &= 
{\rm diag}\left(\overleftarrow O^{(3)}, \overleftarrow O^{(1)}, \overleftarrow O^{(2)} \right)\,,
\\
[\wh B]_{i' i} %(\vec p_{i'},\vec p_{j'}; \vec k_i, \vec k_j)  
&= \tfrac19 B_3 + \delta B_3^{(i' i)}
+ \delta_{i' i} \left[\overline B_{2,L}^{(i)}+ \overline{\delta B}_{2,L}^{(i)}\right]
\,,
\label{eq:BShat}
\\
\wh D'  &= 
\mathbb 1_{\vec p,\vec k}
%\delta_{\vec p_1 \vec k_1} \delta_{\vec p_2 \vec k_2} \delta_{\vec p_3 \vec k_3}
\begin{pmatrix}
D'^{(3)} & D'^{(3)} & D'^{(2)}\\
D'^{(3)} & D'^{(1)} & D'^{(1)}\\
D'^{(2)} & D'^{(1)} & D'^{(2)}
\end{pmatrix}
\,,
\label{eq:Dpmat}
\\
D'^{(i)} &= \overleftarrow O^{(i)} \frac1{4\omega_{p_j} \omega_{p_k}L^6}
 \frac{Z_i(p_i^2)}{p_i^2-m_i^2} \overrightarrow O^{(i)} \,.
 \label{eq:Dp}
\end{align}
We observe that, with this result, we have succeeded in obtaining an expression for
the finite-volume amplitude that is 
similar to the initial matrix form obtained with TOPT, Eq.~(\ref{eq:M23Lres3}).

There are many features of this rather elaborate result that require explanation.
We first discuss the effect of the on-shell projectors that are contained in
the $D^{\prime(i)}$ and also appear as external factors in $\wh \cM^{\prime\,\rm off}_{23,L}$.
When we expand out the geometric series in Eqs.~(\ref{eq:Mp23Lres_off}),
the kernels in $\wh B$ are always projected on both sides,
and thus we need only define the projected kernels.
For $B_3$ and $\delta B_3$ all combinations of projectors can occur, and their
action is exemplified by
\begin{align}
\left[\overrightarrow O^{(1)} B_3 \overleftarrow O^{(3)}\right](\{\vec p\}; \{\vec k\})
&=
B_3(p_1, p_2^{\rm on}, p_3^{\rm on}; k_1^{\rm on},k_2^{\rm on},k_3)\,.
\label{eq:B3proj}
\end{align}

For the two-particle kernels, the possible projections are restricted.\footnote{%
In order that all appearances of $B_2^{(i)}$ have projectors on both sides, we have,
in Eq.~(\ref{eq:Mp23Lres_off}), placed projectors on both ends of the expression.
Strictly speaking this means that $\wh\cM^{\prime\,\rm off}_{3,L}$ 
has external momenta that are only partly off shell, differing from the original definition given above
where all momenta can be off shell. Since we only consider the former quantity in the following,
we have kept the same notation.}
To explain this, we focus on $\overline B_{2,L}^{(1)}$.
Due to the forms of $\wh{\overrightarrow{O}}$, $\wh{\overleftarrow{O}}$,
and $\wh D'$, the projection operators acting on $\overline B_{2,L}^{(1)}$
are either  $\overrightarrow O^{(3)}$ or $\overrightarrow O^{(2)}$ on the left,
and either  $\overleftarrow O^{(3)}$ or $\overleftarrow O^{(2)}$ on the right.
Thus the spectator, with flavor 1, is always on shell, whereas the second on-shell flavor is 
either 2 or 3. The definitions that we need are thus
\begin{align}
\left[\overrightarrow O^{(3)} \overline B_{2,L}^{(1)} \overleftarrow O^{(3)}\right](\{\vec p\}; \{\vec k\})
&=
\delta_{\vec p_1 \vec k_1} 2 \omega_{p_1} L^3 B_2(p_2^{\rm on}, p_3; k_2^{\rm on},k_3)\,,
\label{eq:B21proj}
\\
\left[\overrightarrow O^{(3)} \overline B_{2,L}^{(1)} \overleftarrow O^{(2)}\right](\{\vec p\}; \{\vec k\})
&=
\delta_{\vec p_1 \vec k_1} 2 \omega_{p_1} L^3 B_2(p_2^{\rm on}, p_3; k_2,k_3^{\rm on}) \,,
\label{eq:B22proj}
\\
\left[\overrightarrow O^{(2)} \overline B_{2,L}^{(1)} \overleftarrow O^{(3)}\right](\{\vec p\}; \{\vec k\})
&=
\delta_{\vec p_1 \vec k_1} 2 \omega_{p_1} L^3 B_2(p_2,p_3^{\rm on}; k_2^{\rm on},k_3) \,,
\\
\left[\overrightarrow O^{(2)} \overline B_{2,L}^{(1)} \overleftarrow O^{(2)}\right](\{\vec p\}; \{\vec k\})
&=
\delta_{\vec p_1 \vec k_1} 2 \omega_{p_1} L^3 B_2(p_2, p_3^{\rm on}; k_2,k_3^{\rm on}) \,.
\label{eq:B23proj}
\end{align}
The generalization to other elements of $\overline B_{2,L}^{(i)}$ 
and $\overline{\delta B}_{2,L}^{(i)}$ is straightforward.
We stress that it is only because of the presence of the projection operators 
that we obtain a quantity that depends on three-momenta alone. 

Next we give the definition of $\wh{\overline{\cM}}{}^{\prime\,\rm off}_{2,L}$.
This is a diagonal matrix obtained by keeping the disconnected terms in 
$\wh \cM^{\prime\, \rm off}_{23,L}$,
i.e.~those obtained by keeping only the $\overline B_{2,L}^{(i)}+\overline{\delta B}_{2,L}^{(i)}$ parts of $\wh B$
and the diagonal part of $\wh D'$. Thus one particle spectates for the entire diagram.
The projection rules embedded in the definitions imply that this particle is on shell, and that,
if it has flavor $i$, then the second on-shell particle (which is one of the interacting pair) has the flavor
that follows $i$ cyclically.
The sum of all the diagrams contributing to 
$[\wh{\overline{\cM}}{}^{\prime\,\rm off}_{2,L}]_{ii}$
is simply a rearrangement of the complete set of Feynman diagrams that
describe the interactions of particles with flavors $j$ and $k$.
Additionally, the factors of $2\omega_{p_i} L^3$ cancel in pairs, leaving a single overall such factor.
Thus we find that
$\wh{\overline{\cM}}{}^{\prime\,\rm off}_{2,L}$
has the same form as $\wh{\overline{\cM}}_{2,L}$,  Eq.~(\ref{eq:M2Lhat}), 
except that one each of the incoming and outgoing scattered particles are off shell.
The reason that  $\wh{\overline{\cM}}{}^{\prime\,\rm off}_{2,L}$ is added to
$\wh\cM'_{3,L}$ is the same as in the TOPT analysis: it leads to a quantity, 
$\wh\cM^{\prime\,\rm off}_{23,L}$, that has a simple expression, here 
Eq.~(\ref{eq:Mp23Lres_off}).

One difference between the structure of the results here and those obtained in TOPT is the presence
of the shifts $\delta B_3$ and $\delta B_2$ in the kernels.
As shown in Appendix~\ref{app:Feynman}, these arise from off-shell contributions 
to the energy integrals.
They are associated with particular elements of $\wh B$, and are
not distributed equally like $B_3$ [see Eq.~\eqref{eq:BShat}].
This structure is needed to
ensure that $\wh \cM^{\prime\,\rm off}_{3,L}$ is unchanged, 
and thus, in particular, remains Lorentz invariant. 
We do not have explicit, all-orders expressions for $\delta B_3$ and $\delta B_2$,
but this does not hinder the derivation.

Finally we discuss the form of $\wh D'$, Eq.~(\ref{eq:Dpmat}). This is the analog in the present
derivation of the matrix $\wh D$ defined in Eq.~\eqref{eq:Dhat}.
The difference here is that the elements of the matrix differ, due to the presence of on-shell projectors
and the Feynman propagator for the off-shell particle.
The flavor structure of $\wh D'$ reflects that which appears in the second stage of the TOPT
derivation, namely the decomposition of $\wh D$ given in Eqs.~(\ref{eq:Ddecomp})--(\ref{eq:deltaFGhat}).
It turns out that this decomposition must be introduced at the first stage in the Feynman approach.
The final difference is the presence here of the wavefunction renormalization factor $Z_i$ multiplying
the pole. As noted above, this equals unity on shell, $Z_i(m_i^2)=1$. In the TOPT analysis, 
the corresponding factor is absorbed into the kernels in a preparatory stage, 
as explained in Appendix A of BS1.\footnote{%
The same approach could be used here, but is not necessary, as we can account for the presence
of $Z_i$ in the next step in the analysis.}

We now turn to the second step in the analysis of the Feynman skeleton expansion.
In this step, we project the three-particle state fully on shell using the $\{k_i\ell m\}$ variables
described above. Since we have set up the intermediate states with
two on-shell particles, the on-shell projection involves adjusting momenta so that the third is placed
on shell. This is very similar to the procedure in the TOPT analysis, where we have to adjust momenta
so that the three already-on-shell particles have total energy $E$.
Indeed, as explained in BS1, the on-shell projection in the TOPT case can be done by a small variation
of the method of Ref.~\cite{\HSQCa} used in the Feynman-diagram analysis.
In particular, near the pole in $D'^{(i)}$ [Eq.~(\ref{eq:Dp})], we have
\begin{equation}
p_i^2-m_i^2 \xrightarrow{p_i^0\to \omega_{p_i}} 2\omega_{p_i} (E - \omega_{p_1}-\omega_{p_2}-\omega_{p_3})
+ \cO\left[ (E - \omega_{p_1}-\omega_{p_2}-\omega_{p_3})^2 \right]\,,
\end{equation}
so that the kinematic factor in $D'^{(i)}$ has the same residue at the pole as that in $D$ [Eq.~(\ref{eq:D})].
This allows us to mirror the decomposition of $\wh D$,
Eqs.~(\ref{eq:Ddecomp})--(\ref{eq:deltaFGhat}), and write
\begin{equation}
\wh D' = \wh F_G + \wh{\delta F}{}'_G\,,
\label{eq:Dpdecomp}
\end{equation}
with $\wh F_G$ exactly as in Eq.~(\ref{eq:FGhat}) above (with a technical restriction described below).
The residue matrix $\wh{\delta F}{}'_G$ differs from that in the TOPT analysis
due to both the presence of the $Z_i$ in Eq.~(\ref{eq:Dp}) and the fact that the off-shellness of the
kernels is different.
The former factor can be dealt with by writing it as $1 + [Z_i(p_i^2) -1]$, with the second term canceling the
pole (since $Z_i$ is an analytic function near $p_i^2-m_i^2$) and thus only contributing to a shift in 
the residue matrix $\wh{\delta F}{}'_G$.
The difference in this matrix is not important, however, 
as it does not impact the subsequent algebraic manipulations.

The technical restriction on $\wh F_G$ is that, in order for the final quantization condition
to contain a Lorentz-invariant three-particle K matrix, we must, in the expression for $\wt G^{(ij)}$,
boost to the pair CMF using the original boost of Ref.~\cite{\HSQCa} rather than that introduced in BS1.
This point is explained in detail in Appendix~\ref{app:not}.

To fully justify Eq.~(\ref{eq:Dpdecomp}), we need to explain how the on-shell projection operators
contained in $D'^{(i)}$, Eq.~(\ref{eq:Dp}), lead to the factors of $P_L$ in $\wh F_G$, Eq.~(\ref{eq:FGhat}).
First we note that, when the kernels $B_2^{(i)}$ and $B_3$ are set fully on shell, 
we must choose a convention for the variables $\{k_i\ell m\}$ that are used.
We follow the convention of the TOPT analysis: if the flavor index is $i$,
then the spectator has flavor $i$ and the spherical harmonics are defined relative to the direction of
the momentum of the particle of flavor $j$ (in the pair CMF), where $j$ follows cyclically after $i$.
We note that projectors in $\wh D'$ are set up so that the spectator is always on shell. 
For example, the first row of $\wh D'$ has projectors 
$\overleftarrow{O}^{(3)}$ and $\overleftarrow{O}^{(2)}$, both of which set flavor 1 on shell,
while the third row contains 
$\overleftarrow{O}^{(2)}$ and $\overleftarrow{O}^{(1)}$, both of which set flavor 3 on shell.
What does not always match, however, is the flavor of the particle that determines the spherical harmonic
decomposition.
We discuss this by considering the first row of $\wh D'$ and focusing on the left-hand harmonic
indices.
Considering the first row again, the first element, $D'^{(3)}$, will be replaced after projection with
$\wt F^{(1)}$, in which the harmonics are determined relative to flavor 2, which matches the convention
of the element $[\wh B]_{x1}$ on the left (with $x$ an arbitrary flavor). 
The same is true for the second element, also $D'^{(3)}$,
which will be replaced with $\wt G^{(12)}$, for which the harmonics of the left index are also determined
relative to flavor 2.
However, the third element, $D'^{(2)}$, is replaced by $\wt G^{(13)}$, for which the
harmonics are determined by the particle of flavor 3, which does not match that used for $[\wh B]_{x1}$.
Indeed, the associated projector, $\overleftarrow{O}^{(2)}$, sets flavor 3 on shell first so as to match
the projection enforced by $\wt G^{(13)}$.
The end result is that the projection applied to $[\wh B]_{x1}$ conflicts with the convention defined above.
In order to bring them into agreement, a factor of $(-1)^\ell$ is needed, and this is provided by the
$P_L$ multiplying $\wt G^{(13)}$ on the left in the $[\wh F_G]_{13}$.
A similar analysis explains all other appearances of $P_L$ in $\wh F_G$.

Given the decomposition of Eq.~(\ref{eq:Dpdecomp}), the remaining steps are algebraically identical
to those of the previous section. In particular, if we set the elements of $\wh \cM'_{23,L}$ fully on shell
using the same convention as just described for the kernels, then we obtain
\begin{equation}
i \wh \cM'_{23,L} = i\wh \cK'_{\df,23,L} \frac1{1 - i \wh F_G i \wh \cK'_{\df,23,L}}\,,
\label{eq:Mp23Lres}
\end{equation}
with
\begin{equation}
i \wh \cK'_{\df,23,L}
=
i \wh B \frac1{1 - i\wh{\delta F}{}'_G\, i \wh B}\,,
\label{eq:Kpdf23L}
\end{equation}
where we are implicitly setting the external coordinates of $\wh B$ on shell.
These are identical in form to Eqs.~(\ref{eq:M23Lres}) and (\ref{eq:Kdf23L}), respectively.
From Eq.~(\ref{eq:Mp23Lres}) we immediately obtain the claimed form of the
quantization condition, Eq.~(\ref{eq:QC2}). In this way we have achieved our goal of recasting the
analysis of the Feynman-diagram-based skeleton expansion in a form that mirrors that of the TOPT approach.

\section{Relation of $\wh \cK'_{\df,3}$ to $\cM_3$}
\label{sec:KtoMFeynman}

In this section we derive the relationship of $\wh \cK'_{\df,3}$
to the physical infinite-volume amplitude, $\cM_3$.
Unlike for the TOPT case discussed above, we do so here in complete detail.
The utility of this result is twofold: first, it will be needed in any application of the formalism
derived in this paper that aims to predict $\cM_3$ from the finite-volume spectrum;
second, it allows us to demonstrate that, expressed in the appropriate basis,
the elements of $\wh \cK'_{\df,3}$ are Lorentz invariant.

The method we use follows that first introduced in Ref.~\cite{\HSQCb}, and extended to the TOPT-based
analysis in Appendix E of BS1. Since we have reformulated the Feynman-diagram-based approach to mirror
that using TOPT, many of the results from BS1 can be taken over almost unchanged. 
The main change is the need to take care of the additional flavor indices.

As in the TOPT analysis, we first pull out the divergence-free finite-volume amplitude using
\begin{equation}
\wh \cM'_{23,L} = \wh \cD_{23,L} + \wh \cM'_{\df, 3, L}\,.
\end{equation}
This has the same form as Eq.~(\ref{eq:M23Lres}), 
and includes the same quantity $\wh\cD_{23,L}$---the difference is the presence
of primes on the other two objects.
No primes are needed on $\wh\cD_{23,L}$ because, as can be seen from its definition in
Eq.~(\ref{eq:D23L}), it depends only on the on-shell two-particle scattering amplitude,
and this is the same whether calculated using TOPT or Feynman diagrams.

Starting from the result for $\wh \cM'_{23,L}$ given in Eq.~(\ref{eq:Mp23Lres}),
we then use the same algebraic steps used above to obtain Eq.~(\ref{eq:Mdf3Lhat}).
These are given explicitly in BS1, and lead to 
\begin{align}
i \wh \cM'_{\df,3,L} &=
\left[1+ i \wh \cD_{23,L} i \wh F_G\right]
i \wh \cK'_{\df,3}
\frac1{ 1 - [1 + i \wh F_{G} i \wh \cD_{23,L}] i \wh F_G i \wh \cK'_{\df,3}}
\left[1+ i \wh F_G i \wh \cD_{23,L}\right]\,.
\label{eq:Mpdf3Lhat}
\end{align}
We now take the $L\to\infty$ limit of $\wh \cM'_{\df,3,L}$, using the $i\epsilon$ prescription of Ref.~\cite{\HSQCb}.
This means that sums over spectator momenta with the singular summands contained in $\wh F_G$
go over to integrals with the poles shifted by the usual $i\epsilon$ prescription.
Specifically, since all sums come with associated factors of $(2\omega L^3)^{-1}$,
the integrals that result come with Lorentz-invariant measure
\begin{equation}
\sum_{\vec k} \frac1{2\omega_k L^3} \xrightarrow{L\to\infty}
\int \frac{d^3k}{2\omega_k(2\pi)^3}
 \equiv \int_{\vec k}\,.
\label{eq:measure}
\end{equation}
The sums over flavor and angular momentum indices remain.

Taking the limit in this way, the elements of $\wh\cM'_{\df,3,L}$  go over to functions of momenta,
\begin{align}
\left[\left[\wh \cM'_{\df,3,L} \right]_{i j}\right]_{p_i \ell' m';   k_j \ell m} &\xrightarrow{L\to\infty}
\left[\wh \cM'_{\df,3}\right]_{ij}(\vec p_i,\vec k_j)_{ \ell' m';  \ell m}\,.
\label{eq:Mdf3infty}
\end{align}
Here we have made all matrix indices explicit, including the spectator-flavor indices $i$ and $j$,
and used a nested structure because the choice of spectator momenta depends on the flavor indices.
An analogous limit holds for the elements of $\wh \cD_{23,L}$, which go over to elements of
$\wh \cD_{23}$.
For the elements of the K matrix $\wh\cK'_{\df,3}$, which are already infinite-volume
quantities, one simply replaces discrete momenta with their continuous counterparts, 
leading to a form like the right-hand side of Eq.~(\ref{eq:Mdf3infty}).
We also need the limit 
\begin{align}
\left[\left[\wh{\overline{\cM}}_{2,L}\right]_{ij}\right]_{p_i \ell' m'; k_j\ell m}
&\xrightarrow{L\to\infty}
\left[\wh{\overline{\cM}}_{2}\right]_{ij}(\vec p_i,\vec k_j)_{\ell' m'; \ell m}
\equiv
\delta_{ij} \overline\delta(\vec p_i-\vec k_i) \delta_{\ell' \ell} \delta_{m' m}
\cM_{2,\ell}^{(i)}(q_{2,k_i}^*)\,,
\label{eq:M2hat}
\end{align}
where $\cM_{2,\ell}^{(i)}$ is the $\ell$th partial wave two-particle scattering amplitude for flavors $j$ and $k$,
and
\begin{equation}
\overline\delta(\vec p-\vec k) \equiv 2\omega_k (2\pi)^3 \delta^3(\vec p -\vec k)\,.
\end{equation}
To obtain smooth limits of the elements of $\wh F_G$, we need to introduce the diagonal matrix
$\wh{2\omega}$ with elements
\begin{equation}
\left[\left[\wh{2\omega}\right]_{ij}\right]_{p_i\ell' m'; k_j \ell m}
= \delta_{i j} \delta_{\vec p_i \vec k_j} \delta_{\ell' \ell} \delta_{m' m} 2 \omega_{k_j} \,,
\end{equation}
in terms of which
\begin{equation}
\wh{2\omega} L^3  \wh F_G \wh{2\omega} L^3 
\xrightarrow{L\to \infty}
\wh F_G^\infty \equiv \wh F^\infty + \wh G^\infty\,.
\end{equation}
The nonvanishing elements of the diagonal matrix $\wh F^\infty$ are
\begin{equation}
\left[\wh F^\infty\right]_{ii}(\vec p_i,\vec k_i)_{\ell' m';\ell m} = \overline\delta(\vec p_i- \vec k_i) 
\delta_{\ell' \ell} \delta_{m' m} \wt \rho^{(i)}_{\PV,\ell}(q_{2,k_i}^*)
\label{eq:Fhatinfty}
\end{equation}
with $\wt\rho^{(i)}_{\PV,\ell}$ a modified phase-space factor,
defined by the nondegenerate generalization of Eq.~(B6) of BS1.
The nonvanishing elements of the off-diagonal matrix $\wh G^\infty$ are
\begin{equation}
\left[\wh G^\infty\right]_{ij}(\vec p_i,\vec k_j)_{\ell' m' ; \ell m} =
\frac{\cY_{\ell' m'}( \vec k_j^{*(p_i)})}{q_{2,p_i}^{*\ell'}} 
\frac{H^{(i)}(\vec p_i) H^{(j)}(\vec k_j)}{b_{ij}^2-m_k^2+i\epsilon}
\frac{\cY_{\ell m}(\vec p_i^{*(k_j)} )}{q_{2,k_j}^{*\ell}} 
\,.
\label{eq:Ghatinfty}
\end{equation}
with $i\ne j$ and $b_{ij}= P - p^{\rm on}_i - k^{\rm on}_j$.

With this notation in hand, we can now take the $L\to \infty$ limit of Eq.~(\ref{eq:Mpdf3Lhat}).
We write the results in a compact notation in which all indices, namely $\{j \vec k_j \ell m\}$, are implicit,
and in which internal indices are implicitly either summed (for $\{j \ell m\}$) or integrated 
(for $\{\vec k_j\}$), the latter with measure (\ref{eq:measure}).
First we note that the limit of $\wh \cD_{23,L}$ satisfies 
\begin{equation}
\wh \cD_{23} = \wh{\overline{\cM}}_2 -
\wh{\overline{\cM}}_2 \wh G^\infty  \wh \cD_{23}\,,
\label{eq:D23res}
\end{equation}
which is a set of coupled integral equations.
The core geometric series in the center of the expression (\ref{eq:Mpdf3Lhat}) becomes
an integral equation for a new matrix quantity that we denote $\wh\cT'$, and which has the same
implicit dependencies as $\wh \cM'_{\df,3}$ and $\wh \cK'_{\df,3}$,
\begin{align}
\wh\cT' &=  \wh \cK'_{\df,3} - \wh \cK'_{\df,3} \left[1 - \wh F_G^\infty  \wh\cD_{23}\right]
i\wh F_G^\infty \wh \cT'\,. 
 \label{eq:Tpres}
\end{align}
Combining these ingredients we have
\begin{equation}
\wh \cM'_{\df,3} = \big(1 - \wh \cD_{23}  \wh F_G^\infty\big) \wh \cT' \big(1 -\wh F_G^\infty \wh \cD_{23}\big)\,,
\label{eq:Mpdf3res}
\end{equation}
in which integral operators are applied to both sides of $\wh \cT'$.
Here $1$ is the identity operator in the full matrix space.

To reconstruct the full asymmetric scattering amplitude, we must add 
back in the part that contains the divergences,
\begin{align}
\wh\cM'_3 &=  \lim_{L\to\infty} \wh \cM'_{3,L} = \wh \cM'_{\df,3} + \wh\cD\,,
\label{eq:Mp3fromdf}
\\
\wh \cD &= \lim_{L\to\infty} \wh\cD_L = \wh \cD_{23} - \wh{\overline{\cM}}_2\,,
\label{eq:cDfromD23}
\end{align}
where $\wh\cD_L$ is defined in Eq.~(\ref{eq:DLdef}).
To combine the elements of $\wh \cM'_3$ 
into the full scattering amplitude $\cM_3$, we need first to convert all elements of
this matrix to the same kinematic variables, namely those of Eq.~(\ref{eq:momlist}).
This is done by multiplying by the appropriate spherical harmonics and summing over angular momentum indices:
\begin{equation}
\left[\wh\cM'_3\right]_{i' i}(\{\vec p\};\{\vec k\}) =
\sum_{\ell',m',\ell,m} \sqrt{4\pi} Y_{\ell' m'}(\wh{\vec p}_{j'}^{*(p_{i'})})
\left\{
\left[\wh\cM'_3\right]_{i'i}(\vec p_{i'},\vec k_i)_{\ell'm';\ell m}\right\}
\sqrt{4\pi}Y_{\ell m}(\wh{\vec k}_{j}^{*(k_{i})})\,.
\label{eq:Mp3sph}
\end{equation}
Here $j$ ($j'$) is the flavor that follows $i$ ($i'$) in cyclic order.
We have changed variables on the left-hand side to those in the original frame, and abused notation
by using the same name for the resulting matrix as that on the right-hand side.
The two quantities are distinguished by their argument.
We obtain the full scattering amplitude by summing the elements of the resulting matrix
\begin{equation}
\cM_3(\{\vec p\};\{\vec k\}) = \bra1 \wh \cM_3'(\{\vec p\};\{\vec k\}) \ket1\,.
\label{eq:M3res}
\end{equation}
We note that no prime is needed for $\cM_3$ since one obtains the same result whether
decomposing into TOPT or Feynman diagrams.
We recall that this result holds for the fully on-shell amplitude.

We now return to the issue of the Lorentz invariance of $\wh\cK'_{\df,3}$.
The arguments we give are an elaboration of those first described in Ref.~\cite{\BHSQC}.
By construction, all elements of the flavor matrix $\wh \cM'_3$ are Lorentz invariant, since they
are defined as sums of Feynman diagrams.
This holds only when the amplitude is combined with spherical harmonics,
as in Eq.~(\ref{eq:Mp3sph}).
What we need, however, are the transformation properties of amplitudes
expressed in the $\{\vec k_i \ell m\}$ basis, 
since this is what enters relations such as Eq.~(\ref{eq:Mpdf3res}).
The amplitudes in this basis are not invariant under rotations, 
since they depend on an arbitrary choice of quantization axis (conventionally the $z$ axis).
Instead, they transform under rotations by multiplication by appropriate Wigner D matrices,
due to the standard result
\begin{equation}
Y_{\ell m}(R \hat n) = \sum_{m'} \cD_{m m'}^{(\ell)}(R) Y_{\ell m'}(\hat n)\,.
\end{equation}
This rather trivial dependence also leads to a dependence on boosts, as follows.
Consider a momentum configuration $\{\vec k\}$ and choose 
$\vec k_1$ to be the spectator momentum.
To define the coordinates $\{\vec k_1\ell m\}$ 
we must boost to the $23$ pair CMF and then decompose into harmonics.
Now imagine that we first do an overall boost of the initial configuration, leading to momenta
$\{\vec k'\}$. This time the spectator momentum is $\vec k'_1$. When we boost to the CMF of the $23$ pair,
we end up in the same frame as before, except for an overall rotation.
This is simply because a product of two boosts can be written as a single boost combined with a rotation.
This implies that the elements of $\wh \cM'_3$ in the $\{\vec k_i \ell m\}$ basis will transform with 
Wigner matrices that depend on the choice of flavor index and on the spectator momentum.
In the following, we refer to these transformation properties in as ``standard.''
Any flavor matrix in the $\{\vec k \ell m\}$ basis that has standard transformation properties
will yield a Lorentz-invariant amplitude when combined with harmonics as in Eq.~(\ref{eq:Mpdf3res}).

We now argue that the standard transformation properties of $\wh \cM'_3$ are reproduced
by Eqs.~(\ref{eq:Mpdf3res}) and (\ref{eq:Mp3fromdf})
if the elements of $\wh \cK'_{\df,3}$ themselves transform in the standard way.
First we note that the elements of $\wh{\overline{\cM}}_2$ have standard transformation properties
since the underlying amplitude $\cM_2$ and the quantity $\overline \delta(\vec p-\vec k)$ are both Lorentz
invariant.
Next we argue that the elements of $\wh D_{23}$, given by Eq.~(\ref{eq:D23res}), transform in the standard
way. Iteratively expanding this equation yields an alternating series of factors of $\wh{\overline{\cM}}_2$
and $\wh G^\infty$. From Eq.~(\ref{eq:Ghatinfty}), we see that all quantities in $\wh G^\infty$ are
Lorentz invariant ($q_{2,p_i}^*$, $b_{ij}^2$, $|\vec k_j^{*(p_i)}|$, etc.~and the cutoff functions)
except for the directions of $\vec k_j^{*(p_i)}$ and $\vec p_i^{*(k_j)}$.
These vectors will, in general, be rotated if an overall boost is
first applied, due to the above-discussed properties of successive boosts.
This rotation of the vectors leads to a multiplication of the corresponding $\ell m$ indices by 
appropriate Wigner D matrices.
However, these D matrices cancel those arising from the standard
transformation of the adjacent elements of $\wh{\overline{\cM}}_2$. 
The key point here is that the same rotation appears for contracted indices, since the same boost to the
pair CMF is used.
Due to this cancellation, only the external Wigner D matrices survive---those associated with the
external indices of the factors of $\wh{\overline{\cM}}_2$ on the ends of the chain.
Thus $\wh D_{23}$ indeed has standard transformation properties.

The remainder of the argument follows in a similar way.
The only additional result that we need is the transformation property of the $\wh F^\infty$ part of $\wh F_G^\infty$.
From Eq.~(\ref{eq:Fhatinfty}), we see that the elements of $\wh F^\infty$ are, in fact,
invariant under rotations and boosts. This implies that Wigner D matrices arising from amplitudes
on the two sides of each element of $\wh F^\infty$ cancel.
Together with the result for $\wh G^\infty$ discussed above, this implies that any sequence
of amplitudes with standard transformation properties alternating with factors of $\wh F_G^\infty$ will
itself have standard transformation properties.
Thus, using Eq.~(\ref{eq:Tpres}), if $\wh\cK'_{\df,3}$ has standard transformations, it follows that
$\wh \cT'$ does as well, and, using  Eq.~(\ref{eq:Mpdf3res}), the same holds for $\wh \cM'_{\df,3}$.
Finally, using Eqs.~(\ref{eq:Mp3fromdf}) and (\ref{eq:cDfromD23}), and the standard transformation
properties of $\wh\cD_{23}$ and $\wh{\overline{\cM}}_2$, we find that
$\wh\cM'_3$ transforms in the standard way, which is the desired result.

To complete the discussion we need to show that, if $\wh \cK'_{\df,3}$ does not transform in the
standard way, then neither does $\wh \cM'_3$. This seems highly plausible, since the 
above-described cancellation of  Wigner D matrices would no longer occur. 
Another way of making this argument is to invert the relationship between $\wh \cM'_3$ and $\wh \cK'_{\df,3}$, 
i.e.~to determine the latter from the former.
This can be done, for example,
by first inverting Eq.~(\ref{eq:Mpdf3Lhat}) in finite volume, and then taking the $L\to\infty$ limit.
This leads to an expression involving inverses of integral operators.
By expanding out the inverses in geometric series,
the relationship one obtains always involves sums of 
products of the amplitudes $\wh \cM'_{\df,3}$ and $\wh D_{23}$ alternating with factors of $\wh F_G^\infty$,
and these preserve standard transformation properties.
Thus we claim that $\wh \cK'_{\df,3}$ does transform in the standard way, 
and therefore that, when it is combined with harmonics as in Eq.~(\ref{eq:Mp3sph}),
its elements will be Lorentz invariant.

\section{Symmetric form of the quantization condition}
\label{sec:symmetric}

In this section we describe the derivation of our third and final
form of the quantization condition, Eq.~(\ref{eq:QC3}).
This is written in terms of a single Lorentz-invariant three-particle K matrix,
$\wt \cK_{\df,3}$, which has no flavor indices, and which we thus call symmetric.
To obtain the new form we follow steps analogous to those used in BS1 to
connect the asymmetric and symmetric forms of the quantization condition for identical particles,
with suitable generalizations for nondegenerate particles.
In addition, we provide the integral equations relating $\cM_3$ to $\wt\cK_{\df,3}$,
Eq.~(\ref{eq:M3hatres}).

\subsection{Symmetrization operators}
\label{subsec:symops}

$\wt\cK_{\df,3}$ is obtained by symmetrizing a modified version of $\wh\cK'_{\df,3}$.
We have already encountered symmetrization when constructing $\cM_3$ in
Eq.~(\ref{eq:M3res}), but here we give more details, and introduce some helpful notation.
In particular, we define the symmetrization operators $\overrightarrow{\cS}$ and
$\overleftarrow{\cS}$, which play a central role in the final step of the derivation.

The symmetrization operators act on vectors in flavor space, e.g.~the
row vector $X_j = [\wh \cK'_{\df,3}]_{ij}$ with $i$ fixed. 
In our notation, the index $j$ plays two roles. First, it labels the element of
the vector, and in general the three elements are different. Second, it determines the 
coordinates that are used to describe the on-shell amplitude, with the $j$th element
using coordinates $\{k_j \ell m\}$. Symmetrization acts on the underlying elements, 
but not on the coordinates, and so these two roles of the index must be decoupled.
Here we use $X_j(\{\vec k\})$ to describe the underlying element, 
which depends on the on-shell momenta $\{\vec k\}$,
and make coordinates explicit,
\begin{equation}
\vec X = \left(X_1(\{\vec k\})_{k_1 \ell m}, X_2(\{\vec k\})_{k_2 \ell m}, 
X_3(\{\vec k\})_{k_3 \ell m} \right)\,.
\end{equation}
We recall that the relation between an underlying infinite-volume on-shell quantity 
$X(\{\vec k\})$ and its expression in terms of coordinates $\{k_j \ell m\}$ is given by
Eq.~(\ref{eq:basischange}).
An example of this notation is the expression for the underlying quantity $X_2(\{\vec k\})$
in terms of the coordinates $\{k_1\ell m\}$,
\begin{equation}
X_2(\{\vec k\}) = \sum_{\ell m} 
X_2(\{\vec k\})_{k_1 \ell m} \sqrt{4\pi} Y_{\ell m}(\wh{\vec k}_2^{*(k_1)})\,.
\end{equation}

The left-acting symmetrization operator is defined by
\begin{align}
\vec X \overleftarrow{\cS} &\equiv
\left( 
X_\Sigma(\{\vec k\})_{k_1\ell m}, X_\Sigma(\{\vec k\})_{k_2\ell m}, X_\Sigma(\{\vec k\})_{k_3\ell m}
\right)\,,
\\
X_\Sigma(\{\vec k\}) &= X_1(\{\vec k\}) + X_2(\{\vec k\}) + X_3(\{\vec k\})\,.
\end{align}
The key point is that the same underlying element appears in all positions, but is expressed
in terms of different coordinates.
The right-acting operator $\overrightarrow{\cS}$ is defined analogously for column vectors.
We stress that this definition relies on the fact that the underlying elements are 
infinite-volume functions, defined for all $\{\vec k\}$, rather than finite-volume objects
defined only for momenta in the finite-volume set.

\subsection{Symmetrization identities}

In BS1, three ``asymmetrization" identities 
[Eqs.~(102)-(104) of that work]
were derived and used to convert the
symmetric, identical-particle quantization condition of Ref.~\cite{\HSQCa} into
an asymmetric form. Here we use a generalization of these identities to move in the
other direction, from the asymmetric to the symmetric form. 
Thus we refer to them in this work as symmetrization identities.

These identities apply when factors of $\wh F_G$ lie between two matrix amplitudes,
e.g.~$\wh \cD_{23,L}$ and $\wh\cK'_{\df,3}$.
To simplify the presentation, and without loss of generality, we consider the case where
$\wh F_G$  lies between a row vector $\vec X$ and a column vector $\vec Z$.
The identities are then
\begin{align}
\vec X \wh F_G \vec Z &= \vec X \wh F \overrightarrow{\cS}\vec Z 
+ \vec X \wh{\overrightarrow{\cI}}_G \vec Z\,,
\label{eq:symid1}
\\
&= \vec X \overleftarrow{\cS} \wh F \vec Z
+ \vec X \wh{\overleftarrow{\cI}}_G \vec Z\,,
\label{eq:symid2}
\\
&= \tfrac13 \vec X \overleftarrow{\cS} \wh F \overrightarrow{\cS} \vec Z
+ \vec X \wh \cI_{FG} \vec Z\,.
\label{eq:symid3}
\end{align}
As usual, these hold up to exponentially suppressed corrections.
The key aspect of these results is that the $\wh G$ contribution to $\wh F_G=\wh F+\wh G$ 
on the left-hand side can be replaced by 
one or more symmetrization operators on the right-hand sides, 
aside from integral operators $\wh{\overrightarrow{\cI}}_G$,
$\wh{\overleftarrow{\cI}}_G$, and  $\wh \cI_{FG}$, 
which sew together the two vectors into an extended infinite-volume quantity.

The derivation of these identities is sketched in Appendix~\ref{app:symid}. 
We also provide there the definitions of the integral operators.

\subsection{Applying the symmetrization identities}

We wish to apply the identities to the result obtained above for
$\wh \cM'_{\df,3,L}$, Eq.~(\ref{eq:Mpdf3Lhat}).
The nontrivial aspect of the resulting manipulations is dealing with the integral operators
on the right-hand sides of the identities, 
namely $\wh{\overrightarrow{\cI}}_G$, $\wh{\overleftarrow{\cI}}_G$, and $\wh \cI_{FG}$.
The steps that we follow mirror the approach taken in BS1
[see Eqs.~(105)-(107) and (112)-(113) of that work],
although in that work we were using the identities to asymmetrize a symmetric form,
while here we are working in the opposite direction.

We begin by introducing an intermediate ``decorated" K matrix given by
\begin{equation}
i \wh \cK''_{\df,3} = i \wh Z
\frac1{1 + \left[-i \wh \cI_{FG} + i \wh{\overleftarrow{\cI}}_G 
i \wh{\overline{\cK}}_{2,L} i \wh{\overrightarrow{\cI}}_G\right] i \wh Z}\,,
\label{eq:Kdf3hatpp}
\end{equation}
where
\begin{equation}
i\wh Z \equiv 
\frac1{1-i \wh{\overline{\cK}}_{2,L} i \wh{\overrightarrow{\cI}}_G}
i\wh \cK'_{\df,3}
\frac1{1- i \wh{\overleftarrow{\cI}}_G i \wh{\overline{\cK}}_{2,L}}\,.
\label{eq:Zhat}
\end{equation}
We stress that, although these equations are written in terms of finite-volume
matrices, they are equivalent to infinite-volume integral equations, up to exponentially
suppressed corrections. This is because the decorations themselves involve
integral operators, and because we have chosen a generalized PV prescription
such that the two-particle K matrix $\cK_2$ has no poles.

We now rewrite $\wh \cM'_{\df,3}$ in terms of $\wh \cK''_{\df,3}$.
Using the steps sketched in Appendix~\ref{app:alg}, we find
\begin{equation}
i \wh \cM'_{\df,3,L} = 
\left[1 +i \wh \cD_{23,L} i (\wh F_G - \wh{\overrightarrow{\cI}}_G)\right]
 i\wh \cT_L
\left[1 + i (\wh F_G - \wh{\overleftarrow{\cI}}_G ) i \wh \cD_{23,L}\right]\,,
\label{eq:toshow1}
\end{equation}
where
\begin{equation}
i\wh \cT_L = i\wh \cK''_{\df,3}
\frac1{1 - \left[i(\wh F_G - \wh \cI_{FG})
+ i(\wh F_G-\wh{\overleftarrow{\cI}}_G) i\wh \cD_{23,L}
   i(\wh F_G-\wh{\overrightarrow{\cI}}_G) \right] i \wh \cK''_{\df,3} }
   \,.
   \label{eq:TLhat}
\end{equation}
Using the symmetrization identities (\ref{eq:symid1})-(\ref{eq:symid3}), these can be rewritten as
\begin{equation}
i \wh \cM'_{\df,3,L} = 
\left[1 + i \wh \cD_{23,L} i \wh F \overrightarrow{\cS}\right]
 i\wh \cT_L
\left[1 + \overleftarrow{\cS} i \wh F i \wh \cD_{23,L}\right]\,,
\label{eq:Mpdf3Lhatres2}
\end{equation}
where
\begin{equation}
i\wh \cT_L = i\wh \cK''_{\df,3}
\frac1{1 - \left[\tfrac13 \overleftarrow{\cS} i \wh F \overrightarrow{\cS}
+ \wh{\overleftarrow{\cS}} i \wh F i\wh \cD_{23,L}
   i \wh F \wh{\overrightarrow{\cS}} \right] i \wh \cK''_{\df,3} }
   \,.
   \label{eq:TLhatres2}
\end{equation}
Expanding out the geometric series we see that, except at the ends, $\wh \cK''_{\df,3}$
is sandwiched between two symmetrization operators, and thus fully symmetrized.

\subsection{Quantization condition}
We recall from above that the quantization condition can be obtained from the poles
in $\wh \cM'_{\df,3,L}$. Looking at Eq.~(\ref{eq:Mpdf3Lhatres2}), we see that
poles can only arise from the factors of $\wh F$, $\wh \cD_{23,L}$, or $\wh \cT_L$.
The former only has poles at free energies, which cannot be present in the interacting
spectrum, and must cancel in the full expression. 
Poles arising from $\wh \cD_{23,L}$ do not depend on $\wt\cK_{\df,3}$, and thus 
also must either be absent or cancel, since all finite-volume energies must have some dependence
on the three-particle interaction.
Thus the only source that remains is $\wh \cT_L$.
To determine its poles, we rewrite Eq.~(\ref{eq:TLhatres2}) as
\begin{equation}
i\wh \cT_L = i\wh \cK''_{\df,3}
+ i\wh \cK''_{\df,3} \overleftarrow{\cS}
\frac1{1 - i \wh F_3 i\wh{\wt{\cK}}_{\df,3} }
  i \wh F_3 \overrightarrow{\cS} i\wh \cK''_{\df,3}   \,.
   \label{eq:TLhatres3}
\end{equation}
where
\begin{equation}
\wh F_3 = \tfrac13 \wh F - \wh F \wh \cD_{23,L} \wh F\,,
\label{eq:F3hat}
\end{equation}
and
\begin{equation}
\wh{\wt{\cK}}_{\df,3} =
\overrightarrow{\cS} \wh \cK''_{\df,3} \overleftarrow{\cS}\,.
\label{eq:symmKdf}
 \end{equation}
 Since poles can only arise from the second term in Eq.~(\ref{eq:TLhatres3}),
 we obtain our third and final form for the quantization condition,
 \begin{equation}
 \det\big(1 + \wh F_3 \wh{\wt{\cK}}_{\df,3} \big) = 0\,.
 \label{eq:QC3}
 \end{equation}
We refer to this as the symmetric form of the quantization condition.
 
Comparing to the quantization condition for identical particles derived in Ref.~\cite{\HSQCa},
we see that the nondegenerate result has the same form, but with an additional layer of
matrix indices. This is what one might have naively expected, but, as we have shown,
it is nontrivial to obtain this generalization.
A key property of the matrix $\wh{\wt{\cK}}_{\df,3}$ is that 
{\em it contains the same underlying K matrix in each element},
due to the presence of symmetrization operators on both sides of $\wh \cK''_{\df,3}$ 
in Eq.~(\ref{eq:symmKdf}).
The underlying K matrix is
\begin{align}
\wt{\cK}_{\df,3}(\{\vec p\};\{\vec k\}) 
&= \sum_{i,j} [\wh\cK''_{\df,3}]_{ij}(\{\vec p\};\{\vec k\})
\,,
\end{align}
where, on the right-hand side, each element of
$\wh \cK''_{\df,3}$ has been converted from the $\{k\ell m\}$ basis to the
momentum basis, using the appropriate generalization of Eq.~(\ref{eq:basischange}),
and then summed.
The difference between the elements of the matrix
$\wh{\wt{\cK}}_{\df,3}$ arises only because $\wt\cK_{\df,3}$
is expressed in different coordinates,
\begin{align}
\left[ \wh{\wt{\cK}}_{\df,3}\right]_{ij}
&= \wt{\cK}_{\df,3}(\{\vec p\};\{\vec k\})_{p_i\ell' m'; k_j \ell m}\,,
\end{align}

We stress that the complicated nature of the relation between $\wt \cK_{\df,3}$
(which appears in our final quantization condition)
and the elements of $\wh \cK'_{\df,3}$ 
[which appear in the previous form, Eq.~(\ref{eq:QC2})]
is not a practical concern, because we are
simply replacing one set of unknown quantities with another.
In fact, as already stressed above, the final form of the condition, Eq.~(\ref{eq:QC3}) has
the great advantage of requiring the parametrization of only a single K matrix, rather than nine.

The form of $\wh F_3$, Eq.~(\ref{eq:F3hat}), is also the same as that in Ref.~\cite{\HSQCa},
although here the matrix structure has more content. In particular, the entries of the
diagonal flavor matrix $\wh F$ are different, as they correspond to a different choice of
spectator flavor. Similarly, the factors of $\wh G$ contained in $\wh \cD_{23,L}$ have
a nontrivial matrix structure. 
Since this matrix version of $\wh F_3$ is a quantity not previously considered, we
note that it can be written as 
\begin{align}
\wh F_3 &= \tfrac13 \wh F - \wh F \frac1{\wh{\overline{\cK}}{}_{2,L}^{-1} + \wh F_G} \wh F
\\
%&= \tfrac13 \wh F - \wh F \frac1{\wh{\overline{\cK}}_{2,L}^{-1} + \wh F + \wh G } \wh F\,,
%\\
&= \wh F \left[ -\frac23 + \frac1{1+ (1 + \wh{\overline{\cK}}_{2,L} \wh G)^{-1}
\wh{\overline{\cK}}_{2,L} \wh F } \right]\,,
\end{align}
which are generalizations of forms that have been used for identical particles.

\subsection{Relating $\wt \cK_{\df,3}$ to $\cM_3$}
\label{sec:KtoMsymm}

The final ingredient in the symmetrized form of the formalism for nondegenerate particles
is to relate $\wt \cK_{\df,3}$ to $\cM_3$. The approach we take has already
described in detail in Sec.~\ref{sec:KtoMFeynman}, so here we provide only a summary.

We begin by noting that $\wh\cM'_{\df,3,L}$ cannot be
written in terms of $\wt \cK_{\df,3}$ alone, because the ``$1$" terms in square brackets
in Eq.~(\ref{eq:Mpdf3Lhatres2}) do not involve symmetrization operators.
However, since $\cM_3$ is itself symmetrized [as in Eq.~(\ref{eq:M3res})],
it can be written in terms of a symmetrized version of $\wh\cM'_{\df,3,L}$ 
which itself {\em can} be written solely in terms of $\wt \cK_{\df,3}$.
To explain this it is useful to introduce a matrix version of $\cM_{3}$, whose elements are
\begin{equation}
\left[\wh \cM_{3}\right]_{ij} \equiv
\cM_{3}(\vec p_i;\vec k_j)_{\ell' m';\ell m}\,.
\label{eq:M3matrix}
\end{equation}
In other words, all elements are given by
the same underlying quantity, but expressed in different coordinates.
Equations (\ref{eq:Mp3fromdf})-(\ref{eq:M3res}) can then be rewritten as
\begin{equation}
\wh \cM_3 = \lim_{L\to \infty}\left(\overrightarrow \cS \wh \cM'_{\df,3,L} \overleftarrow\cS\right)
+ \overrightarrow \cS \wh\cD \overleftarrow \cS\,,
\end{equation}
where $\cD$ is defined in Eq.~(\ref{eq:cDfromD23}), and
the infinite-volume limit is taken using the $i\epsilon$ pole prescription.
Here we are using the fact
that the symmetrization operators work equally well on infinite-volume quantities.
Using the properties
\begin{equation}
\overrightarrow\cS \overrightarrow\cS = 3 \overrightarrow\cS 
\,,
\qquad
\overleftarrow\cS \overleftarrow\cS = 3 \overleftarrow\cS\,,
\end{equation}
we obtain
\begin{equation}
\overrightarrow \cS \wh \cM'_{\df,3,L} \overleftarrow\cS
=
\overrightarrow\cS
\left[\tfrac13 + i \wh \cD_{23,L} i \wh F\right] 
\wh{\wt{\cK}}_{\df,3} 
\frac1{1-i \wh F_3 i\wh{\wt{\cK}}_{\df,3}}
\left[\tfrac13 + i \wh F i \wh \cD_{23,L}\right] \overleftarrow\cS\,,
\label{eq:symMpdf3L}
\end{equation}
which indeed depends only on the symmetrized $\wt\cK_{\df,3}$.

These results can be written as integral equations using results from
Sec.~\ref{sec:KtoMFeynman}. The equation for $\wh \cD_{23}= \lim_{L\to\infty} \wh \cD_{23,L}$
is unchanged, Eq.~\eqref{eq:D23res}; from this and the result
for $\wh{\overline{\cM}}_2$, Eq.~(\ref{eq:M2hat}), we obtain 
$\wh\cD=\wh\cD_{23}-\wh{\overline{\cM}}_2$.
The central geometric series in Eq.~(\ref{eq:symMpdf3L}) is solved by the integral equation
\begin{equation}
\wh{\wt{\cT}} =  \wh{\wt{\cK}}_{\df,3}
- \wh{\wt{\cK}}_{\df,3}
\left[\tfrac13 - \wh F^\infty \wh\cD_{23}\right] \wh F^\infty \wh{\wt{\cT}} \,,
\label{eq:Thattilderes}
\end{equation}
which is the symmetric version of the equation for $\wh \cT'$, Eq.~(\ref{eq:Tpres})
Despite its matrix form, this is an integral equation for a single function
$\wt \cT(\{\vec p\};\{\vec k\})$
which is packaged into the matrix $\wh{\wt{\cT}}$
in the same manner as in Eq.~(\ref{eq:M3matrix}).
We next apply integral operators to $\wh{\wt{\cT}}$, combine with $\wh\cD$, and symmetrize
to obtain the final result
\begin{equation}
\wh\cM_3 = \overrightarrow \cS
\left\{ \left[\tfrac13 - \wh\cD_{23} \wh F^\infty\right] \wh{\wt{\cT}}
\left[\tfrac13 - \wh F^\infty \wh\cD_{23}\right]
+ \wh\cD \right\}
\overleftarrow \cS\,.
\label{eq:M3hatres}
\end{equation}
Again, the matrix form is somewhat deceptive, as one needs only to calculate
a single element of $\wh \cM_3$, since all elements contain the same function expressed
in different coordinates.

\subsection{Symmetrizing the TOPT quantization condition}
\label{sec:symTOPT}

The steps we have taken to obtain the symmetrized quantization condition starting
from the result for $\wh\cM'_{\df,3,L}$, Eq.~(\ref{eq:Mpdf3Lhat}),
can also be applied to the TOPT result for
$\wh \cM_{\df,3,L}$, Eq.~(\ref{eq:Mdf3Lhat}).
Since these two equations have the same form, differing only by the version of
$\wh \cK_{\df,3}$ that enters, the final results of the symmetrization process will
also have the same form.
In particular, we obtain a TOPT-based quantization condition having the form
of Eq.~(\ref{eq:QC3}),
and an equation for $\wh\cM_3$ having the same form as Eq.~(\ref{eq:M3hatres}),
except in both cases we are starting from $\wh \cK_{\df,3}$
rather than $\wh \cK'_{\df,3}$.

We now argue, however, that these new forms of the final results are
actually {\em exactly the same}.\footnote{%
This only holds if the HS boost is used in the TOPT approach.}
 In other words, although we start with different
versions of $\wh\cK_{\df,3}$ in the two cases, one Lorentz invariant and the other not,
we claim that, after the manipulations involved in symmetrization, the final resulting
symmetrized quantity $\wh{\wt{\cK}}_{\df,3}$ is the same.
Our argument for this is the same as that we used for identical particles in BS1.
The key point is that, after symmetrization, one ends with an equation for the
same quantity, $\wh\cM_3$, in both cases.
This is because symmetrization corresponds to summing all diagrams that contribute,
and this results in the full scattering amplitude irrespective of 
whether one uses TOPT or Feynman diagrams.
If we assume that the relation between $\cM_3$
and $\wt{\cK}_{\df,3}$ given by Eqs.~(\ref{eq:Thattilderes}) and (\ref{eq:M3hatres}) 
is invertible,
then it must be that the symmetrized K matrix is the same for
both TOPT and Feynman approaches.
In more physical terms, the assumption is that any changes to the K matrix 
(which is simply a short-distance three-particle interaction) will lead
to a change in the full scattering amplitude.

If we accept this argument, then we can obtain the symmetrized form of the
quantization condition and the relation between $\wt\cK_{\df,3}$ and $\cM_3$
{\em without using the Feynman-diagram approach as an intermediate step}.

\section{Summary and Outlook}
\label{sec:conc}

In this paper we have generalized the relativistic three-particle quantization condition,
and the relation of the intermediate three-particle K matrix to $\cM_3$,
to the case of non-degenerate particles.
We have derived three versions of the quantization condition:
two asymmetric forms---Eqs.~(\ref{eq:QC1}) and (\ref{eq:QC2})---each involving a flavor matrix
composed of nine three-particle K matrices, and a symmetric form---Eq.~(\ref{eq:QC3})---involving
only a single K matrix. The latter two versions of the quantization condition involve 
Lorentz-invariant K matrices. These three quantization conditions are the generalizations of 
those for identical particles obtained in BS1 (the first two) and Ref.~\cite{\HSQCa} (the final
form).

The main new feature that arises for nondegenerate particles is the need to introduce an
additional flavor index on the matrices, at least at intermediate steps.
This corresponds to the different choices for the flavor of the external spectator particles.
Even though the symmetric form of the quantization condition involves a K matrix,
$\wt \cK_{\df,3}$,
that has been symmetrized with respect to these indices, the quantization condition
must still be written in terms of flavor matrices because of the different kinematical
factors arising from the different choices of spectator flavor.
Aside from this extra layer of indices, the form of the quantization conditions is
essentially unchanged from those for identical particles.

The path that we have taken to derive the final, symmetric, form of the quantization condition
has been rather lengthy and indirect. We first use TOPT, where the derivation 
is relatively straightforward, but involves an asymmetric and  Lorentz-noninvariant K matrix.
Then we revert to a Feynman-diagram expansion of the amplitudes,
and develop a new all-orders approach that yields expressions that mirror those from TOPT,
and leads to a quantization condition that involves a K matrix that is Lorentz invariant,
although still asymmetric.
Finally, we use symmetrization identities to obtain a quantization condition involving
a K matrix that is both Lorentz invariant and symmetric.
A natural question is whether there is a shorter path to the final result, especially since,
as already noted, it has a very similar form to that derived in Ref.~\cite{\HSQCa} for
identical particles. 
For example, could one not simply generalize every step in the
derivation of Ref.~\cite{\HSQCa}? We think that this is almost certainly possible,
but have not followed that path as the derivation of Ref.~\cite{\HSQCa} is itself very
lengthy and does not lead to explicit expressions for the K matrices and other quantities.
The approach followed here is explicit at every stage, so that, for example,
we have given a chain of expressions relating $\wt\cK_{\df,3}$ back to the Feynman
or TOPT Bethe-Salpeter amplitudes.
Now that we have done the groundwork, we expect
that the present approach can be simply generalized to other cases of interest.

Furthermore, as discussed in Sec.~\ref{sec:symTOPT}, if the relation between
$\cM_3$ and $\wt\cK_{\df,3}$ is invertible, then we can derive the final form of the
quantization condition without the need for the intermediate step involving Feynman diagrams.
Instead, we need only symmetrize the expressions that result from the TOPT approach.
Although we have not demonstrated the necessary invertibility, we think that this is a physically
reasonable assumption. Thus, although here we have provided the longer path to the final result,
in which no assumptions are needed, we think that 
the shorter, two-step path can be used for future generalizations.

The theory that we consider in our derivations, 
which has a $\mathbb Z_2$ symmetry for each of the real scalar fields, 
is somewhat artificial, and has no direct application in QCD. 
It is clear from the derivations, however, 
that all that matters for the validity of the final results is that the kinematic constraints are such that 
the only on-shell intermediate state consists of one particle of each flavor.
One example, already discussed in Sec.~\ref{sec:setup}, is the
$D_s^+ D^0 \pi^- $ system, which is chosen such that each of the
three particles has a different total flavor. Here the $U(1)$ flavor symmetries are playing a similar
role to the $\mathbb Z_2$ symmetries in our standard theory.
Another example is the $D_s^+ D^0 D^+$ system, and
similar examples can be constructed containing $B$ mesons.

Since there are few direct applications, and also because this paper is quite lengthy,
we have reserved discussion of issues related to practical implementation,
as well as various cross checks, for a follow-up article. 
For example, a threshold expansion of $\wt \cK_{\df,3}$ needs to be developed,
along the lines of Ref.~\cite{\dwave}.
Also of interest is the degenerate limit of our formalism,
which can be related to the recently developed generalization of the symmetric
quantization condition of Ref.~\cite{\HSQCa} 
to three pions of arbitrary isospin in isosymmetric QCD~\cite{\isospin}.
The $I=0$ case can be described by both formalisms, because this only has contributions from
$\pi^+\pi^0\pi^-$ intermediate states, with no mixing with the $3\pi^0$ state.
Another issue we aim to address is the relation between the degenerate limit of
our formalism and the results for identical particles obtained in Ref.~\cite{\HSQCa} and BS1.

We also intend, in this follow-up work, to present the generalization of the
formalism that will allow application to systems
of greater phenomenological interest.
A simple extension is to ``2+1'' systems like $K^+\pi^+\pi^+$, 
with two identical particles and a third that is different.
Cases with multiple three-particle channels are also of interest, for example 
$\pi^+\pi^-\pi^0\leftrightarrow 3\pi^0$ with $m_u\ne m_d$,
and the $D_s^+ D^0 \pi^- \leftrightarrow D^0 D^0 K^0$ system mentioned above.
Another ``2+1'' system of great interest is $N\pi\pi$, given its relevance to the Roper resonance,
but in this case one needs also to include the $N\pi$ channel, requiring a combination
of the methods introduced here and those of Refs.~\cite{\BHSQC,\largera}.
We also note that the quantization condition for the $DDK$ system has recently been determined
in the s-wave approximation using NREFT \cite{Pang:2020pkl}.

As already observed, we expect that the symmetric form of the quantization condition, 
Eq.~(\ref{eq:QC3}), will be most useful for
practical applications, since it requires parametrizing only a single three-particle K matrix.
Nevertheless, the asymmetric, Lorentz-invariant form, Eq.~(\ref{eq:QC2}),
may be useful in order to determine the relation
to the finite-volume unitarity (FVU) approach to deriving the
quantization condition~\cite{\MD,\MDpi}. 
In the case of identical particles, we have recently shown that
the asymmetric RFT quantization condition, when written in terms of the
R matrix introduced in Refs.~\cite{\Maiisobar,\isobar}
(an alternate version of the three-particle K matrix),
is equivalent to the FVU quantization condition~\cite{\BSequiv}.
We expect that this equivalence can be extended to the nondegenerate case,
where the R matrix is  extended to a flavor matrix.

\section*{Acknowledgments}

We thank Ra\'ul Brice\~no, Andrew Jackura, 
and Fernando Romero-L\'opez for comments on the manuscript.
This work was supported in part 
by the U.S. Department of Energy contract DE-SC0011637.

\appendix
\section{Technical details}
\label{app:not}

In this appendix we collect some technical details relevant for the discussion of the main text.

First we discuss the smooth cutoff functions $H^{(i)}(\vec p_i)$ that enter into $\wt G^{(ij)}$.
Aside from one feature, these are straightforward generalizations of the cutoff introduced in Ref.~\cite{\HSQCa} and
used in all subsequent RFT works.
These functions smoothly cut off the sums or integrals over the spectator momentum $\vec p_i$
once the flavor $j+k$ pair lies far below its threshold.
To implement this, we use the quantity $\sigma_i= (P-p_i)^2$, which equals $(m_j+m_k)^2$ at the pair threshold,
and decreases as one drops below this threshold.
For identical particles, $\sigma_i$ is the same as the 
quantity $E_{2,k}^{*2}$ used in the definition of the cutoff function in  Ref.~\cite{\HSQCa}.
If we use the same boost as in that work, which is discussed below,
then we are restricted to the range $\sigma_i>0$, since the boost becomes singular when $\sigma_i=0$.
Another constraint, described in BS1,
is that the cutoff function should equal unity for some range below the pair threshold.
This ensures that all terms which are dropped in the derivation are exponentially suppressed.
Finally, we also need the function to be Lorentz invariant.
One choice that satisfies these requirements is
\begin{align}
H^{(i)}(\vec p_i) &= J(z_i)\,,\quad
z_i = (1+\epsilon_H) \frac{\sigma_i}{(m_j+m_k)^2}\,,
\\
J(z) &= \left\{ 
\begin{array}{ll} 0, & z \le 0 \\ 
\exp\left(-\frac1z\exp\left[-\frac1{1-z}\right]\right), & 0 < z < 1\\
1, & 1 \le z \,. \end{array}\right.
\end{align}
Here choosing $\epsilon_H > 0$ ensures that the cutoff function reaches unity 
below threshold, at the point where $\sigma_i=(m_j + m_k)^2/(1+\epsilon_H)$.
We expect that, in practice, a value $\epsilon_H\sim 0.1$ should be sufficiently large.
We stress that the choice of cutoff functions is not unique, and the choice above is simply
an example, albeit one that is close to those that have been used in previous numerical 
implementations~\cite{\BHSnum,\dwave,\largera,\HHanal}.

We next give some details concerning the two boosts that have been mentioned in the main text.
These enter into both the on-shell projections of kernels and amplitudes, 
and in the definitions of $\wt F^{(i)}$ and $\wt G^{(ij)}$.
To explain the boosts, we use the example of the quantity $\vec p_j^{*(p_i)}$ that 
enters in $\wt F^{(i)}$, Eq.~(\ref{eq:Ft}).
The setup is that the four-momenta $p_i=(\omega_{p_i},\vec p_i)$ and
$p_j=(\omega_{p_j},\vec p_j)$ are on shell. The former is the spectator momentum, and
the latter is the pair momentum relative to which the decomposition into spherical harmonics is defined
(after boosting to the pair CMF).
We note that, in situations where we use the boosts, these two particles are always on shell,
either because (in the TOPT approach) all intermediate particles are on shell,
or (in the Feynman-diagram approach) we have set them on shell in preparation for full on-shell projection,
as discussed in Appendix \ref{app:Feynman}.
For the boost of Ref.~\cite{\HSQCa}, referred to subsequently as the ``HS boost",
we obtain $\vec p_j^{*(p_i)}$ by boosting $p_j$ using boost velocity
\begin{equation}
\boldsymbol\beta_{\rm HS} = - \frac{\vec P-\vec p_i}{E-\omega_{p_i}}\,.
\end{equation}
The transformation is
\begin{equation}
p_j \longrightarrow p_j^* = \Lambda(\boldsymbol\beta_{\rm HS}) p_j \equiv
\left(\omega_{p_j}^{*(p_i)}, \vec p_j^{*(p_i)}\right)\,,
\end{equation}
where $\Lambda$ is the corresponding Lorentz transformation matrix.
By construction, this transforms the pair four-momentum $P-p_i$ to its CMF,
\begin{equation}
\Lambda(\boldsymbol\beta_{\rm HS}) (P-p_i) = (\sqrt{\sigma_i}, \vec 0)\,, \qquad
\sigma_i = (P-p_i)^2\,.
\end{equation}
For the Wu boost, the setup is the same, but the boost velocity is changed to
\begin{equation}
\boldsymbol\beta_{\rm Wu} = - \frac{\vec P-\vec p_i}{\omega_{p_j}+\omega_{p_k}}\,,
\end{equation}
where $\vec p_k=\vec P-\vec p_i-\vec p_j$ is the momentum of the third particle.
Thus the result for $\vec p_j^{*(p_i)}$ is in general different.
The exception is if all three particles are on shell \textit{and} total four-momentum is conserved,
for then $E-\omega_{p_i}=\omega_{p_j}+\omega_{p_k}$, and the boost velocities are the same.
Because of this, it does not matter which boost we use when defining on-shell projected
quantities such as the elements of $\wh \cK_{\df,3}$, $\wh \cM_3$, etc.

When using TOPT, as in Secs~\ref{sec:derivation} and \ref{sec:KtoM},
one should initially use the Wu boost, as explained in BS1.
This is because the boost is designed to apply for three on-shell particles.
However, once one has obtained results in terms of on-shell projected quantities,
e.g.~$C_{3,L}$ in  Eq.~(\ref{eq:C3Lf}), one can change to the HS boost in the definitions of the 
elements of $\wh F_G$. For $\wt F^{(i)}$ the change is exponentially suppressed,
while for $\wt G^{(ij)}$ the change is nonsingular (since the boosts agree at the pole) and can be
absorbed into a shift in $\delta \wt G^{(ij)}$. 
In this way, one finds that the same form for $C_{3,L}$ holds, but with redefined quantities $\wh \cK_{\df,3}$,
$\wh A_F$, and $\wh A'_F$. 
The only caveat is that the cutoff functions $H^{(i)}$ must be chosen 
such that the HS boost is well defined, which requires $\sigma_i > 0$, as discussed above.
By contrast, for the Wu boost there is no such constraint on $\sigma_i$.
The conclusion, already noted in BS1 for identical particles,  is that the TOPT forms of the quantization
condition and the relation of $\cM_3$ to $\wh \cK_{\df,3}$ are valid with both boosts, although 
$\wh \cK_{\df,3}$ will, of course, depend on the choice.

The situation is reversed in the Feynman-diagram approach. 
In the initial derivation, the third particle is off shell, and so the natural choice is the HS boost.
Once we obtain the result for, say, $\wh \cM'_{3,L}$, however, we could switch to using the Wu boost.
We could also raise the cutoff in the functions $H^{(i)}$ to allow $\sigma_i < 0$, since that only changes
the elements of $\wh F_G$ away from the pole.
However, there is a clear reason not to use the Wu boost, which is that only with the HS boost can one obtain
a Lorentz-invariant $\wh \cK'_{\df,3}$, i.e.~one that is Lorentz invariant when combined with
spherical harmonics as in Eq.~(\ref{eq:Mp3sph}).
We close this section by explaining this result.

The context for the discussion is the infinite-volume relation between $\cM_{3}$ and
$\wh \cK'_{\df,3}$ derived in Sec.~\ref{sec:KtoMFeynman}.
As described in that section, the only nontrivial part of the demonstration of Lorentz invariance
concerns the transformation properties of $[\wh G^\infty]_{ij}$, Eq.~(\ref{eq:Ghatinfty}).
If we Lorentz transform the momentum arguments
$p_i=(\omega_{p_i},\vec p_i)$ and $k_j =(\omega_{k_j},\vec k_j)$, then the arguments of
the spherical harmonics, $\vec k_j^{*(p_i)}$ and $\vec p_i^{*(k_j)}$, should only be rotated.

To show that this is the case with the HS boost, we focus on $\vec k_j^{*(p_i)}$. This is defined by
\begin{equation}
k^*_j = \Lambda(\boldsymbol\beta_{\rm HS}) k_j \equiv \left(\omega_{k_j}^{*(p_i)}, \, \vec k_j^{*(p_i)} \right) \,.
\label{eq:kjstar}
\end{equation}
Now we apply an arbitrary global transformation $\Lambda_0$ to all momenta, so that
\begin{equation}
p_j \to p'_i = \Lambda_0 p_i\,, 
\ \ 
k_j \to k'_j = \Lambda_0 k_j\,,
\ \
P \to P' = \Lambda_0 P\,,
\ \ {\rm etc.}\,,
\end{equation}
and then boost to the pair CMF, which now requires the boost velocity
\begin{equation}
\boldsymbol \beta'_{\rm HS} = - \frac{\vec P'-\vec p'_i}{E'-\omega_{p'_i}}\,.
\end{equation}
Thus we find a new value of $\vec k_j^{*(p_i)}$, given by the spatial part of
\begin{equation}
k'^*_j = \Lambda(\boldsymbol\beta'_{\rm HS}) k'_j =
\Lambda(\boldsymbol\beta'_{\rm HS}) \Lambda_0 k_j \,.
\label{eq:kpjstar}
\end{equation}
To proceed, we note that, by definition, the boosts satisfy the following relations:
\begin{align}
\Lambda(\boldsymbol\beta_{\rm HS}) (P-p_i) = \left(\sqrt{\sigma_1},\vec 0\right)
= \Lambda(\boldsymbol\beta'_{\rm HS}) (P'-p'_i) 
= \Lambda(\boldsymbol\beta'_{\rm HS}) \Lambda_0 (P-p_i) \,,
\end{align}
from which follows the standard result that
\begin{equation}
\Lambda(\boldsymbol\beta'_{\rm HS}) \Lambda_0 =
\Lambda({\rm rot}) \Lambda(\boldsymbol\beta_{\rm HS}) \,,
\label{eq:boostrot}
\end{equation}
where $\Lambda({\rm rot})$ implements a rotation. The precise form of the rotation can, of course,
be determined, but is not needed here.
Inserting this into Eq.~(\ref{eq:kpjstar}) we find 
\begin{equation}
k'^*_j = \Lambda({\rm rot}) \Lambda(\boldsymbol\beta_{\rm HS}) k_j = \Lambda({\rm rot}) k'_j
\,,
\end{equation}
which yields the claimed result that $\vec k_j^{*(p_i)}$ and $\vec k'^{*(p_i)}_j$ are related by a rotation.
A completely analogous argument holds for $\vec p_i^{*(k_j)}$.

Now we show why this argument fails for the Wu boost.
The definitions of $k_j^*$ and $k'^*_j$ take the same form as above,
except that $\boldsymbol\beta_{\rm HS}$ in Eq.~(\ref{eq:kjstar})
and $\boldsymbol\beta'_{\rm HS}$ in Eq.~(\ref{eq:kpjstar}),
are replaced, respectively, by 
\begin{equation}
\boldsymbol\beta_{\rm Wu} =  - \frac{\vec P-\vec p_i}{\omega_{b}+\omega_{k_j}}\,,\quad
\boldsymbol\beta'_{\rm Wu} =  - \frac{\vec P'-\vec p'_i}{\omega_{b'}+\omega_{k'_j}}\,,
\end{equation}
where $\vec b = \vec P - \vec p_i - \vec k_j$ 
and $\vec b' = \vec P' - \vec p'_i - \vec k'_j$.
These boosts are defined by the properties
\begin{align}
\Lambda(\boldsymbol\beta_{\rm Wu}) (k_j+b) &=
\left(\sqrt{(\omega_b+\omega_{k_j})^2- |\vec P-\vec p_i|^2}, \vec 0\right)\,,
\label{eq:Wu1}
\\
\Lambda(\boldsymbol\beta'_{\rm Wu}) (k'_j+b') &=
\left(\sqrt{(\omega_{b'}+\omega_{k'_j})^2- |\vec P'-\vec p'_i|^2}, \vec 0\right)\,,
\label{eq:Wu2}
\end{align}
where $b=(\omega_b,\vec b)$ and $b'=(\omega_{b'},\vec b')$.
This brings up the obstruction to continuing the argument as for the HS boost,
namely that
\begin{equation}
b' \ne \Lambda_0 b\,.
\end{equation}
The quantities that do transform in this manner are $b_{\rm off} = P - p_i - k_j$
and $b'_{\rm off}=P'-p'_i-k'_j$, but, in general, $b\ne b_{\rm off}$ and $b'\ne b'_{\rm off}$.
A related obstruction to the argument is that the right-hand sides of
Eqs.~(\ref{eq:Wu1}) and (\ref{eq:Wu2}) are not, in general, equal.
Because of these issues, the analog of Eq.~(\ref{eq:boostrot}) for Wu boosts is not valid,
and $\vec k_j^{*(p_i)}$ and $\vec k'^{*(p_i)}_j$ are not, in general, related by a rotation.
Thus the arguments of Sec.~\ref{sec:KtoMFeynman} concerning Lorentz invariance do not hold.

\section{Details of derivation using Feynman diagrams}
\label{app:Feynman}

In this appendix we explain how Eqs.~(\ref{eq:Mp23Lres_off})-(\ref{eq:Dp}) 
are obtained from the initial skeleton expansion by doing the energy integrals over independent momenta.

As discussed in Ref.~\cite{\HSQCa}, when doing an integral over the energy component of
a four-momentum associated with a propagator,
we can close the contour in the complex plane so as
to pick up the contribution from the particle pole, and
exclude the antiparticle pole contribution. The integral will also contain contributions from other poles
(e.g.~corresponding to a three-particle cut across the dressed propagator, or the antiparticle pole
contributions from another propagator), but these can never lead to an
on-shell cut of the entire diagram. Thus we can separate the contributions from each such integral into
that from the particle pole, which we refer to as the ``on-shell" contribution, and the remainder,
denoted ``off-shell." If any of the contributions from the three propagators in a given cut
(of which only two are integrated, due to overall momentum conservation) are off shell, then we know that
the cut is nonsingular, and momentum sums can be replaced by integrals.
The contributions from off-shell propagators can then be absorbed into shifts in the Bethe-Salpeter kernels,
leading to a reshuffled skeleton expansion in which all independent momenta are now on shell.
Important features of this reshuffling are that it maintains Lorentz invariance of the kernels, and that
it does not mix up the elements of $\wh\cM^{\prime\,\rm off}_{3,L}$.

 \begin{figure}[tb]
\begin{center}
\vspace{-10pt}
\includegraphics[width=\textwidth]{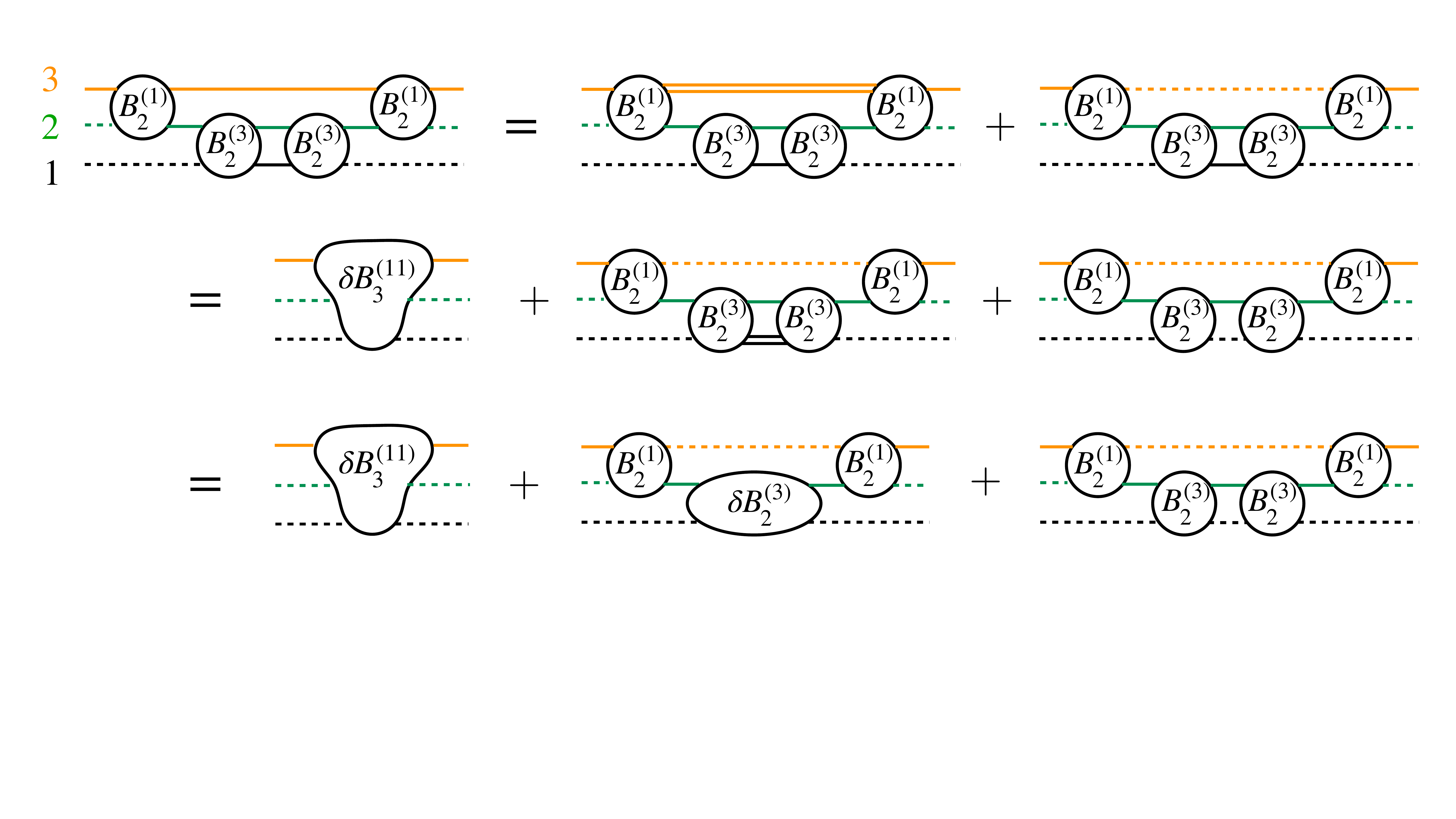}
\vspace{-1.4truein}
\caption{
An example of the first step in the analysis for a contribution to $[\wh \cM^{\prime\,\rm off}_{3,L}]_{11}$
containing only $B_2$ kernels.
Colors and labels indicate flavors as in earlier figures.
Solid lines indicate fully dressed propagators, prior to integration over energy,
except for the external lines, where a solid propagator indicates only that it is has an off-shell momentum.
Dashed lines indicate on-shell propagators, which come with a factor of $1/(2\omega)$ 
when internal (due to the energy integration picking up the particle pole), or are amputated when external.
Double solid line propagators indicate that only the off-shell contributions from the energy integral
are kept, as described in the text. These cannot lead to a three-particle on-shell cut.
In the first step,  the energy of the flavor 3 (upper, orange) spectator propagator is integrated, 
leading to the off- and on-shell terms as shown. 
The former can be absorbed into a shift in the appropriate element of the three-particle Bethe-Salpeter kernel,
as shown by the first term on the second line, and is not manipulated further.
In the second step, the flavor 1 (lower, black) propagator is integrated
over its energy, again leading to on- and off-shell contributions for both diagrams.
In the final step, the two $B_2^{(3)}$ kernels connected by an off-shell
(double-line) propagator (which can now be integrated) are absorbed into a shift  in that kernel.
This leads to the three diagrams shown on the bottom line, in which all
independent internal propagators are on shell. The remaining propagators (which in this case are the middle,
green, flavor 2 lines) have four-momenta that are determined by the momenta flowing through the other lines
and  in general are off shell.
\label{fig:M23decomp1}
}
\end{center}
\end{figure}

To explain this in more detail, as well as to highlight some technical features, we work through three examples.
We begin with one involving only $B_2$ kernels, shown in Fig.~\ref{fig:M23decomp1}.
A detailed explanation of the steps is given in the caption, and we comment here on how the procedure 
is generalized to all such diagrams. The first step is to integrate over the energies
of all internal spectator momenta. In the example shown, there is one such momentum (the upper, orange
line). In general there are multiple spectator momenta, in different loops. 
We find that the order of integration of their energies does not matter,
since all lead to the same final decomposition in terms of shifted kernels.
The result of these first integrals are on- and off-shell terms, one of each in the example.
For the off-shell terms, all spatial momenta associated with the spectator can now be integrated,
including any involving the interacting pair.
Thus, in the example, the lower (black) propagator between the middle $B_2^{(3)}$s can also be fully integrated
when the off-shell part of the upper (orange, spectator) propagator is taken.
This results in an infinite-volume quantity that can be absorbed by a shift in the $(11)$ element of $B_3$,
as shown by the first term of the second and third lines of the figure.

In the second step, the energies of all remaining independent momenta are integrated.
These lie in what we call F-type cuts,
i.e.~those which, in the TOPT analysis described in the previous section, would lead to factors of $\wt F$.
Here we must choose which of the interacting pair to integrate, and, as in the TOPT analysis,
we pick that whose flavor follows cyclically from the flavor of the spectator.
In our example the spectator has flavor 3, so we integrate the energy of the flavor 1 propagator in the lower pair.
The integrations over all F-type cuts in such diagrams can be done independently, with each leading
to on- and an off-shell contributions, as shown in the second line in the figure.
We then absorb any sequences of interacting pairs connected by off-shell propagators into
shifts in the corresponding $B_2$ kernels, since the loops within them can be fully integrated.
The final result is a set of diagrams containing (possibly shifted) kernels in which all 
propagators with independent momenta are on shell, and with only one 
propagator per cut whose momentum can be off shell.

At this stage we can already understand most of the factors present in Eqs.~(\ref{eq:Mp23Lres_off})-(\ref{eq:Dp}),
in particular those involving $B_2^{(i)}$ and $\delta B_2^{(i)}$.
Each cut is associated with two on-shell and one off-shell propagator, 
and with two loop sums, leading to the kinematical factors in $D'^{(i)}$, Eq.~(\ref{eq:Dp}).
The on-shell projectors that enter the diagonal elements of $\wh D'$ correspond to F-type
cuts and are determined by the cyclical flavor rule. For example, if the spectator has flavor 3,
as in Fig.~\ref{fig:M23decomp1},
then the on-shell flavors are 3 and 1, requiring on-shell projectors $\overleftarrow{O}^{(2)}$
and $\overrightarrow{O}^{(2)}$. Thus the 33 element of $\wh D'$ is $D'^{(2)}$.
The off-diagonal elements of $\wh D'$ correspond to G-type cuts in which the spectator is switched,
in which case the two on-shell flavors are simply those of the spectators on either side. 
Thus, for example, the 12 element of $\wh D'$ is $D'^{(3)}$, which places flavors 1 and 2 on shell.
The geometric series in Eq.~(\ref{eq:Mp23Lres_off}) then produces all possible orderings of F-type and G-type cuts.

Another feature to be explained is the presence of factors of $2\omega L^3$ in the projected $\overline B_{2,L}^{(i)}$,
Eqs.~(\ref{eq:B21proj})-(\ref{eq:B23proj}).
These are needed because, in general, a spectator propagator spans multiple F-type cuts, but should only
appear once in the final expression. Each F-type cut comes with a factor of $1/(2\omega L^3)$---contained
in the corresponding $D'^{(i)}$---and all but one are canceled by the factors of
$2\omega L^3$ in the numerators of the $\overline B_{2,L}^{(i)}$s.
This structure, which follows exactly the pattern of the TOPT result, yields the correct propagator factors,
and factors of $L^3$, for each diagram.

 \begin{figure}[tb]
\begin{center}
\vspace{-10pt}
\includegraphics[width=\textwidth]{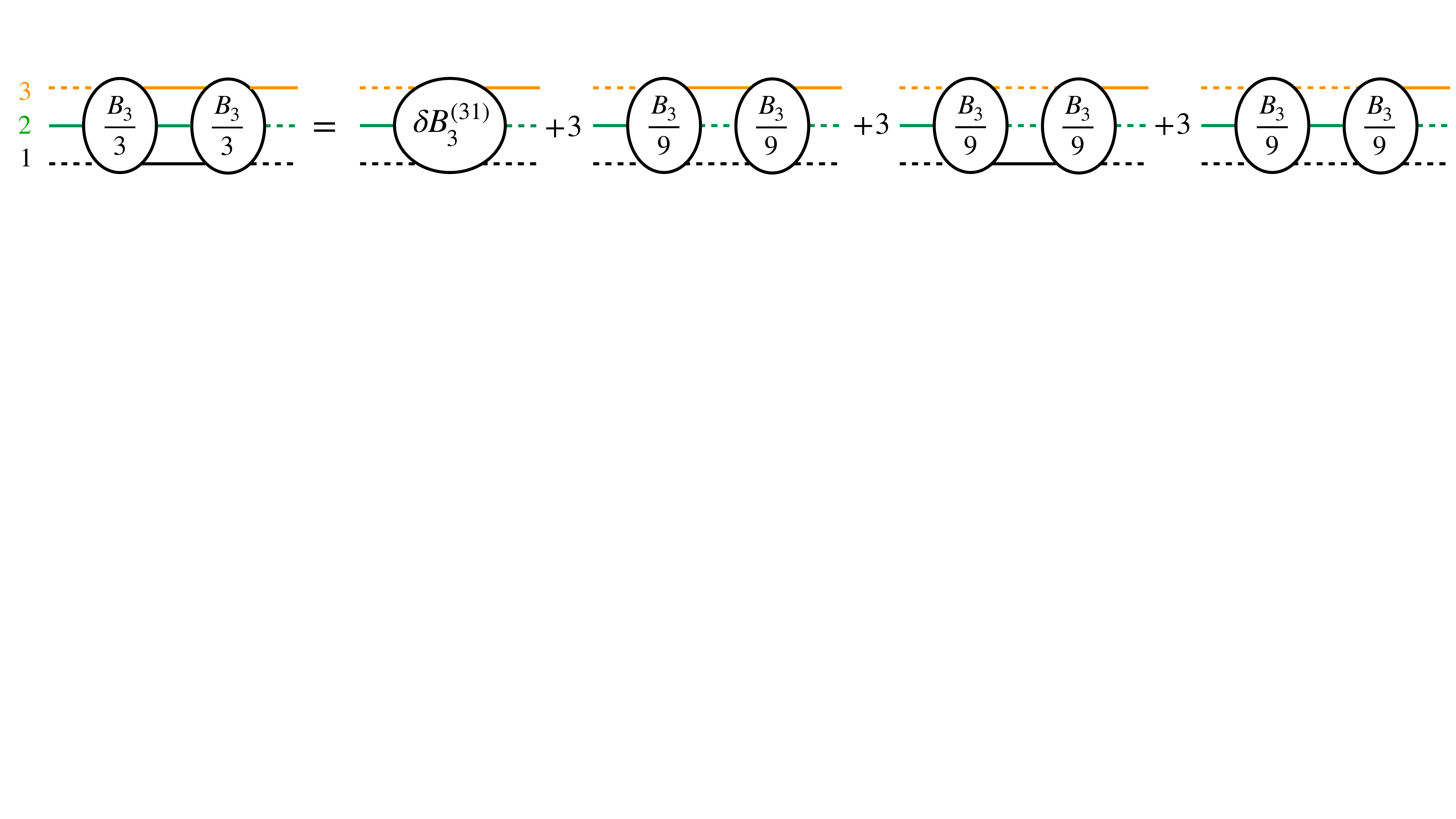}
\vspace{-3.2truein}
\caption{
Example of the first step in the analysis for a diagram that contributes to $[\wh \cM^{\prime\,\rm off}_{3,L}]_{31}$
and contains two $B_3$ kernels. 
Notation as in Fig.~\ref{fig:M23decomp1}.
Only the final result is shown. See text for further discussion.
\label{fig:M23decomp3}
}
\end{center}
\end{figure}

We next consider diagrams involving only $B_3$ kernels, the simplest nontrivial example of which is
analyzed in Fig.~\ref{fig:M23decomp3}. Since we are considering a contribution to one of the nine
elements of the matrix, $[\wh \cM^{\prime\,\rm off}_{3,L}]_{31}$, there is an overall factor of $1/9$,
which we partition equally between the two external $B_3$ kernels in the initial expression.
We now have to choose for which two of the propagators we do the energy integrals.
The choice needed to match the form of Eqs.~(\ref{eq:Mp23Lres_off})-(\ref{eq:Dp}) is to pick
equal weights for all three pairs, which leads to the last three terms in the equation in the figure.
This matches the result obtained 
by expanding out Eqs.~(\ref{eq:Mp23Lres_off})-(\ref{eq:Dp}) 
and evaluating the term corresponding to that in the figure,
\begin{equation}
\left(\ket1 \frac{B_3}9 \bra1\right)
\wh D'
\left(\ket1 \frac{B_3}9 \bra1\right)
=
\ket 1 \frac{B_3}9 \mathbb 1_{\vec p,\vec k}
%\delta_{\vec p_1 \vec k_1} \delta_{\vec p_2 \vec k_2} \delta_{\vec p_3 \vec k_3} 
(3 D'^{(1)}+ 3 D'^{(2)}+3 D'^{(3)}) \frac{B_3}9 \bra1\,,
\end{equation}
where $\mathbb 1^{\vec p}$ is defined in Eq.~(\ref{eq:onep}).
All contributions containing an off-shell propagator, which are not shown
explicitly in the figure, are included in the shift to the three-particle kernel.
The extension of this analysis to multiple adjacent $B_3$ kernels is straightforward.

 \begin{figure}[tb]
\begin{center}
\vspace{-10pt}
\includegraphics[width=\textwidth]{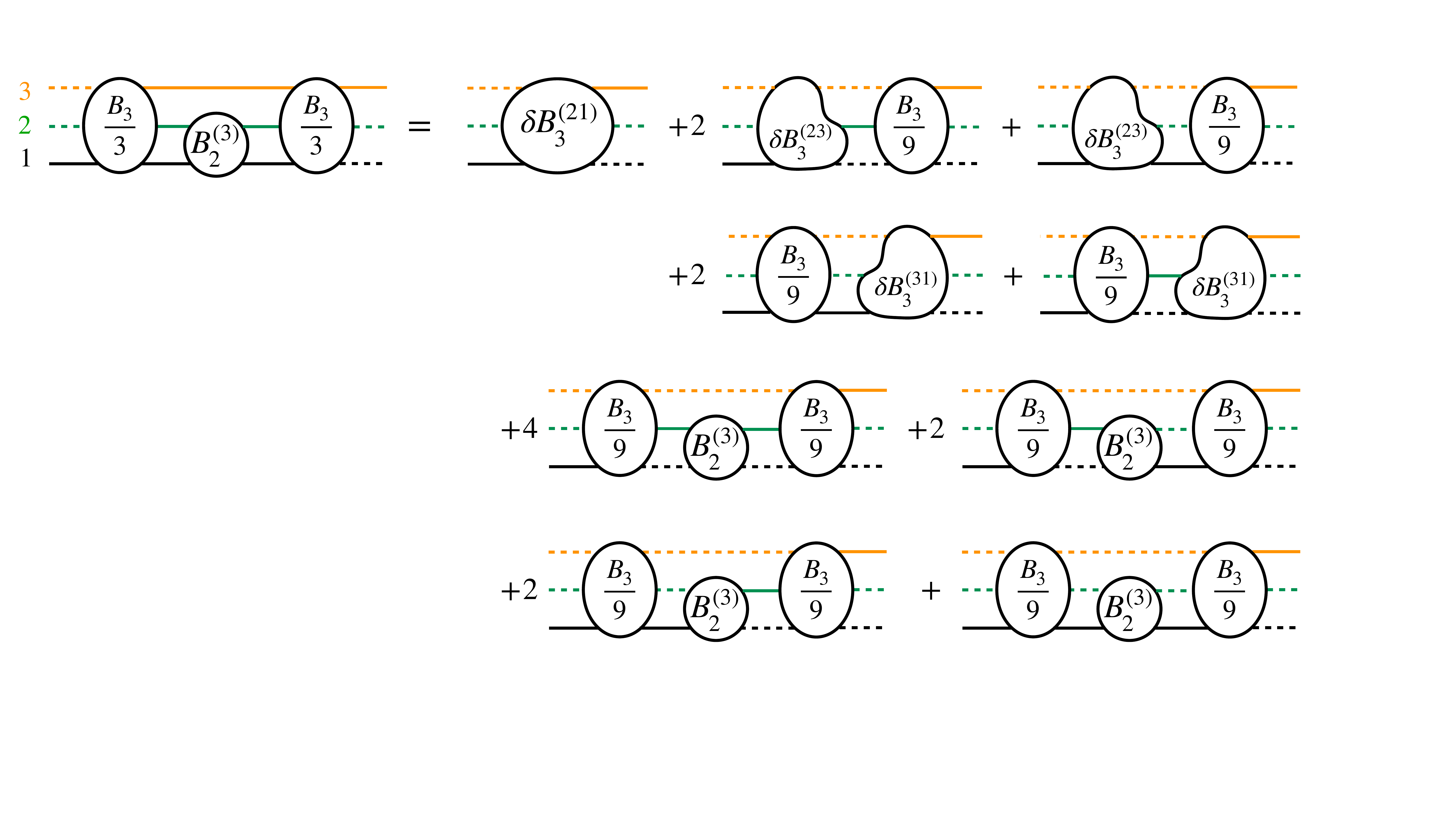}
\vspace{-1.truein}
\caption{
Example of the first step in the analysis for a diagram that contributes to $[\wh \cM^{\prime\,\rm off}_{3,L}]_{21}$
and contains both $B_3$ and $B_2$ kernels.
Notation as in Fig.~\ref{fig:M23decomp1}.
Only the final result is shown. See text for further discussion.
\label{fig:M23decomp2}
}
\end{center}
\end{figure}

Our final example, shown in Fig.~\ref{fig:M23decomp2}, is of a case with both $B_3$ and $B_2$ kernels.
The factors of $1/3$ contained in the $B_3$ kernels in the original expression appear for the same reason
as in the previous example.
The ordering of energy integrals is as in the first example: spectators first (here the top, flavor 3, orange line), 
and then those lying between $B_3$ and $B_2$ kernels. 
For the latter we again have a choice, and we pick two-thirds of the contribution to have the lower, flavor 1,
black line to be integrated, with the remaining one-third having the middle, flavor 2, green line integrated.
Subsuming cuts with off-shell propagators into shifts in the kernels leads to the result shown in the figure.
The resulting factors match those that arise from the desired equations;
for example, 
\begin{equation}
\left(\ket1 \frac{B_3}9 \bra1\right)
\wh D' 
\wh B\big|_{B_2^{(3)}\ \rm part}
\wh D'
\left(\ket1 \frac{B_3}9 \bra1\right)
=
\ket 1 \frac{B_3}9
\mathbb 1_{\vec p,\vec k} (D'^{(1)}+ 2 D'^{(2)})
\overline{B}_{2,L}^{(3)}
\mathbb 1_{\vec p,\vec k} (D'^{(1)}+ 2 D'^{(2)})
 \frac{B_3}9 \bra1
\end{equation}
agrees with the factors in the final two lines in the figure.

Diagrams with multiple $B_2$ kernels, including switches, 
between $B_3$s are analyzed by a combination of the above-described methods.
First all spectator energies are integrated, then those in the F-type cuts 
(choosing which flavor to integrate as in the first example),
and finally those between $B_3$s and adjacent $B_2$s (with flavor choices as in our third example).
This results in  a unique decomposition into diagrams in which all independent internal propagators on shell.
To obtain a simple final expression, we add in the disconnected contributions with a single
spectator line (of, say, flavor $i$) and one or more $B_2^{(i)}$s,
which build up the quantity $\wh{\overline{\cM}}{}^{\prime\,\rm off}_{2,L}$ discussed in the main text.
The combination is then given by Eq.~(\ref{eq:Mp23Lres_off}).

A final technical issue is whether the same total shifts to the kernels, i.e.~$\delta B_3^{(ij)}$ and 
$\overline{\delta B}^{(i)}_2$, appear in all resulting diagrams.
To demonstrate this we note that every diagram that results from the energy integrations is
of the same form as one of the unintegrated diagrams, except with some propagators on shell and
some kernels replaced by their shifts. No new topologies arise.
Now we consider one topology, and insert every possible choice for the shift in each of the kernels.
Each of the diagrams that results can be uniquely traced back to an allowed original unintegrated diagram.
Thus every energy-integrated diagram that should appear does appear.

\section{Derivation of symmetrization identities}
\label{app:symid}

In this appendix we derive the identities (\ref{eq:symid1})-(\ref{eq:symid3}).
We lean heavily on the derivation of the analogous identities for identical particles
given in Appendix D of BS1, although we have made some improvements to the argument.

The first identity replaces the $\wh G$ elements of $\wh F_G$ with factors of 
$\wh F \overrightarrow{\cS}$ plus integral operators. We demonstrate this result by considering
a representative example,
\begin{multline}
X_1(\{\vec p\})_{p_1\ell' m'} \wt G^{(12)}_{p_1\ell' m';k_2 \ell m} (-1)^\ell
Z_2(\{\vec k\})_{k_2\ell m}
-
X_1(\{\vec p\})_{p_1\ell' m'} \wt F^{(1)}_{p_1\ell' m';k_1 \ell m} 
Z_2(\{\vec k\})_{k_1\ell m}
\\=
X_1(\{\vec p\})_{p_1\ell' m'} \Big[\wh{\overrightarrow{\cI}}_{G,12}\Big]_{p_1\ell' m';k_2 \ell m} 
Z_2(\{\vec k\})_{k_2\ell m}
\,,
\label{eq:appD1}
\end{multline}
where there is an implicit sum over repeated indices (including over $\ell$ which appears
thrice in one term).
Here we are using the notation of Sec.~\ref{subsec:symops}, and stress in particular
the difference between $Z_2(\{\vec k\})_{k_2 \ell m}$ and $Z_2(\{\vec k\})_{k_1 \ell m}$.
Both involve the same underlying infinite-volume on-shell quantity $Z_2$,
but the former is expressed in the coordinates in which the spectator has flavor 2 
(with momentum $\vec k_2$), and the spherical harmonics are defined
relative to the flavor 3 particle, while the latter quantity is expressed in coordinates
in which the flavor 1 particle is the spectator (with momentum $\vec k_1$), 
and the harmonics are defined relative to the particle of flavor 2.
The precise definition of the decompositions 
into spherical harmonics is given in Eq.~(\ref{eq:basischange}).

Using the definitions of $\wt G^{(ij)}$ and $\wt F^{(i)}$, Eqs.~(\ref{eq:Gt}) and (\ref{eq:Ft}),
the left-hand side of Eq.~(\ref{eq:appD1}) can be written as
\begin{multline}
\sum_{\vec p_1 \ell' m'} \sum_{\vec k_2 }
X_1(\{\vec p\})_{p_1\ell' m'} \frac{H^{(1)}(\vec p_1)}{2\omega_{p_1}L^3}
\frac{\cY_{\ell'm'}(\vec k_2^{*(p_1)})}{q_{2,p_1}^{*\ell'}}
\frac{H^{(2)}(\vec k_2)}{b_{12}^2-m_3^2} \frac1{2\omega_{k_2}L^3} 
\\
\times \left\{ \sum_{\ell m} \left[
\frac{\cY_{\ell m}( \vec p_1^{*(k_2)})}{q_{2,k_2}^{*\ell}}
(-1)^\ell Z_2(\{\vec k\})_{k_2\ell m}
-
\frac{\cY_{\ell m}( \vec k_2^{*(p_1)})}{q_{2,p_1}^{*\ell}}
Z_2(\{\vec k\})_{p_1\ell m}\right]
\right\}
+\textrm{integral term from}\ \wt F^{(1)}\,.
\end{multline}
Here we have made several emendations to $\wt F^{(1)}$:
used $H^{(2)}(\vec k_2)$ as the UV regulator,
converted to the relativistic form of the denominator,
labeled the dummy variable in the sum as $\vec k_2$,
and used the Kronecker delta to set $\vec k_1=\vec p_1$.
We have also made the sums explicit.
The key observation is that, if $\vec p_1$ and $\vec k_2$ are chosen such that
$b_{12}=P-p_1^{\rm on}-k_2^{\rm on}$ is on shell, i.e.~$b_{12}^2=m_3^2$,
then the term in curly braces vanishes.
This is because both contributions to this term become equal to
$Z_2(\{\vec p_1,\vec k_2, \vec k_3\})$, where $\vec k_3=\vec P-\vec p_1-\vec k_2$.
This in turn is because, at the on-shell point, we have
\begin{equation}
|\vec p_1^{*(k_2)}| = q_{2,k_2}^* \ \ {\rm and}\ \
|\vec k_2^{*(p_1)}| = q_{2,p_1}^*\,,
\end{equation}
so that the ratios involving harmonic polynomials reduce to spherical harmonics, e.g.
\begin{equation}
\frac{\cY_{\ell m}( \vec k_2^{*(p_1)})}{q_{2,p_1}^{*\ell}}
\longrightarrow
\sqrt{4\pi} 
Y_{\ell m}( \hat k_2^{*(p_1)})\,.
\end{equation}
The sums over $\ell$ and $m$ can then be done.
The only remaining subtlety is in the first term in curly braces, where the $(-1)^\ell$
is needed to convert the spherical harmonics from being defined with respect to flavor 1
to being defined relative to flavor 3, so as to match the convention used in
$Z_2(\{\vec k\})_{k_2\ell m}$.

Since the term in curly braces vanishes on shell, it cancels the pole at $b_{13}^2=m^2$,
so that the sum over $\vec k_2$ can be replaced by an integral, up to exponentially suppressed
contributions. If we introduce a PV pole prescription, then this integral can be split into two terms,
one for each of the terms in curly braces. The second of these simply cancels the integral
term from $\wt F^{(1)}$, leaving an integral over the first term in curly braces, i.e.~the term that
originated from $\wt G^{(12)}$. For this remaining term
the sum over $\vec p_1$ can also be converted into an integral, 
due to the PV pole prescription used for the integral over $\vec k_2$.
Thus the left-hand side of Eq.~(\ref{eq:appD1}) becomes
\begin{equation}
\int_{\vec p_1} \sum_{\ell' m'} \PV \int_{\vec k_2} \sum_{\ell m}
X_1(\{\vec p\})_{p_1\ell' m'} 
\frac{\cY_{\ell'm'}(\vec k_2^{*(p_1)})}{q_{2,p_1}^{*\ell'}}
\frac{H^{(1)}(\vec k_1) H^{(2)}(\vec k_2)}{b_{12}^2-m_3^2} 
\frac{\cY_{\ell m}( \vec p_1^{*(k_2)})}{q_{2,k_2}^{*\ell}}
(-1)^\ell Z_2(\{\vec k\})_{k_2\ell m}\,,
\label{eq:intop}
\end{equation}
which defines the integral operator on the right-hand side of
Eq.~(\ref{eq:appD1}). Here we are using the Lorentz-invariant measure
for the momentum integrals, defined in Eq.~(\ref{eq:measure}).

We now introduce a compact notation for the result Eq.~(\ref{eq:appD1}),
\begin{equation}
[X_1]_1 [\wh G]_{12} [Z_2]_2 - [X_1]_1 [\wh F]_{11} [Z_2]_1
= [X_1]_1 [\wh{\overrightarrow{\cI}}_G]_{12} [Z_2]_2\,.
\end{equation}
By essentially the same argument, we can extend this result 
to all nonvanishing elements of $\wh G$, obtaining
\begin{equation}
[X_i]_i [\wh G]_{ij} [Z_j]_j - [X_i]_i [\wh F]_{ii} [Z_j]_i
= [X_i]_i [\wh{\overrightarrow{\cI}}_G]_{ij} [Z_j]_j\,, \qquad i\ne j \ \textrm{(no sum)}\,.
\label{eq:id1}
\end{equation}
This fills out the off-diagonal matrix of integral operators $\wh{\overrightarrow{\cI}}_G$,
and completes the derivation of the first symmetrization identity.

Using the same arguments, but with the roles of $X$ and $Z$ interchanged, we obtain
\begin{equation}
[X_i]_i [\wh G]_{ij} [Z_j]_j - [X_i]_j [\wh F]_{jj} [Z_j]_j
= [X_i]_i [\wh{\overleftarrow{\cI}}_G]_{ij} [Z_j]_j\,, \qquad i\ne j  \ \textrm{(no sum)}\,,
\label{eq:id2}
\end{equation}
where in the new integral operator the order of integrals is interchanged
compared to that in Eq.~(\ref{eq:intop}).
This result leads immediately to the second 
symmetrization identity, Eq.~(\ref{eq:symid2}).

We now turn to the third identity, Eq.~(\ref{eq:symid3}). In our new notation
this is equivalent to
\begin{equation}
\Delta \equiv 3 \sum_{i,j} [X_i]_i [\wh F_G]_{ij} [Z_j]_j
-
\sum_i [X_1+X_2+X_3]_i [\wh F]_{i i} [Z_1+Z_2+Z_3]_i
=
3 \sum_{i,j} [X_i]_i [\wh \cI_{FG}]_{ij} [Z_j]_j\,.
\label{eq:symid3a}
\end{equation}
To demonstrate this (and define the integral operator $\wh \cI_{FG}$) we break the left-hand side
into four parts
\begin{align}
\Delta &= \Delta_1+\Delta_2+\Delta_3+\Delta_4 \,,
\\
\Delta_1 &= 
\sum_{i\ne j} \left(
[X_i]_i [\wh F_G]_{ij} [Z_j]_j - [X_i]_i [\wh F]_{ii} [Z_j]_i\right)\,,
\label{eq:Delta1}
\\
\Delta_2 &= 
\sum_{i\ne j} \left(
[X_i]_i [\wh F_G]_{ij} [Z_j]_j - [X_i]_j [\wh F]_{jj} [Z_j]_j\right)\,,
\label{eq:Delta2}
\\
\Delta_3 &= 
\sum_{i\ne j} \left(
[X_j]_j [\wh F]_{jj} [Z_j]_j - [X_j]_i [\wh F]_{ii} [Z_j]_i\right)\,,
\label{eq:Delta3}
\\
\Delta_4 &= 
\sum_{i\ne j} \left(
 [X_i]_i [\wh F_G]_{ij} [Z_j]_j
- [X_i]_k [\wh F]_{kk} [Z_j]_k\right)\,,
\label{eq:Delta4}
\end{align}
where in the final equation $k$ is the third flavor index, i.e. $k\ne i$ and $k\ne j$.
The first two parts can be evaluated using Eqs.~(\ref{eq:id1}) and (\ref{eq:id2}), respectively.
For the remaining two, we need additional work.

$\Delta_3$ involves the difference between choosing two different spectator
flavors for an F cut. To evaluate this difference we focus on the case where $i=1$ and $j=2$,
and also replace $X_j$ and $Z_j$ with $X$ and $Z$, respectively, since the flavor of these
quantities plays no role in the argument.
Thus we consider
\begin{equation}
\Delta_3^{(21)} 
= [X]_2 [\wh F]_{22} [Z]_2 - [X]_1 [\wh F]_{11} [Z]_1\,.
\label{eq:Delta312}
\end{equation}
Considering first the sum parts of the $\wt F^{(i)}$s, we have
\begin{multline}
\Delta_{3\Sigma}^{(21)} = 
\frac1{L^6}\sum_{\vec k_2 \ell' m'} H^{(2)}(\vec k_2) 
 \sum_{\vec k_3 \ell m}
X_{k_2\ell' m'} \frac{\cY_{\ell' m'}(\vec k_3^{*(k_2)})}{q_{2,k_2}^{*\ell'}} 
H^{(1)}(\vec k_1)D_{123}
\frac{\cY_{\ell m}(\vec k_3^{*(k_2)})}{q_{2,k_2}^{*\ell}} Z_{k_2\ell m}
\\
- \frac1{L^6}\sum_{\vec k_1 \ell' m'} H^{(1)}(\vec k_1) 
\sum_{\vec k_2 \ell m}
X_{k_1\ell' m'} \frac{\cY_{\ell' m'}(\vec k_2^{*(k_1)})}{q_{2,k_1}^{*\ell'}}
H^{(2)}(\vec k_2)D_{123}
\frac{\cY_{\ell m}(\vec k_2^{*(k_1)})}{q_{2,k_1}^{*\ell}} Z_{k_1\ell m}\,,
\end{multline}
where we have chosen the nonrelativistic form for the denominator (which is the same for
both terms), i.e.
\begin{equation}
D_{123} \equiv \frac1{8 \omega_{k_1}\omega_{k_2}\omega_{k_3} 
(E-\omega_{k_1}-\omega_{k_2}-\omega_{k_3})}\,,
\label{eq:D123}
\end{equation}
and regulated the UV in the $\vec k_3$ sum in the first term with $H^{(1)}(\vec k_1)$, 
where $\vec k_1=\vec P-\vec k_2-\vec k_3$, and in the $\vec k_2$ sum in the second term
with $H^{(2)}(\vec k_2)$.
If we now change the summation variable in the second term from $\vec k_1$ to $\vec k_3$,
we obtain
\begin{multline}
\Delta_{3\Sigma}^{(21)} =
\frac1{L^6} \sum_{\vec k_2\vec k_3} H^{(1)}(\vec k_1) H^{(2)}(\vec k_2) D_{123}
\bigg\{ \sum_{\ell' m' \ell m} \bigg[
X_{k_2\ell' m'} \frac{\cY_{\ell' m'}(\vec k_3^{*(k_2)})}{q_{2,k_2}^{*\ell'}} 
\frac{\cY_{\ell m}(\vec k_3^{*(k_2)})}{q_{2,k_2}^{*\ell}} Z_{k_2\ell m}
\\
-X_{k_1\ell' m'} \frac{\cY_{\ell' m'}(\vec k_2^{*(k_1)})}{q_{2,k_1}^{*\ell'}}
\frac{\cY_{\ell m}(\vec k_2^{*(k_1)})}{q_{2,k_1}^{*\ell}} Z_{k_1\ell m}
\bigg]\bigg\}\,,
\end{multline}
where the order of summations is immaterial.
Again the term in curly braces vanishes on shell, although here this requires
summing over both sets of spherical harmonic indices.
Thus we can replace the double momentum sum by a double integral.
Combining this with the integral parts from the $\wt F^{(i)}$s in 
$\Delta_{3}^{(12)}$ the result can be brought into the form
\begin{align}
\Delta_3^{(21)}  &= 
\left[ \int_{\vec k_2} \!\!\PV\!\! \int_{\vec k_3} - \int_{\vec k_3}\!\!\PV \!\!\int_{\vec k_2}\right]
\sum_{\ell' m'\ell m}
X_{k_2\ell' m'} \frac{\cY_{\ell' m'}(\vec k_3^{*(k_2)})}{q_{2,k_2}^{*\ell'}} 
\frac{H^{(1)}(\vec k_1) H^{(2)}(\vec k_2)}
{2\omega_{k_1} (E-\omega_{k_1}-\omega_{k_2}-\omega_{k_3})}
\frac{\cY_{\ell m}(\vec k_3^{*(k_2)})}{q_{2,k_2}^{*\ell}} Z_{k_2\ell m}\,.
\end{align}
By a similar argument, one can show that $\Delta_3^{(23)}$ is given
by the same expression.
This shows that the integral operator can be written such that
it only depends on the flavor 2. Thus we use the following notation for the final result
\begin{equation}
\Delta_3^{(21)} = \Delta_3^{(23)} \equiv [X]_2 [\wh \cI_{F}]_{22} [Z]_2\,,
\end{equation}
where $\wh \cI_F$ is a diagonal matrix of integral operators.

Finally we consider $\Delta_4$, where we only sketch the argument.
The sum part of $\wt F^{(k)}$ cancels with the summand of $[\wh F_G]_{ij}$ at the
on-shell point, so that, for this combination, the sums can be replaced by integrals.
The pole prescription can be chosen so that the integral over the $\wt F^{(k)}$ summand
cancels the integral part of this quantity, leaving an integral over the $\wt G$ term.
The end result is that
\begin{equation}
\Delta_4 = \sum_{i\ne j}  [X_i]_i [\wh \cI_G]_{ij} [Z_j]_j\,,
\end{equation}
where the integral operator is given, for example, by
\begin{equation}
 [X_1]_1 [\wh \cI_G]_{12} [Z_2]_2 
 =
\int_{\vec k_1}\!\!\PV\!\! \int_{\vec k_2} \sum_{\ell' m'\ell m}
X_{k_1\ell' m'} \frac{\cY_{\ell' m'}(\vec k_2^{*(k_1)})}{q_{2,k_1}^{*\ell'}} 
\frac{H^{(1)}(\vec k_1) H^{(2)}(\vec k_2)}
{2\omega_{k_3} (E-\omega_{k_1}-\omega_{k_2}-\omega_{k_3})}
\frac{\cY_{\ell m}(\vec k_1^{*(k_2)})}{q_{2,k_2}^{*\ell}} (-1)^\ell Z_{k_2\ell m}\,.
\end{equation}
We define the diagonal elements of $\wh \cI_G$ to vanish.

Combining these results we obtain
\begin{equation}
\Delta =  {\vec X} \left(
\wh{\overrightarrow{\cI}}_G + \wh{\overleftarrow{\cI}}_G + 2 \wh \cI_F + \wh \cI_G  
\right) {\vec Z}
\equiv {\vec X} 3\wh \cI_{FG} {\vec Z}\,,
\end{equation}
which completes the demonstration of Eq.~(\ref{eq:symid3a}).

\section{Derivation of Eqs.~(\ref{eq:toshow1}) and (\ref{eq:TLhat}) }
\label{app:alg}

In this appendix we show that the two forms of $\wh\cM'_{\df,3,L}$,
Eqs.~(\ref{eq:Mpdf3Lhat}) and (\ref{eq:toshow1}), are equivalent.
We find it easiest to start from the latter relation and work back to the former.

We will need the following results for $\wh\cD_{23,L}$, 
\begin{align}
i\wh \cD_{23,L} 
%&= \frac1{\wh{\overline{\cK}}_{2,L}^{-1}+ \wh F_G}\,, \\
&= \frac1{1- i \wh{\overline{\cK}}_{2,L}i\wh F_G} i\wh{\overline{\cK}}_{2,L}\,,
\\
&= (1 +i  \wh \cD_{23,L} i\wh F_G) i\wh{\overline{\cK}}_{2,L}\,,
\label{eq:appE3}
\\
&=i\wh{\overline{\cK}}_{2,L} 
+ i\wh{\overline{\cK}}_{2,L}  (1 + i \wh F_G i\wh \cD_{23,L}) i \wh F_G 
i \wh{\overline{\cK}}_{2,L}\,,
\label{eq:appE4}
\end{align}
which follow from Eqs.~(\ref{eq:D23L}) and (\ref{eq:Mhat2L}).

We first note that, using Eq.~(\ref{eq:appE3}),
\begin{align}
\left[1+ i \wh \cD_{23,L} i (\wh F_G - 
\wh{\overrightarrow{\cI}}_G )\right]
&=
\left[1 + i \wh \cD_{23,L} i\wh F_G\right] \wh L_G\,,
\\
\wh L_G &\equiv 
\left[1 -i \wh{\overline{\cK}}_{2,L}  i\wh{\overrightarrow{\cI}}_G \right]
\,,
\end{align}
and, similarly,
\begin{align}
\left[1+ i (\wh F_G -  \wh{\overleftarrow{\cI}}_G) i \wh \cD_{23,L} )\right]
&=
\wh R_G
\left[1 + i\wh F_G i \wh \cD_{23,L}\right] \,,
\\
\wh R_G &\equiv \left[1 - i\wh{\overleftarrow{\cI}}_G i \wh{\overline{\cK}}_{2,L} \right]
\,.
\end{align}
Given these results, Eq.~(\ref{eq:toshow1}) can be rewritten
\begin{equation}
\wh \cM'_{\df,3,L} = \left[1 + i \wh \cD_{23,L} i\wh F_G\right] 
%\left[1 -i \wh{\overline{\cK}}_{2,L}  i\wh{\overrightarrow{I_G}} \right]
\wh L_G \wh \cT_L \wh R_G
%\left[1 - i\wh{\overleftarrow{I_G}} i \wh{\overline{\cK}}_{2,L} \right]
\left[1 + i\wh F_G i \wh \cD_{23,L}\right] \,.
\end{equation}
Comparing to Eq.~(\ref{eq:Mpdf3Lhat}), this implies that what we need to show is that
\begin{equation}
%\left[1 -i \wh{\overline{\cK}}_{2,L}  i\wh{\overrightarrow{I_G}} \right] 
\wh L_Gi \wh \cT_L \wh R_G
%\left[1 - i\wh{\overleftarrow{I_G}} i \wh{\overline{\cK}}_{2,L} \right]
=
i\wh \cK'_{\df,3}
\frac1{ 1 - [1 + i \wh F_{G} i \wh \cD_{23,L}] i \wh F_G i \wh \cK'_{\df,3}}
\,.
\label{eq:desired}
\end{equation}

To proceed, we rewrite the definition of $\wh\cT_L$, Eq.~(\ref{eq:TLhat}), as
\begin{align}
[i\wh \cT_L]^{-1}   
&= [i \wh \cK''_{\df,3}]^{-1}
-i(\wh F_G - \wh \cI_{FG})
- i(\wh F_G-\wh{\overleftarrow{\cI}}_G) i\wh \cD_{23,L}
   i(\wh F_G-\wh{\overrightarrow{\cI}}_G) 
\\
&= [i \wh \cK''_{\df,3}]^{-1}
+ i \wh \cI_{FG} 
- i \wh{\overleftarrow{\cI}}_G i\wh{\overline{\cK}}_{2,L} i \wh{\overrightarrow{\cI}}_G
-
\wh R_G
(1 + i \wh F_G i \wh\cD_{23,L}) i \wh F_G
\wh L_G
\,,
\\
&= [i\wh Z]^{-1} -
\wh R_G
(1 + i \wh F_G i \wh\cD_{23,L}) i \wh F_G
\wh L_G\,,
\end{align}
where to obtain the second line we have used Eqs.~(\ref{eq:appE3}) and (\ref{eq:appE4}),
and to obtain the third line we have used Eq.~(\ref{eq:Kdf3hatpp}).
Thus
\begin{align}
\wh L_G i\wh \cT_L \wh R_G
&= \wh L_G i\wh Z
\frac1{1 - \wh R_G (1+ i \wh F_G i\wh \cD_{23,L}) i \wh F_G \wh L_G i \wh Z} \wh R_G
\\
&= \wh L_G i\wh Z \wh R_G
\frac1{1 - (1+ i \wh F_G i\wh \cD_{23,L}) i \wh F_G \wh L_G i \wh Z \wh R_G}
\,,
\end{align}
which gives the desired result (\ref{eq:desired}) since we can rewrite Eq.~(\ref{eq:Zhat}) as 
$\wh L_G \wh Z \wh R_G = \wh \cK'_{\df,3}$.

\bibliography{ref} %%% ref.bib file

\end{document}